\documentstyle[emulateapj]{article}

\def\hii{\protect{\sc H\thinspace ii}}
\def\hei{\ion{He}{1}}
\def\heii{\ion{He}{2}}
\def\oiii{[\ion{O}{3}]}
\def\oii{[\ion{O}{2}]}
\def\oi{[\ion{O}{1}]}
\def\siii{[\ion{S}{3}]}
\def\sii{[\ion{S}{2}]}
\def\neiii{[\ion{Ne}{3}]}
\def\ariii{[\ion{Ar}{3}]}
\def\etap{$\eta^\prime$}
\def\etal{et\thinspace al.~}
\def\Ha{H$\alpha$}
\def\Hb{H$\beta$}
\def\Hg{H$\gamma$}
\def\Hd{H$\delta$}

\def\hb{\rm H\beta}

\def\Rsol{{\rm\,R_\odot}}
\def\tstar{$T_\star$}
\def\Llyc{L_{\rm Lyc}}
\def\lam{$\lambda$}
\def\cc{{\rm\,cm^{-3}}}
\def\ergs{{\rm\,erg\ s^{-1}}}
\def\kms{{\rm\,km\ s^{-1}}}

\received{5 November 1999}
\accepted{14 December 1999}
\slugcomment{Accepted 14 Dec 1999 to The Astrophysical Journal Supplement Series}

\lefthead{Oey, Dopita, Shields, and Smith}
\righthead{Calibration of Nebular $T_\star$ Diagnostics}

\begin{document}

%\title{{\rm\small *****
%DRAFT version \today\ --- please do not circulate
%***** \\
%For submission to ApJ } \\ \bigskip
\title{Calibration of Nebular Emission-Line Diagnostics: 
	I.  Stellar Effective Temperatures}

\author{M. S. Oey}
\affil{Space Telescope Science Institute, 3700 San Martin Drive,
	Baltimore, MD   21218, USA; oey@stsci.edu}
\author{M. A. Dopita}
\affil{Research School of Astronomy and Astrophysics, Institute of 
	Advanced Study, The Australian National University, \\
	Private Bag, Weston Creek P.O., ACT 2611,
	Australia; mad@mso.anu.edu.au.}
\author{J. C. Shields}
\affil{Ohio University, Dept. of Physics and Astronomy, Clippinger
	Research Labs. 251B, Athens, OH   45701-2979, USA; 
	shields@helios.phy.ohiou.edu}
\author{R. C. Smith}
\affil{Cerro Tololo Inter-American Observatory\altaffilmark{1}, 
	Casilla 603, La Serena, Chile; csmith@noao.edu}
\altaffiltext{1}{National Optical Astronomy Observatories, operated by the
	Association of Universities for Research in Astronomy, Inc.,
	under cooperative agreement with the National Science Foundation}

\begin{abstract}

We present a detailed comparison of optical \hii\ region spectra to
photoionization models based on modern stellar atmosphere models.
We examine both spatially resolved and integrated emission-line
spectra of the \hii\ regions DEM L323, DEM L243, DEM L199, and DEM
L301 in the Large Magellanic Cloud.  The published spectral
classifications of the dominant stars range from O7 to WN3, and 
morphologies range from Str\"omgren sphere to shell structure.  Two of the 
objects include SNR contamination.  The overall agreement with the
predictions is generally within 0.2 dex for major
diagnostic line ratios.  An apparent pattern in the remaining discrepancies
is that the predicted electron temperature is $\sim 1000$ K
hotter than observed.  \neiii\ intensities are also slightly
{\it over}predicted, which may or may not be related.  We model the shock
emission for the SNR-contaminated objects, and find excellent
agreement with the observations for composite shock and photoionized
spectra.  DEM L301's emission apparently results from both shocks and
density-bounded photoionization.  The existence of contaminating
shocks can be difficult to ascertain in the spatially integrated spectra. 

Our analysis of the complex DEM L199 allows a nebular emission-line
test of unprecedented detail for WR atmospheres.  Surprisingly, we
find no nebular \heii\ \lam4686 emission, despite the fact that both of the
dominant WN3 stars should be hot enough to fully ionize \hei\ in their
atmospheres.  The nebular diagnostics are again in excellent agreement
with the data, for stellar models not producing He$^+$-ionizing photons.  The
optical diagnostics are furthermore quite insensitive to the ionizing
energy distribution for these early WR stars.  

We confirm that the \etap\ emission-line parameter is not as useful as
hoped for determining the ionizing stellar effective temperature, \tstar.  Both
empirically and theoretically, we find that it is insensitive for
$T_\star \gtrsim 40$ kK, and that it also varies spatially.  The
shock-contaminated objects show that \etap\ will also yield a
spuriously high \tstar\ in the presence of shocks.  It is furthermore
sensitive to shell morphology.  We suggest
\neiii/\Hb\ as an additional probe of \tstar.  Although it is
abundance-dependent, \neiii/\Hb\ has higher sensitivity to
\tstar, is independent of morphology, and is insensitive to shocks in
our objects.  These observations should be useful data points for a first
empirical calibration of nebular diagnostics of \tstar, which we
attempt for LMC metallicity.

\end{abstract}


\section{Introduction}

Extragalactic \hii\ regions are a vital probe of physical conditions
in distant star-forming regions.  The luminosities and distribution of
\hii\ regions are widely used to infer the magnitude and distribution
of massive star formation.  Likewise, the ratios of prominent emission
lines in nebular spectra are commonly used as diagnostics to infer
properties such as chemical abundances, gas density, electron
temperature, and characteristics of the ionizing stellar population.
The emergent nebular spectrum depends primarily on three parameters:
stellar ionizing effective temperature ($T_\star$), ionization
parameter ($U$), and metal abundance ($Z$).  Hence, we need
adequate constraints on two of these quantities to estimate the third.
In practice, all three parameters are relatively complicated to obtain
accurately, especially in distant extragalactic \hii\ regions.  Nebular
photoionization codes (e.g., reviewed by Ferland {\etal}1995) are therefore 
extensively used to interpret the observed emission-line spectra and
constrain these properties.  Yet, in spite of our heavy dependence on
such codes, there has been little comparison of model \hii\ regions
with spatially resolved, spectrophotometric observations of objects
having clearly determined stellar ionizing sources.  The prototypical
\hii\ region for such comparisons is the Orion nebula (e.g., Baldwin
{\etal}1991; Walter {\etal}1992), and studies of planetary nebulae
(e.g., Liu \& Danziger 1993) have also been useful in this
regard.  However, there have been almost no studies exploring the
parameter space of ordinary \hii\ regions with known stellar ionizing sources.

This omission is due primarily to the difficulty in determining
accurate nebular parameters in Galactic \hii\ regions, which suffer
from large and variable extinction, distance uncertainties, and
line-of-sight confusion.  The problem is compounded by large angular
size, and similar difficulties in obtaining reliable and complete
stellar censuses for these regions.  However, at a distance of
$\sim 50$ kpc, the Large Magellanic Cloud (LMC) offers a great variety of
\hii\ regions that subtend a few arcminutes, are at a well-determined
distance, and have low foreground extinction.  For many of these 
nebulae, the associated massive star population has been examined in
detail and the dominant stars classified with spectroscopic
observations (e.g., Massey {\etal}1995; Oey 1996a).  We have therefore
exploited these conditions to examine the emission-line diagnostics of
four bright, LMC \hii\ regions with respect to their known properties.  
The angular sizes are small enough that we are able to obtain
spatially integrated observations by scanning a long slit across the
objects, thereby reproducing observations of spatially unresolved
\hii\ regions in more distant galaxies.  We have also obtained
unscanned, spatially-resolved observations.  Together, these
data will clarify the dominant effects that are relevant to interpreting
spatially unresolved observations, and test how well photoionization
models reproduce actual nebular conditions.
In this paper, we present the observations and examine the nebular
diagnostic line ratios with respect to the stellar ionizing sources.  A
companion paper (Oey \& Shields, in preparation; Paper~II) will examine the
diagnostics with respect to the chemical abundances.

We mentioned above that \hii\ region emission-line spectra are
determined primarily by the three parameters, $T_\star,\ U$, and $Z$.
In practice, there are complicating factors associated with 
each of these.  In the case of $T_\star$, the main problem is
uncertainty in the actual energy distribution emitted in the Lyman
continuum; we discuss this in greater detail below.  With respect to
$U$, the nebular density, morphology, and geometric relation to
the ionizing stars will significantly affect the ionization parameter
across the nebula.  We will examine this effect in both this work and
in Paper~II.  Finally, complications for $Z$ are primarily ionization
correction factors and variations in elemental ratios.  Issues affecting
the abundance determinations will be addressed primarily in Paper~II.

\section{Observations}

In order to empirically explore the \tstar\ parameter space, we
selected LMC \hii\ regions whose spectral types vary from O7 to WN3.
The sample is summarized in Table~\ref{sample}:  column~1 identifies
the object from the Davies, Elliott, \& Meaburn (1976; DEM) \Ha\
catalog of the LMC; column~2 gives the corresponding identification
from the Henize (1956) catalog; column~3 lists the ionizing  Lucke \&
Hodge (1970) OB association; column~4 identifies the dominant
ionizing stars; column~5 gives the corresponding spectral types;
and column~6 gives the reference for the star ID and spectral types.
Figures~\ref{figd323}$a$ -- \ref{figd301}$a$ show the \Ha\ CCD images of the
sample objects, obtained with the CTIO/U. Michigan 0.6/0.9-m Curtis
Schmidt Telescope as part of the Magellanic Clouds Emission-Line
Survey (Smith {\etal}1998).  Panels $b$ and $c$ of the same figures
show the continuum-subtracted ratio maps of \oiii$\lambda5007$/\Ha\ and
\sii$\lambda6724$/\Ha, with darker regions corresponding to higher
values.  The spatial resolution of the images is 2.3\arcsec\ px$^{-1}$.

\begin{figure*}
\epsscale{2.0}
%\plotone{DEM323.ps}
\vspace*{0.7in}
\caption{\protect\Ha\ image of DEM L323, with the dominant
O3--4 stars identified.  White lines indicate apertures on stationary slit
positions and black lines show the end points of scans for scanned
observations.  North is up and east to the left.  The pixel scale is
marked along the edges, with 2.3\arcsec\ px$^{-1}$.  Panel $b$ shows
continuum-subtracted 
\protect\oiii/\protect\Ha; panel $c$ shows continuum-subtracted \protect\sii/\protect\Ha.  The
dominant stars are marked in each image.
\label{figd323}}
\end{figure*}

\begin{figure*}
\figurenum{1$b$}
\epsscale{2}
%\plotone{d323oiii_ha.ps}
\caption{}
\end{figure*}
\begin{figure*}
\figurenum{1$c$}
\epsscale{2}
%\plotone{d323sii_ha.ps}
\caption{}
\end{figure*}

\begin{figure*}
\epsscale{2.0}
%\plotone{DEM243.ps}
\vspace*{0.7in}
\caption{\protect\Ha\ image of DEM L243, with the dominant O7 stars
identified.  Line types are as in Figure~\ref{figd323}.  The black
circle indicates the approximate boundary of the supernova blastwave
seen in X-ray and radio continuum.  Panel $b$ shows continuum-subtracted
\protect\oiii/\protect\Ha; panel $c$ shows continuum-subtracted \protect\sii/\protect\Ha.
\label{figd243}}
\end{figure*}

\begin{figure*}
\figurenum{2$b$}
\epsscale{2}
%\plotone{d243oiii_ha.ps}
\caption{}
\end{figure*}
\begin{figure*}
\figurenum{2$c$}
\epsscale{2}
%\plotone{d243sii_ha.ps}
\caption{}
\end{figure*}

\begin{figure*}
\epsscale{2.0}
%\plotone{DEM199.ps}
\vspace*{0.7in}
\caption{\protect\Ha\ image of DEM L199, with the dominant WR stars and O3--4
star identified (see Table~\ref{sample}).  Line types are as in
Figure~\ref{figd323}.  Panel $b$ 
shows continuum-subtracted \protect\oiii/\protect\Ha; panel $c$ shows
continuum-subtracted \protect\sii/\protect\Ha. 
\label{figd199}}
\end{figure*}

\begin{figure*}
\figurenum{3$b$}
\epsscale{2}
%\plotone{d199oiii_ha.ps}
\caption{}
\end{figure*}
\begin{figure*}
\figurenum{3$c$}
\epsscale{2}
%\plotone{d199sii_ha.ps}
\caption{}
\end{figure*}

\begin{figure*}
\epsscale{2.0}
%\plotone{DEM301.ps}
\vspace*{0.7in}
\caption{\protect\Ha\ image of DEM L301, with the dominant O3--5 stars
identified.  Line types are as in Figure~\ref{figd323}. 
Panel $b$ shows continuum-subtracted
\protect\oiii/\protect\Ha; panel $c$ shows continuum-subtracted \protect\sii/\protect\Ha.
\label{figd301}}
\end{figure*}

\begin{figure*}
\figurenum{4$b$}
\epsscale{2}
%\plotone{d301oiii_ha.ps}
\caption{}
\end{figure*}
\begin{figure*}
\figurenum{4$c$}
\epsscale{2}
%\plotone{d301sii_ha.ps}
\caption{}
\end{figure*}

As can be seen in the figures, the sample
also includes a range of nebular morphology:  DEM L323
(Figure~\ref{figd323}) has beautiful, spherical morphology, and can be
seen to be emerging from a neutral cocoon; DEM L243
(Figure~\ref{figd243}) shows a slightly more irregular morphology, but
with no apparent shells or filaments; DEM L199
(Figure~\ref{figd199}) has a complex morphology
that is a composite of large blobs, filaments, and shell structure; and
DEM L301 (Figure~\ref{figd301}) shows a strongly filamentary shell
structure.  To constrain the role of morphology, we chose DEM L323 for its
``classical'', spherical, Str\"omgren sphere morphology, in contrast
to the extreme shell structure of DEM L301, while maintaining the same
O3 spectral type for the ionizing stars.

Embedded or superimposed supernova remnants (SNRs) can strongly affect the
emission-line properties in a predominantly photoionized \hii\
region (e.g., Peimbert, Sarmiento, \& Fierro 1991).  Since the massive
stars that ionize these nebulae expire as core-collapse supernovae, we
do expect such associated SNRs to occasionally be found in \hii\
regions.  DEM L243 encompasses a well-known SNR (e.g., Shull 1983;
Westerlund \& Mathewson 1966), whose spatial area 
is delineated by the black circle in
Figure~\ref{figd243}.  This nebula will be used to study
the effect of an SNR on the emission-line properties of the
surrounding \hii\ region.  In addition, it is likely that DEM L301
also has had recent SNR activity (Chu \& Mac Low 1990; Oey 1996b).

We obtained spectroscopic observations of the \hii\ regions in the
sample using the 2.3-m telescope at Siding Spring Observatory of the
Australian National University.  The data were obtained during the
nights of 1997 January 2 -- 4.  The Double-Beam Spectrograph permitted
simultaneous observations of the blue and red spectral regions,
providing a total wavelength coverage over the range 3500 -- 9200\AA.
The dispersing element was a 300 line mm$^{-1}$ grating in the blue,
and a 316 line mm$^{-1}$ grating in the red, with the beam split by a
dichroic centered at 6200\AA.  We recorded the data with two SITE 1752
$\times$ 532 CCDs, whose 15$\mu$ pixels yield a spatial resolution of
0.91$\arcsec\ \rm px^{-1}$, and spectral resolution of about 5.5\AA.
The long slit length of $6.^\prime 4$ was well-suited to sizes of the
targets, which subtend several arcminutes in diameter.

Our standard stars for flux calibration were EG~21 and LTT~4364
(Stone \& Baldwin 1983; Hamuy {\etal}1992); observing conditions were
not strictly photometric.  Total exposure times ranged between one and
two hours, depending on target surface brightness, with individual
frames exposed for 900 -- 1200 s.  We observed all objects in the
sample with at least two stationary slit positions, and three of the
four targets were also observed by scanning the slit across the
nebula.  In order to avoid irregularities in the scan rate, the rate
was set such that the desired region was scanned in a single pass.
The data were reduced and extracted with {\sc Iraf}\footnote 
{{\sc iraf} is distributed by NOAO, which is operated by AURA,
Inc., under cooperative agreement with the National Science Foundation.}
using standard longslit reduction techniques.  Figures~\ref{figd323}$a$
-- \ref{figd301}$a$ show the slit positions of the observations;
stationary positions are shown in white, and the endpoints of scanned
positions are shown in black.  The individual extracted apertures are
indicated with their ID numbers.
The scanned observations are integrated over the slit length
where the nebular emission is detected, to produce a spectrum
integrated over most of the nebular area, as described in \S~4.
All apertures extracted from the same slit observation had the
background level determined from the same region.  This corresponds
to the lowest emission level in the aperture, and we caution that
there is likely to be very low-level, diffuse nebular emission over
these regions as well.  This increases the uncertainty in background
subtraction, but disproportionately affects the low ionization
species.

For each extracted spectrum, we derived reddening corrections from
the Balmer decrements \Ha/\Hb, \Hg/\Hb, and \Hd/\Hb, and obtained a
weighted mean correction $c$, where
\begin{equation}
\frac{I(\lambda)}{I({\hb})} = \frac{I_0(\lambda)}{I_0({\hb})}\ 
	10^{-c(f_\lambda - f_{\hb})} \quad .
\end{equation}
The ratios $\frac{I(\lambda)}{I({\hb})}$ and $\frac{I_0(\lambda)}{I_0({\hb})}$
are the observed and intrinsic line intensities with respect to \Hb,
and $f_\lambda$ and $f_{\hb}$ describe the reddening law, for which we
adopt that of Savage \& Mathis (1979).  We then assigned three or four
reddening zones for each object, based on the range of derived
values.  The adopted $c$ for each aperture is then taken as that of
the zone with the closest reddening value.  These adopted reddenings
are listed in Table~\ref{lines} with the dereddened line intensities.  
Our adoption of the Galactic reddening law 
rather than LMC should not introduce significant errors; we note that
an error of 0.05 in $c$ corresponds to 4\%, or $< 0.02$ dex difference
across the interval \Hb\ to \Ha.

We examined the possibility of contamination in the Balmer lines by
absorption in underlying stellar continuum.  This effect should be
unimportant in the stationary slit positions, since the \hii\ regions
and stellar populations are fully resolved.  However, the scanned
observations include many stars.  We treated these exactly as is done
with observations of unresolved distant nebulae, and solved
simultaneously for the reddening and a single absorption equivalent
width (EW).  The only observations that showed significant
contamination by absorption were the scanned observations for DEM~L243
and DEM~L199.  For DEM~L199, we found an absorption EW of 2.4 \AA, and
for DEM L243 we found 1.1 \AA.  As described below, we extracted an
additional scanned observation for DEM L243 which excludes the
SNR-contaminated region; for this aperture we found an absorption EW
of 1.4 \AA.

Table~\ref{lines} lists the reddening-corrected line intensities with
respect to \Hb.  We estimate that uncertainties in background
subtraction, reddening correction, and flux calibration are typically
around 5\%; the listed errors in Table~\ref{lines} are obtained from $\surd
\overline{0.05^2 + (\rm S/N)^{-2}}$, where the S/N is that of the given
line.  Note that for the individual 
lines, \Hb\ is simply used as a scaling factor, therefore the errors
do not represent the error in the line ratio with respect to \Hb, but
are uncertainties for the individual line intensities only.  However,
we also list the diagnostic quantities $R23$
%\equiv$ ([\ion{O}{2}]$\lambda 3727$ +
%[\ion{O}{3}]$\lambda\lambda$4959,5007)/\Hb\
(Pagel {\etal}1979), $S23$
% \equiv$ ([\ion{S}{2}]$\lambda\lambda$6717,6732 + 
%[\ion{S}{3}]$\lambda\lambda$9069,9532)/\Hb\ 
(Christensen, Petersen, \& Gammelgaard 1997), and $\eta^\prime$
%\equiv$ ([\ion{O}{2}]$\lambda$3727/
%[\ion{O}{3}]$\lambda\lambda$4959,5007)/([\ion{S}{2}]$\lambda\lambda$6717,6732/
%[\ion{S}{3}]$\lambda\lambda$9069,9532)
(V\'\i lchez \& Pagel 1988) given by equations~\ref{etap} --
\ref{eqS23} in \S 3.1.2 below;
the listed errors for these do represent those of the given line ratio. 
% NOTE THE ERRS ARE NOT COMPUTED DIRECTLY...15% FUDGE

We take the intensity of [\ion{S}{3}]$\lambda$9532 to be
2.5$\times$ [\ion{S}{3}]$\lambda$9069, as set by the relative
transition probabilities (Osterbrock 1989).  It is important to note
that both of these near-IR lines are susceptible to absorption
by telluric water vapor.  The wavelength and magnitude of this
absorption varies depending on the earth's motion relative to the
target.  Stevenson (1994) emphasizes the insidiousness of the
problem, especially because the water vapor lines are usually
unresolved by typical instrumental resolution.  Therefore extreme care
should be exercised in applying correction techniques such as the
commonly-used procedure of dividing the target by a reference star.
Stevenson also shows that it is unclear that \lam 9532 should always
be expected to be more strongly absorbed than \lam 9069, as is
often assumed.  Fortunately, our observations of \lam 9069 appear
to be clear of the telluric absorption, but this issue is worth
bearing in mind.  As will be apparent below, the diagnostic line
ratios involving [\ion{S}{3}] do not show any evidence of errors
resulting from telluric absorption.

\section{Spatial comparison to models}

The spatially resolved, stationary slit observations permit a detailed
comparison with photoionization models of \hii\ regions.  How well do
these idealized models reproduce the spectra of real objects, and how
accurate is the current generation of stellar atmosphere models for
hot stars?  We now compare various diagnostic line ratios predicted by
the photoionization code {\sc Mappings~II} (Sutherland \& Dopita 1993)
with the observed line ratios across the nebulae.  For the O stars,
our default stellar atmosphere models are the ``CoStar'' models of
Schaerer \& de Koter (1997).  These represent the first models
that combine stellar structure and evolution models with atmosphere
models.  The emergent energy distributions account for non-LTE
radiation transfer, line blanketing, and wind blanketing.  For the
earliest spectral types, the CoStar models are somewhat harder than
previous stellar atmosphere models, since all of the effects mentioned
tend to obstruct radiative cooling.   Schaerer \& de Koter (1997) do not
provide models at LMC metallicity, but the variation from
Galactic to SMC abundances yields variations in the energy
distribution of typically less than 0.1 dex, in the sense that the
lower metallicity models are slightly hotter.  We adopted the
models for SMC metallicity, but note that the results are not
sensitive to this variation in metallicity.  

\begin{figure*}
\epsscale{1.7}
\plotone{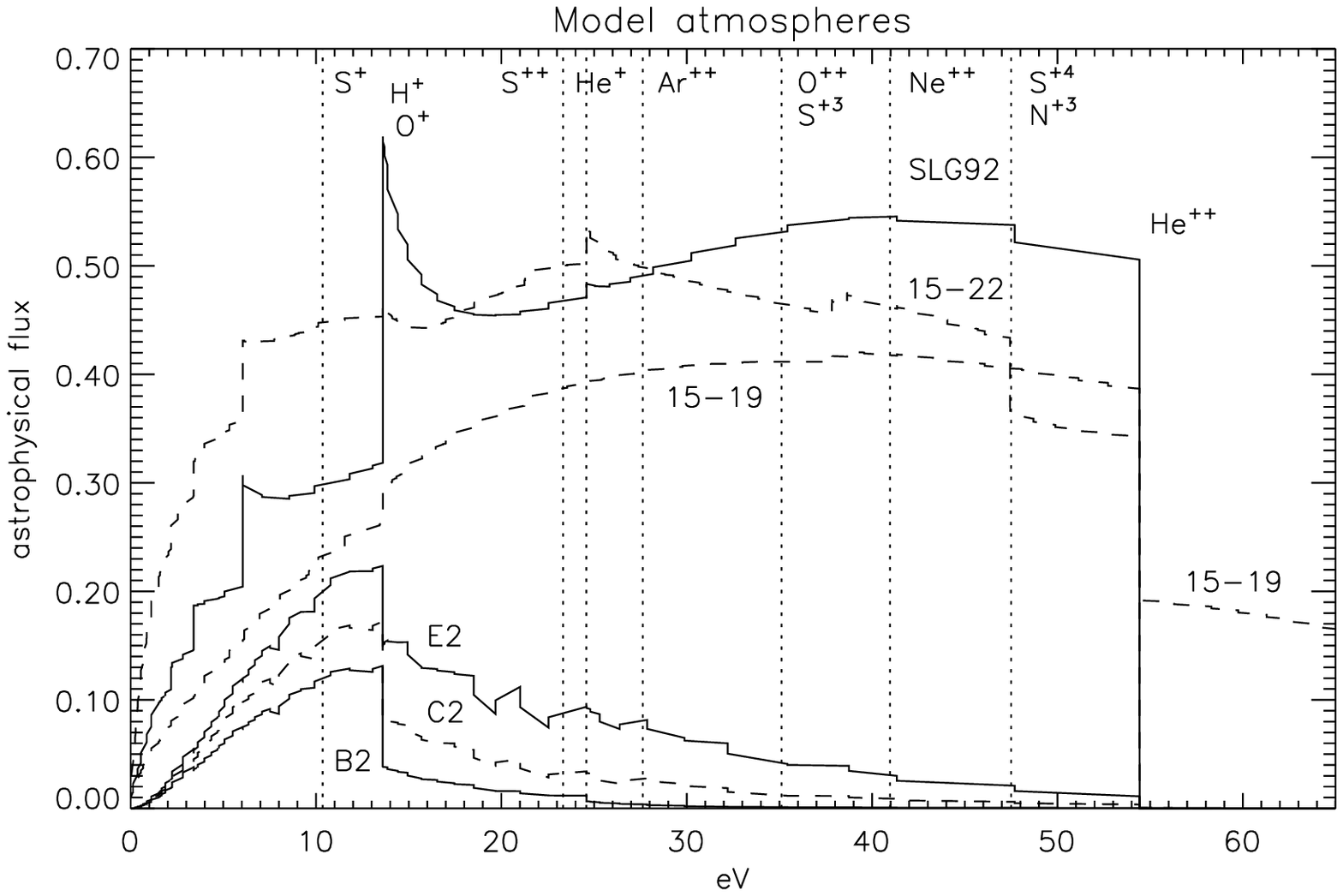}
\vspace*{-3.7in}
\caption{Examples of energy distributions from model stellar
atmospheres in $\rm \ergs\ cm^{-2}\ Hz^{-1}$ (arbitrary scaling).  
The bottom three are CoStar models
B2, C2, and E2 (Schaerer \& de Koter 1997), corresponding to
$T_\star\sim36$, 42, and 49 kK; the top solid line shows a WR model for
$\log T_\star/{\rm K} = 5.1$ by Schmutz {\etal}(1992); and the top two
dashed lines show WR models for the same \tstar\ by Hamann \& 
Koesterke (1998), with model \#15-19 producing He$^+$-ionizing
emission, and \#15-22 without.  See \S 3.3.1 for discussion of the WR models.
We also indicate the ionization potentials required to produce
the named ions.  
\label{atm}}
\end{figure*}

In Figure~\ref{atm}, we show the CoStar model atmospheres B2, C2, and
E2, which correspond roughly to dwarf spectral types O8--O9, O6--O7, and
O3--O4; or stellar effective temperature \tstar\ of 36, 42, and 49 kK, 
respectively.  The correspondence and accuracy of the spectral types
and \tstar\ is subject to uncertainties in the calibration of these
parameters.  For \tstar\, the uncertainty is perhaps $\pm 2$ kK (e.g.,
Schaerer \& de Koter 1997).
DEM~L199 is ionized by early Wolf-Rayet (WR) stars, so we also show
model WR atmospheres from Schmutz {\etal}(1992) and Hamann \&
Koesterke (1998).  These will be discussed in detail in \S~3.3 below.
%  The atmospheres are not shown to scale. **
The vertical, dotted lines indicate the ionization potentials (IP) required
to produce the given ions; these nebular species are thus useful
probes of the ionizing spectrum.  The ionization edges of H$^+$
(13.60 eV) and He$^{++}$ (54.42 eV) are clearly apparent.

\subsection{DEM L323:  O3--4}

We first examine DEM~L323 in detail, which we will take as a reference
object on account of its simple morphology and high S/N observations
obtained for it.
This \hii\ region has the most uniform, Str\"omgren sphere structure
in the sample, and can be seen in Figure~\ref{figd323} to be emerging
from a cocoon of presumably neutral material.  The ratio of
[\ion{S}{2}]\lam6716/\lam6731 suggests electron densities $n$ of a few
10's $\cc$ in all apertures.  The parent OB association is LH~117,
whose members have 
been classified by Massey {\etal}(1989).  We use their identifications
for the three stars, \#118, \#140, and \#214, that dominate the
nebular ionization.  These stars are all of spectral type O3--O4
(Table~\ref{sample}), and their positions are indicated in 
Figure~\ref{figd323}.  Two of the dominant stars are close to the
center in projection, while star \#214 is closer to the edge.

We have two stationary slit positions for DEM~L323, both of which
bisect the \hii\ region, perpendicular to each other
(Figure~\ref{figd323}).  Figure~\ref{d323dgn1} shows a series of
diagnostic line ratios from the apertures in these stationary
observations, as a function of position in arcsec, across the nebula.
The circular points show the data from position D323.C1, and the
crosses are from position D323.C2.  The y-axis error bars are the
formal errors for the line ratio derived from the uncertainties listed
in Table~\ref{lines}.  The lighter, x-axis line segments for each
point represent the spatial extent of each aperture; note that they do
not indicate positional uncertainties.  Since both slit positions pass
through the center of the object, these plots should be representative
cross-sections, projected through the line of sight.  

\begin{figure*}
%\epsscale{1.7}
\plotone{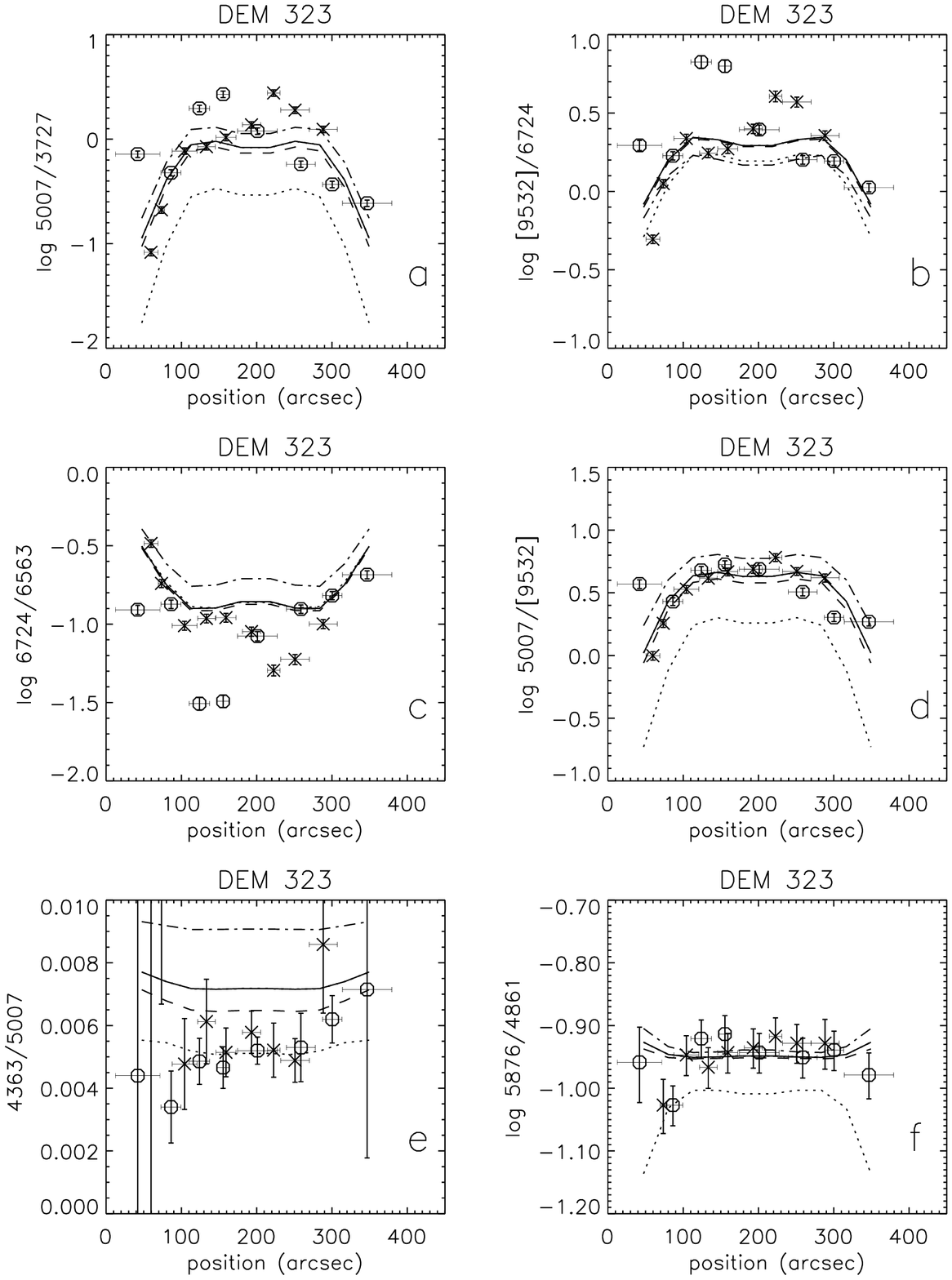}
%\vspace*{-3.7in}
\caption{Projected spatial variation of diagnostic line ratios across DEM~L323.
The data points for position D323.C1 (circles) and D323.C2 (crosses)
are superimposed, aligned at their intersection
(Figure~\ref{figd323}).  The light, horizontal segments show the
spatial extent of the aperture associated with each data point, while
the vertical segments are error bars on the data.  Dotted, dashed, and
solid lines show photoionization models using CoStar atmospheres B2,
C2, and E2, respectively; the dot-dashed model uses the WR atmosphere
of SLG92.  \protect\siii\protect\lam9532 is denoted in brackets to indicate that it is
derived from the observed value of \protect\lam9069 (see text).
\label{d323dgn1}}
\end{figure*}

\begin{figure*}
\figurenum{\ref{d323dgn1}}
%\epsscale{1.7}
\plotone{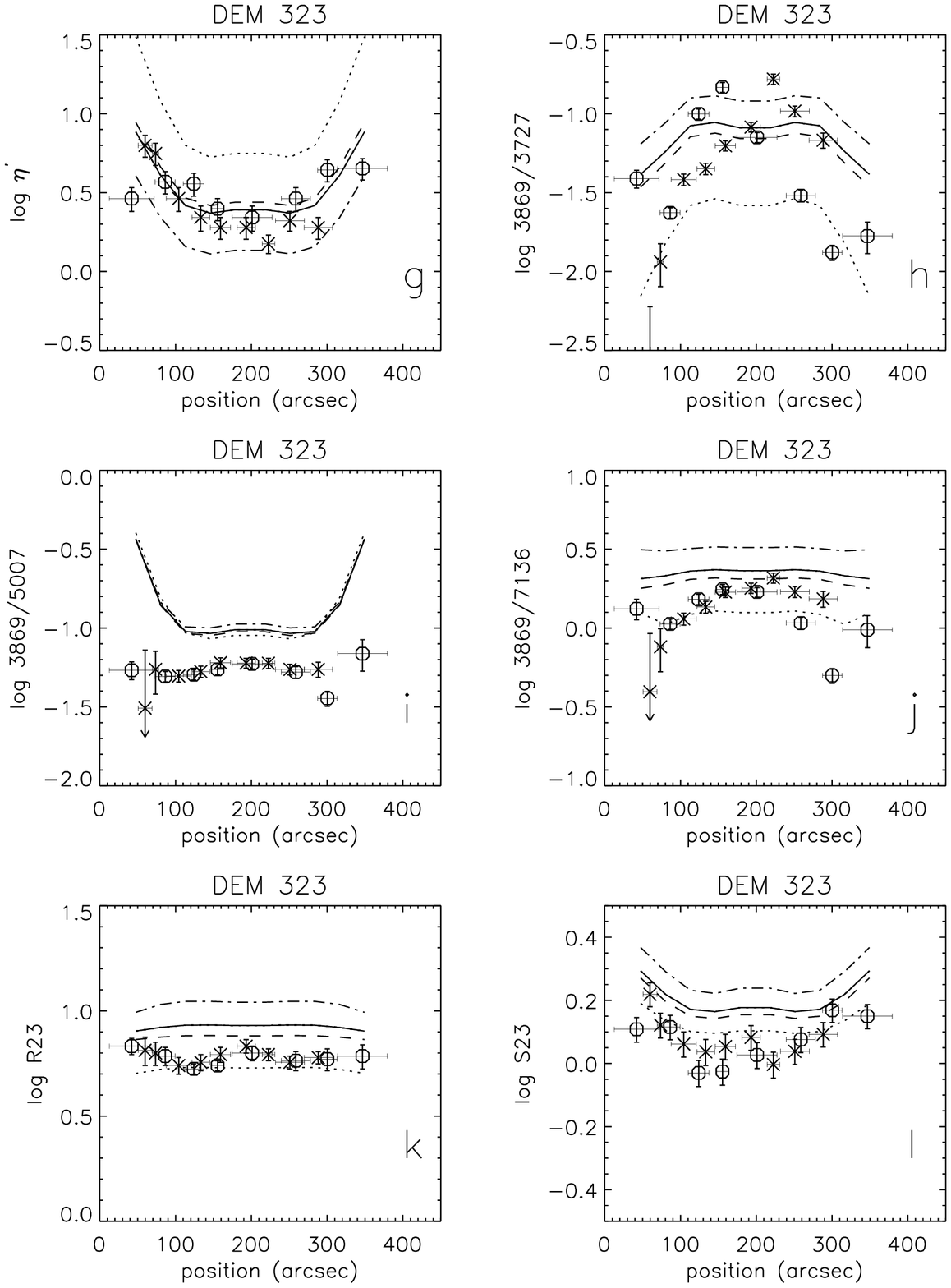}
\caption{{\it continued}}
\end{figure*}

The superimposed lines in Figure~\ref{d323dgn1} show the {\sc
Mappings} model predictions for different stellar atmospheres.  The
nebular models are divided into five radial zones, as seen in
projection by the observer, so we are able to predict the observed
cross-section of a spherically symmetric \hii\ region.  
For most of our models, we take $n=10\ \cc$ as the default
electron density, with a filling factor $f= 0.1$.  The elemental
abundances are weighted averages from the observations, as
reported for this object in Paper~II, namely, $\log X$/H = ($-1.09,
-5.15, -3.63, -4.52, -5.34, -5.93$) for (He, N, O, Ne, S, and Ar).
We adopt the values for $\log$ C/H and $\log$ Si/H of --4.1 and --5.3 from
Garnett (1999), that were measured from \hii\ regions in the LMC.  For
the refractory elements Mg and Fe, we adopted the solar values of
Anders \& Grevesse (1989), reduced by a factor of --1.6 dex to account
for LMC metallicity and grain depletion.  This yields $\log$ Mg/H and
$\log$ Fe/H of --6.0 and --5.9, respectively.  Since 
the dominant stars have essentially the same spectral type, we 
use a single atmosphere model, scaled to the ionizing luminosity
$\log\Llyc/\ergs = 38.7$, implied by the observed \Ha\ luminosity
reported by Oey \& Kennicutt (1997).  In that work, the nebular
luminosity was found to be consistent with the expected stellar ionizing  
fluxes for this object.  Figure~\ref{d323dgn1} presents models generated
with the CoStar atmospheres shown in Figure~\ref{atm}, and we also
show one model with a WR atmosphere, using that of Schmutz {\etal}(1992).
For the observed O3--4 spectral types present, we would therefore
predict the model generated with CoStar atmosphere E2 to most closely
reproduce the observed ionization structure.  The \Ha\
profiles suggest a central cavity of perhaps 40\% the Str\"omgren
radius, which we adopt in the models.  

It is important to bear in mind that the photoionization models
themselves are subject to uncertainty.  We have compared the
{\sc Mappings} results to the Lexington benchmarks (Ferland
{\etal}1995) and single-star photoionization models by Stasi\'nska \&
Schaerer (1997).  For the two blackbody models of the Lexington
benchmarks, at solar metallicity ($Z_\odot$), we find excellent
agreement.  The Stasi\'nska \& Schaerer models use the CoStar
atmospheres, and we find generally good agreement for both $Z_\odot$
and $0.25 Z_\odot$ models, including element ionization fractions, 
electron temperatures, and line strengths.  However, there are slight
discrepancies in the sense that the {\sc Mappings} models produce
somewhat lower optical line strengths in, e.g., \oiii, \oii, \siii,
and \sii.  For atmospheres used in this study, the differences in the
models could cause up to $\sim0.2$ dex variation in the line ratio
predictions. 

\subsubsection{U-sensitive diagnostics}

In general, the E2 model (solid line) does a great job
of reproducing the observed diagnostics and their spatial dependence on
position across the nebula.  O and S are the principal elements with
strong emission lines of more than one ion that are readily observable
in this wavelength range.  The ratios of \oiii\lam5007/\oii\lam3727
and \siii\lam9532/\sii\lam6724 (Figure~\ref{d323dgn1}$a$ and $b$) 
agree well with the E2 model and are within the intrinsic scatter 
of the different 
apertures.  (The densities of our objects are sufficiently low that
\lam6717 is largely unaffected by collisional de-excitation; we
therefore use the sum of the two red S lines, \lam6716 +
\lam6732, designating this as \lam6724, 
analogous to the unresolved components of \lam3727.)  These ratios are
primarily indicators of the ionization parameter, which can be globally
defined for an \hii\ region as,
\begin{equation}\label{U}
U = Q/(4\pi R_{\rm S}^2 n c) \propto (n f^2 Q)^{1/3}\quad, 
\end{equation}
where $Q$ is the ionizing photon emission rate, and $R_{\rm S}$ is the
Str\"omgren radius.  The concept of a local $U$ is commonly described as
the relative photon to gas density, but it is helpful to bear in mind
that the global $U$ increases with both these quantities, as is apparent 
from the right side of equation~\ref{U}.  The data show a large range
in excitation across the nebula, corresponding to a large range in the
local $U$, which is reproduced well by the models.  The models are almost
degenerate for the \siii\ excitation; this results from the lower
ionization potentials of the S ions, which are therefore less
sensitive to high \tstar.  For the hottest model, the WR atmosphere
(dot-dashed line),
there is actually a decrease in \siii/\sii\ since the ionization
balance is redistributed to S$^{+4}$ at IP = 47.30 eV.

We can see a distinct difference in the degree of intrinsic scatter for line
ratios that include the lower ionization species, \oii\ and \sii\, as
compared to higher ionization species.  As contrasted between
Figure~\ref{d323dgn1}$a-c,h$ and Figure~\ref{d323dgn1}$d,i,j$, the
scatter in line intensities about the mean 
is about 0.3 dex for \oii\ and \sii\ and is $\lesssim 0.1$ dex for the
higher ionization lines.  This difference is apparent in the data for
all our objects.  The lower ionization species dominate in the outer
regions of the nebulae (e.g., Figure~\ref{figd323}$c$), where perhaps
there may be more structural and 
density variations, simply as a matter of greater surface area.  The
line intensities originating in this zone are sensitive to such
variations, especially when compared in a single line of sight to
emission from the inner, higher ionization zone.  Our data show that
individual, spatially resolved line ratios involving lower ionization
species should be interpreted with great care, since the large-scale
ionization structure has intrinsic order-of-magnitude variation 
and also a substantial scatter superimposed.  This caveat would apply, for
example, to the use of \neiii/\oii\ as an indicator of the He
ionization correction factor, as suggested by Ali {\etal}(1991).

Historically, predictions for the \siii/\sii\ ratio have sometimes been
greater than observed values (e.g., Garnett 1989;
Dinerstein \& Shields 1986).  Various causes have been suggested, such
as problems with assumed photoionization cross sections and charge
exchange reactions.  The dielectronic recombination rates could be
suspect (Ali {\etal}1991).  Figure~\ref{d323dgn1}$c$ and $d$ show our data
for \sii\lam6724/\Ha\ and \oiii\lam5007/\siii\lam9532, which again
show good agreement between observations and predictions for model
E2.  These results are reassuring, and imply that there is in fact no
theoretical discrepancy.  Furthermore, there is no suggestion of the \siii\
intensities being attenuated by telluric absorption.

\subsubsection{\tstar-sensitive diagnostics}

We now examine some line ratios that are more sensitive to
\tstar.  The E2 and C2 atmospheres yield almost identical models
and resulting diagnostic line ratios, which results from the fact that
their energy distributions are quite similar (Figure~\ref{atm}).
The implication is that it will be 
difficult to distinguish spectral types hotter than about 40 kK.
One of our objectives is to identify diagnostics that can discern
between different stellar temperatures in this regime.
Figure~\ref{d323dgn1}$f$ shows the recombination line ratio
\hei\ \lam5876/\Hb.  The models are degenerate for the hottest
atmospheres since the \hei\ Str\"omgren sphere is fully ionized, but
this is not the case for the coolest model, B2.  As we shall see below
in Figure~\ref{d199dgn1}$f$, \lam5876/\Hb\ also decreases when He is
ionized to He$^{++}$.

In Figure~\ref{d323dgn1}$e$ we show the \oiii\ ratio \lam4363/\lam5007,
which is the well-known indicator of electron temperature $T_e$.
Atmosphere E2 results in a model that overpredicts this line ratio
and produces a model nebula that is about 850 K hotter than indicated
by the data.  The variation between actual $T_e({\rm
O^{++}})$ and $T_e$([\oiii]) is negligible, $\ll 1$\%.
A similar discrepancy in $T_e$ is seen in most of 
our objects.  We confirmed that this result is not peculiar to {\sc
Mappings} by running a similar model using the {\sc Cloudy} code (e.g.,
Ferland {\etal}1998), yielding a similarly high $T_e$.  Since the
other line intensities and abundances (Paper~II) are generally in good
agreement with the models, the discrepancy in $T_e$ is puzzling.
Restoring the abundances of refractory elements to undepleted values
does not resolve the problem.  We also note that this discrepancy is in the
{\it opposite} sense of that expected from the well-known effect of
spatial temperature fluctuations (Peimbert 1967).  In the presence of
fluctuations, the characteristic measured $T_e$ should be higher than
predictions, since the line emissivity favors the hotter regions.
However, in our case the observed $T_e$ is lower than the prediction.
We will discuss the nebular temperature structure in more detail in Paper~II.

A quantity that is widely used as an observable diagnostic of \tstar\
is the ``radiation softness parameter'' (V\'\i lchez \& Pagel 1988),
\begin{equation}\label{etap}
\eta^\prime \equiv \rm \frac{[O\thinspace II]\lambda3727/
	[O\thinspace III]\lambda\lambda4959,5007}
	{[S\thinspace II]\lambda6724/
	[S\thinspace III]\lambda\lambda9069,9532}
 \quad .
\end{equation}
Given the excellent agreement between model E2 and the observations
in Figure~\ref{d323dgn1}$a-d$, it is no surprise that
Figure~\ref{d323dgn1}$g$ also shows a beautiful concurrence.  It is
disappointing, however, that the models for E2 and C2 are virtually
indistinguishable so that \etap\ can no longer discriminate between
O7 and O3 stars.  As we show below, this was not the case for earlier
generations of model atmospheres.  Another potential problem with
\etap\ is that it is fairly sensitive to $U$, as is apparent in the
variation by about 0.6 dex across the \hii\ region.  Thus, extreme
care should be used in estimating \tstar\ from \etap\ for spatially
resolved objects.
% THIS CONTRADICTS EARLIER WORK...

The line \neiii\lam3869 is of particular interest because it requires
the highest IP (40.96 eV) below that of \heii, while producing readily
observable optical emission.  It is thus an important probe for hot
stellar atmospheres.  Historically, photoionization models
have tended to underpredict \neiii\ emission (e.g., Mathis 1985;
Simpson {\etal}1986; Garnett {\etal}1997), and it was unclear whether
this ``\neiii\ 
problem'' was caused by shortcomings in the stellar atmosphere models, 
incorrect Ne abundances resulting from incorrect ionization correction
factors, or inaccuracies in the atomic data for Ne.  The stellar
atmospheres have been considered to be the most likely problem, and
Sellmaier {\etal}(1996) have recently shown convincing evidence that 
this is the case.  Their model atmospheres that include NLTE effects
and explicit, comprehensive line-blocking and wind effects produce
harder energy distributions that are able to reproduce the observed
\neiii\ intensities. 

The CoStar models do not include explicit treatment of millions of
metal lines, although they do incorporate an approximation to account
for line blanketing.  Figure~\ref{d323dgn1}$h$ and $i$ show the nebular
\break \neiii\lam3869/\oii\lam3727 and \neiii\lam3869/\oiii\lam5007.  Given
the large intrinsic scatter in \neiii/\oii, model E2 yields excellent agreement
with the observations.  However, for \neiii/\oiii, the models are
actually {\it over}predicting the observations.  There is only a short
baseline (5.85 eV; Figure~\ref{atm}) in IP for the two ions, so the
discrepancy probably corresponds to small-scale features in the
stellar energy distribution, or the depth of the Ne$^{++}$ edge.  In
Figure~\ref{d323dgn1}$j$ we show \neiii\lam3869/\ariii\lam7136, which
probes a baseline in the stellar spectrum intermediate to 
those of the O$^+$ and O$^{++}$ ions.  The \neiii/\ariii\ ratio also
shows a slight overprediction for the E2 atmosphere.
%  We note that
%this line ratio does not exhibit the degree of scatter seen in 
%\neiii/\oii, and its predicted range with nebular radius does
%not vary as much as that for \etap, for example.  The \neiii/\ariii\
%ratio may hold promise as a useful indicator of \tstar.
To be sure, it is possible to obtain better agreement with the models by
adjusting the Ne abundance downward by 0.2 -- 0.3 dex.  Our adopted
value of log Ne/H = --4.52 was derived from the standard assumption
that Ne$^{++}$/O$^{++}$ is constant (e.g., Peimbert \& Costero 1969;
see Paper~II), and Simpson {\etal}(1995) have shown that this
assumption appears robust for Galactic objects observed in the mid-IR
transitions of Ne$^+$ and Ne$^{++}$.  It would therefore appear that
the overestimate of \neiii\ intensities is still most likely caused by
the stellar atmosphere models.  We also note that reducing the Ne
abundance has a negligible effect in resolving the overestimated $T_e$
in the models.

Another widely-used parameter is (Pagel {\etal}1979),
\begin{equation}
R23 \equiv \rm\frac{[O\thinspace II]\lambda3727 + 
	[O\thinspace III]\lambda\lambda4959,5007}{\hb} \quad .
\end{equation}
This line ratio is normally used as an indicator of O abundance, but
it also varies with \tstar.  It is therefore important to
examine its behavior with \tstar\ to constrain abundances, and this has
been discussed in the literature by many authors (e.g., McGaugh
1991; McCall {\etal}1985; Oey \& Kennicutt 1993).
Figure~\ref{d323dgn1}$k$ shows $R23$ across the nebula, along with the
models for different \tstar.  We indeed see a strong differentiation
between the different atmospheres, and the data are in reasonable
agreement with model E2, to within $\sim0.1$ dex.   

A similar parameter has recently been introduced for S (Christensen
{\etal}1997; D\'\i az \& P\'erez-Montero 1999):
\begin{equation}\label{eqS23}
S23 \equiv \rm\frac{[S\thinspace II]\lambda6724 + 
	[S\thinspace III]\lambda\lambda9069,9532}{\hb} \quad .
\end{equation}
Figure~\ref{d323dgn1}$l$ shows the behavior of $S23$ across the
nebula.  We again see reasonable agreement between the data and model
E2, noting that there is spatial variation in this parameter.
This is related to a non-negligible ionization fraction of S$^{+3}$,
whereas O$^{+3}$ will not be present in our objects, as evidenced by
the lack of \heii\ \lam4686.  The models for $S23$ also show a
somewhat smaller range of mean values, perhaps $\sim 0.15$ dex, in
contrast to the range in mean $R23$ of $\sim 0.4$ dex, for the stellar
atmospheres used.  For high \tstar, $S23$ will therefore be less
sensitive to the stellar population.  A more complete discussion of the
abundance applications of these parameters will be given in Paper~II.

\subsubsection{Comparison to Hummer \& Mihalas atmospheres}

\begin{figure*}
%\plotone{d323dgn2.ps}
\plotfiddle{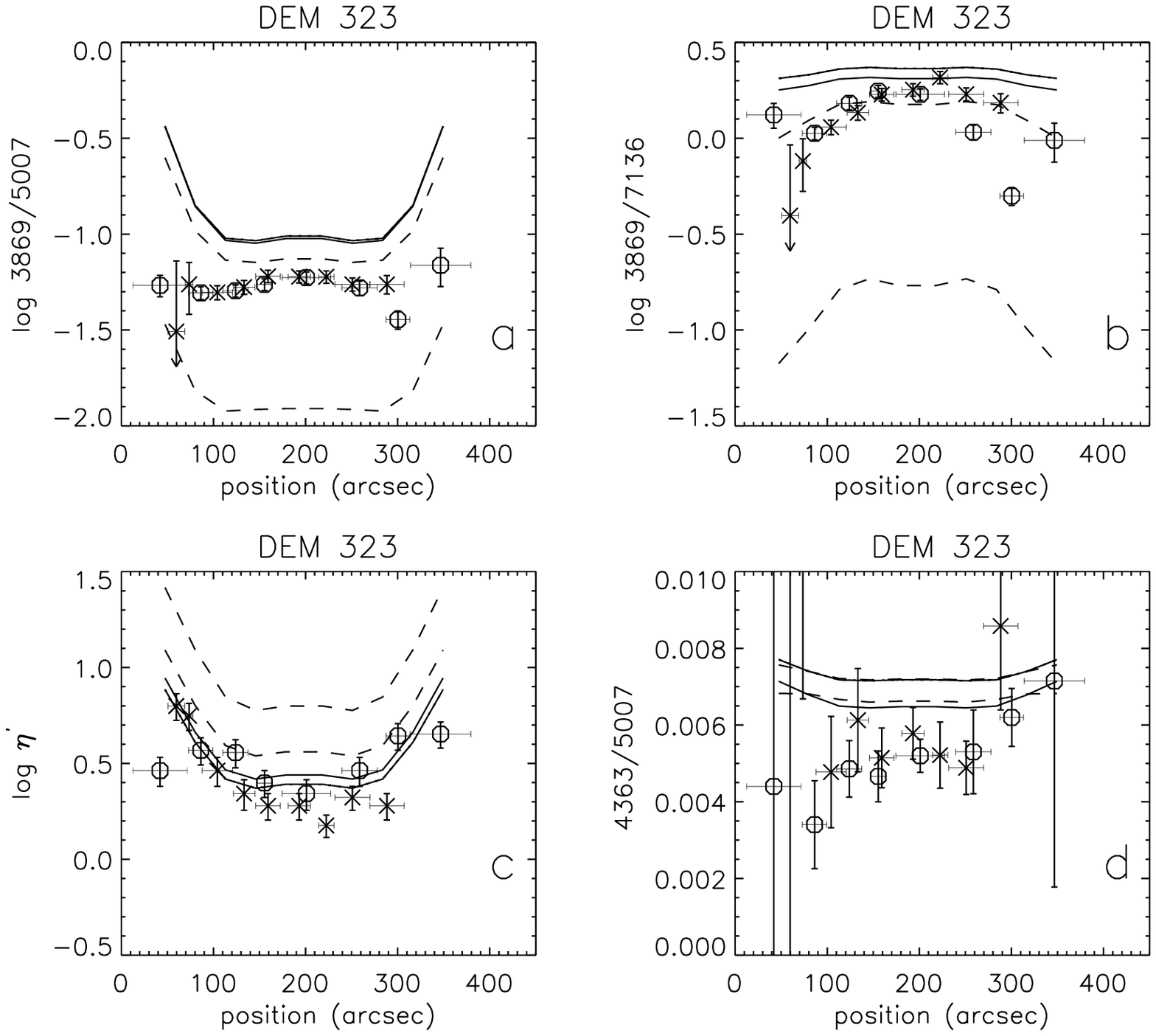}{7in}{0}{80}{80}{-250}{0}
%\vspace*{-3in} FOR AASPP4
\vspace*{-2in}
\caption{Same conventions as Figure~\ref{d323dgn1}, but now comparing
models using CoStar atmospheres E2 and C2 (solid
lines) with those using the corresponding Hummer \&
Mihalas (1970) atmospheres for 49 and 42 kK (dashed lines).  
\label{d323dgn2}}
\end{figure*}

Overall, the diagnostic line ratios presented in Figure~\ref{d323dgn1}
show good agreement between the data for DEM~L323 and the
model predictions for the CoStar atmosphere E2, which corresponds most
closely to the observed dominant spectral type O3 -- O4.  The variation
between the observations and predictions is typically $\lesssim 0.1$
dex.  Figure~\ref{d323dgn2} demonstrates the degree of improvement
shown by the CoStar models over the Hummer \& Mihalas (1970; hereafter
HM70) LTE atmospheres.  The variation between the CoStar and Kurucz (1991)
LTE atmospheres is discussed by Stasi\'nska \& Schaerer (1997), and
should be similar to a comparison with the HM70 atmospheres (Evans
1991).  Figure~\ref{d323dgn2} plots some of the same line ratios as
before, showing 
models using CoStar atmospheres E2 and C2 in solid lines, and Hummer
\& Mihalas atmospheres for 49 kK and 42 kK in dashed lines.  The two sets
of atmospheres should therefore correspond to the same stellar \tstar.  
The ratios of \neiii/\oiii\ and \neiii/\ariii\
(Figure~\ref{d323dgn2}$a$ and $b$) dramatically demonstrate that the
\neiii\ problem was indeed caused by stellar energy distributions that
were not hard enough in earlier generation atmospheres.  The \neiii\
line ratios for the HM70
atmosphere at 42 kK are reduced by about an order of magnitude in
comparison to the CoStar C2 model, because the HM70 energy distribution is
not hard enough to produce significant Ne$^{++}$.  The ``\neiii\
problem'' was identified in the Orion Nebula and regions ionized by
\tstar\ $\lesssim$ 40 kK, so the earlier atmospheres at those \tstar\ would
indeed cause that underprediction of \neiii.  It is also reassuring
that the CoStar models compare favorably to the more detailed models of,
e.g., Sellmaier {\etal}(1997) in reproducing the nebular ionization
properties, and yield a good approximation to the stellar energy
distributions.

Figure~\ref{d323dgn2}$c$ shows the predictions for the radiation
softness parameter \etap, using the CoStar and HM70 atmospheres.  Here we
see a strong differentiation between the HM70 42 and 49 kK models, but
almost none for the CoStar models.  Thus, although previously \etap\ had
been predicted to be a good diagnostic of \tstar\ (e.g., Skillman
1989), unfortunately the new 
atmospheres now suggest that it cannot effectively discriminate
\tstar\ for stars hotter than about 40 kK.  The data are clearly in
better agreement with the CoStar models.

We also show in Figure~\ref{d323dgn2}$d$ the model predictions for the
$T_e$ diagnostic, \oiii\lam4363/\lam5007.  The discrepancy with
observed temperatures is not resolved by using HM70 atmospheres.  We
note that Baldwin {\etal}(1991), in their study of the Orion Nebula,
did match 4363/5007, but allowed \tstar\ to be a free parameter.
They arrived at $T_\star = 39.6$ kK for the O6--7 V star $\theta^1$ Ori C, a
temperature about 2 kK cooler than the current calibration of that
spectral type for the CoStar models.  Their work was based on
Kurucz (1991) LTE atmospheres, which are also systematically softer
than the CoStar atmospheres.

\subsection{DEM L243:  O7}

For DEM L243, the dominant ionizing sources are stars \#2 and \#5 from
Oey (1996a), having spectral types O7~V and O7~I, respectively
(Table~\ref{sample}).  Their positions are identified in
Figure~\ref{figd243}.  The \hii\ region has a fairly uniform
morphology with a couple higher density knots
(Figure~\ref{figd243}$a$). 
% ** THE SNR POSITIONS ARE NOT IN LO-DENS LIMIT...DOES SHOCK
% HEATING AFFECT THE RATIO?  ELSE REAL.
DEM~L243 is of special interest because of the prominent SNR on
the eastern side; this is almost circular and is outlined in black in
Figure~\ref{figd243}.  We defer a discussion of the SNR to \S~3.2.2 below,
and first consider only the photoionized region of the nebula.

\subsubsection{Photoionization models}

For our photoionization models, we follow a similar prescription to
that for DEM~L323.  The \Ha\ profile does not show clear evidence of a
central cavity, so for DEM~L243 we set the inner nebular radius to be
10\% of the Str\"omgren 
radius.  The observed \Ha\ emission (Oey \& Kennicutt 1997) implies
an ionizing luminosity $\log\Llyc/\ergs = 38.0$, and we scale the input
O7 stellar atmosphere accordingly.  For DEM~L243, we show the effect 
of varying $U$, by changing the gas density between $n = 1$
and $100\ \cc$ (equation~\ref{U}).  Given the existence of actual
fluctuations in gas density and filling factor, local variations in
$U$ in this range would be unsurprising.  We again adopt abundances
that were measured for this object in Paper~II, $\log X$/H = ($-1.18,
-5.18, -3.85, -4.77, -5.51, -6.19$) for (He, N, O, Ne, S, and Ar), with
refractory elements as before.
There are three stationary slit positions for DEM~L243:  one centered
on each of the dominant stars, and one passing through the knot
associated with the SNR (Figure~\ref{figd243}).  Since these slit
positions do not share a common center, the direct superposition of
the data as a function of slit position will not yield a strictly
radial cross-section of the nebula, as we had in the case of DEM~L323.  
However, the \hii\ region does not have the simple,
spherical morphology of DEM~L323, and the two dominant stars are not
located close to each other within the nebula.  Therefore, we
feel that a simple superposition of the slit data still
provides an adequate spatial representation of the nebular structure.
It is important to bear in mind that the nebular geometry now differs
from the spherical symmetry assumed in the models, and spatial discrepancies
in Figure~\ref{d243dgn1} should not necessarily be considered
significant.  As an additional 
caveat, there is no information on the stellar population north of the
prominent dust lane, so it is possible that a star in that northern
extension is also an important ionizing source.

Figure~\ref{d243dgn1} presents diagnostic line ratios for DEM~L243,
using the same conventions as before.  The triangles, squares, and
crosses show apertures from slit positions D243.2S, D243.5S, and
D243.30S, respectively.  The solid lines show the {\sc Mappings}
models for CoStar atmosphere E2, at $n = 1$ and $100\ \cc$.  The
dashed and dotted lines show the corresponding models for atmospheres
C2 and B2, respectively.  For each \tstar, the
lower-density model corresponds to lower ionization parameter
(equation~\ref{U}), so with the exception of Figure~\ref{d243dgn1}$e$,
the $n=1$ models have lower values on the ordinate.  For the observed
dominant spectral type O7, we now predict atmosphere C2 (dashed line) to
most closely reproduce the nebular observations.  An overall
comparison of the models in Figure~\ref{d243dgn1} with those in
Figure~\ref{d323dgn1} for DEM~L323 also demonstrates how the range in
the given line ratios can change depending on morphology, i.e.,
size of the central cavity.

\begin{figure*}
\plotone{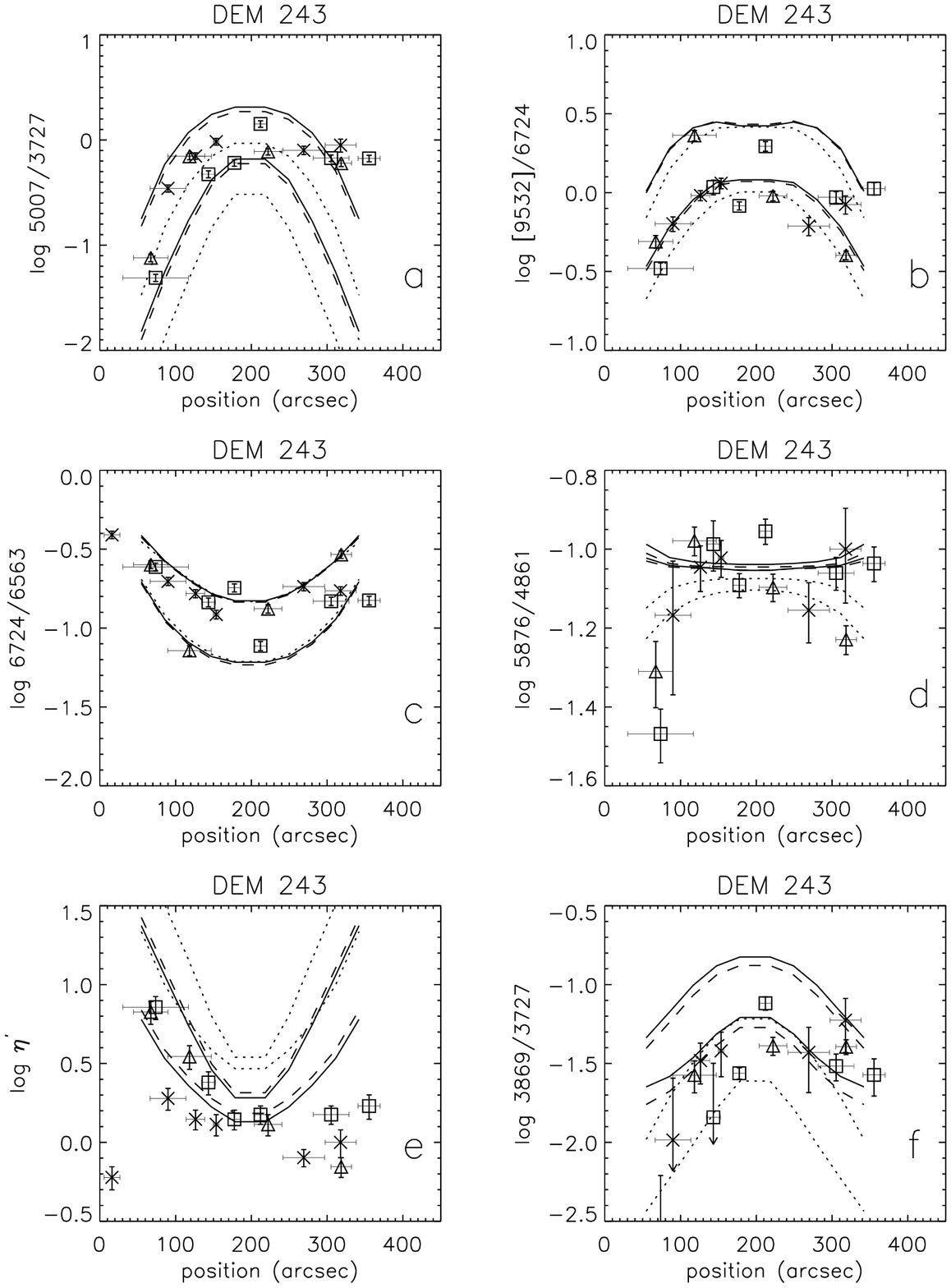}
\caption{Projected spatial variation of diagnostic line ratios across
DEM~L243.  The data points for all slit positions are superimposed.
Triangles, squares, and crosses show positions D243.2S, D243.5S, and
D243.30S, respectively (Figure~\ref{figd243}).  Dotted, dashed, and
solid lines show photoionization models using CoStar atmospheres B2,
C2, and E2, respectively, for densities of $n = 1$ and $n = 100\ \cc$
(see text).
\label{d243dgn1}}
\end{figure*}

\begin{figure*}
\figurenum{\ref{d243dgn1}}
\plotone{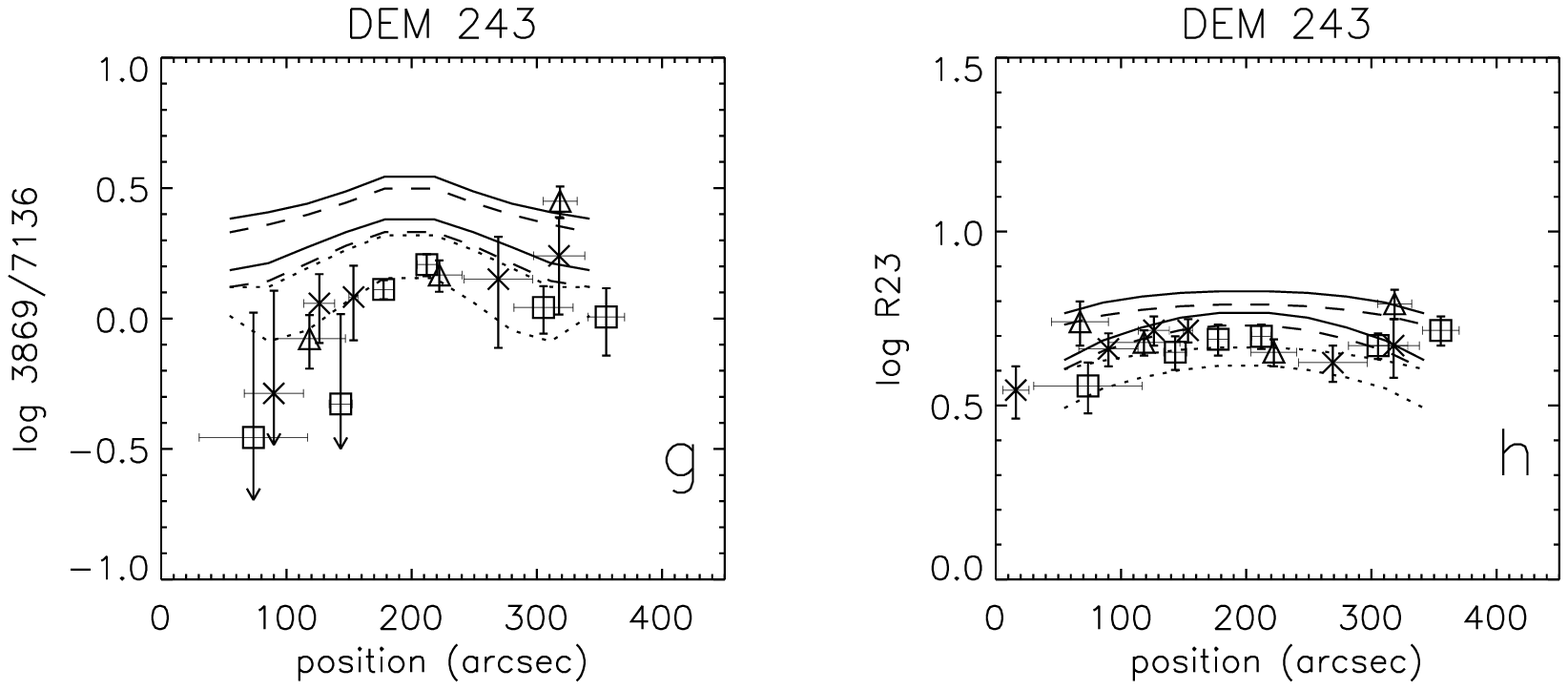}
\vspace*{-5in}
\caption{{\it continued}}
\end{figure*}

Figure~\ref{d243dgn1}$a$ and $b$ show the $U$ diagnostics for DEM
L243.  We again see the strong degeneracy of the E2 and C2 models, and
excellent agreement with the data for \oiii/\oii.  For \siii/\sii, the
models appear to overpredict the data by about 0.2 dex.  This appears to
result in part from a slight underprediction in \sii\lam6724, as is
apparent from Figure~\ref{d243dgn1}$c$.  The offset between models and
data is in the reverse sense from that seen for DEM L323.

The observations of \hei/\Hb\ are interesting (Figure~\ref{d243dgn1}$d$).
There is a large scatter in comparison to the models, and some
values are significantly below the predictions.  This scatter is not
seen in our other objects (Figures~\ref{d323dgn1}$f$, and
\ref{d199dgn1}$f$).  It would therefore appear that He is not
uniformly fully ionized in DEM~L243, in contrast to the 
other objects in this sample.  This suggests that the CoStar
atmospheres may be slightly too hard around this temperature of
$T_\star = 42$ kK.  The slight overprediction of S ionization described
above is also consistent with an overly hard stellar atmosphere model,
although there is some uncertainty in the spectral type -- \tstar\ conversion.
It is worth noting that the transition from full to partial He ionization
apparently occurs around this O7 spectral type or slightly earlier.

The radiation softness parameter is plotted in
Figure~\ref{d243dgn1}$e$.  Its behavior is similar to that found for
DEM~L323, and the data are again in good agreement with the model
prediction for atmosphere C2.  Figure~\ref{d243dgn1}$f$ and $g$ show
\neiii/\oii\ and \neiii/\ariii.  In contrast to DEM~L323
(Figure~\ref{d323dgn1}$h$), \neiii/\oii\ is overpredicted in the
models by about 0.3 dex.  \neiii/\ariii\ shows about the same $\sim
0.2$ dex overprediction for atmosphere C2 as seen for E2 in DEM L323.

The abundance parameter $R23$ is shown in Figure~\ref{d243dgn1}$h$.
The data are in good agreement with the C2 atmosphere, 
about 0.1 dex below the prediction.

\subsubsection{The SNR}

The SNR in DEM L243 (N63 A in the Henize 1956 catalog) has been the
subject of previous detailed studies 
(e.g., Shull 1983; Dickel {\etal}1993; Levenson {\etal}1995).  As
noted by several authors, the bright optical remnant is offset to
the west side of the circular region of X-ray and radio emission,
which is delineated by the black circle in Figure~\ref{figd243}.  
Part of the SNR boundary is also visible in the \oiii/\Ha\ ratio map
in Figure~\ref{figd243}$b$.  This shows the
area ionized by weak photoionization from the shock precursor.
Such precursors are known to generate large \oiii/\Hb\ ratios (Dopita
\& Sutherland 1995, 1996), and so this
region can be seen in the \oiii/\Ha\ ratio map.
The three-lobed structure seen in published images of the optical remnant 
(e.g., Mathewson et al. 1983; Shull 1983) is barely resolved in our
images, and corresponds to the bright knot traversed by slit position
D243.30S.  Four associated bright knots of shocked gas are
clearly apparent in the \sii/\Ha\ ratio map (Figure~\ref{figd243}$c$),
and indicate the locus of the blast wave.

\begin{figure*}
\plotone{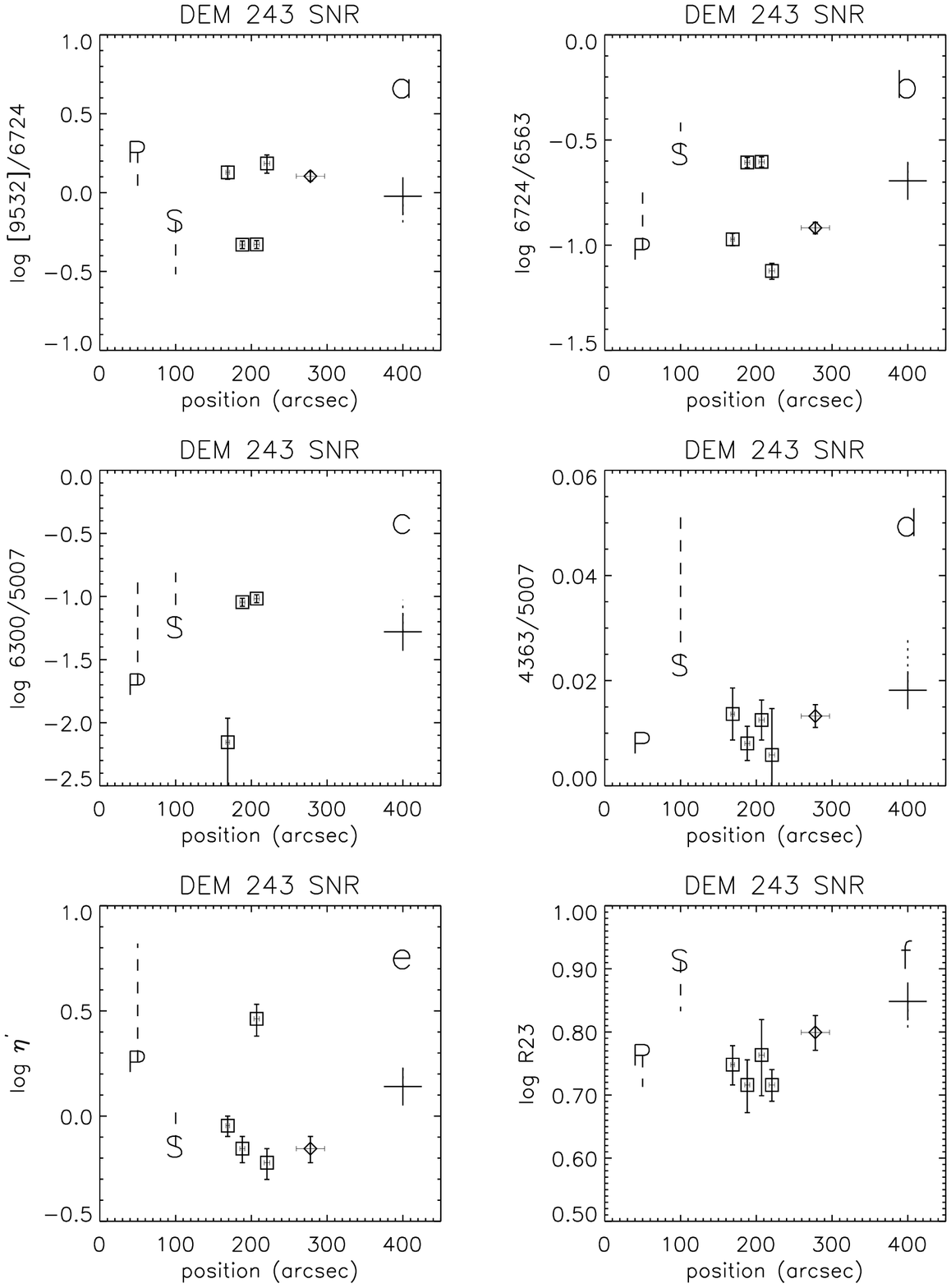}
\caption{Diagnostics for apertures affected by the SNR shock.  The
diamond and squares correspond to slit positions D243.2S and D243.30S,
respectively.  The `P' symbol represents a model photoionized by
CoStar atmosphere C2, with the dashed line indicating the range of
values for distances up to $0.8 R_{\rm S}$.  The `S' symbol represents
a shock model with $B/n^{0.5} = 3.16\ \ \rm\mu G\ cm^{3/2}$ and shock
velocity 200 $\kms$, 
with the dashed line extending to a model at 100 $\kms$.  The cross
shows the sum of the photoionized and 200 $\kms$ shock model, with the
dotted line extending to a composite with the 100 $\kms$ model.  
The spatial positions for the models are arbitrary.
\label{d243dgn2}}
\end{figure*}

In Figure~\ref{d243dgn2}, we show some diagnostics for observations
affected by the SNR.  The diamond corresponds to aperture
D243.2S-12, and squares show values for slit position D243.30S.
As before, the horizontal lines denote the spatial extent of the
aperture, while the error bars are indicated vertically.
A photoionization model for CoStar atmosphere C2 with $n = 10\ \cc$
is shown with the `P' symbol.  Since these apertures are all
localized in one area of the nebula, their relative positions are not
significant, so the models are not shown spatially resolved as
before.  Instead, the vertical dashed line shows the range of values
for projected distances up to 0.8 $R_{\rm S}$, with the `P'
symbol at the locus of the central zone.  

We also generated models of plane-parallel shock
emission with {\sc Mappings} v1.1.2, also for $n = 10\ \cc$, and
magnetic field strength $B = 10\ \mu$G, thereby corresponding to
magnetic parameter $B/n^{0.5} = 3.16\ \rm\mu G\ cm^{3/2}$, which
corresponds to equipartition between the magnetic and gas thermal
pressures in the ionized pre-shock plasma.  We take an input ionization 
balance with H 99\% ionized, and include contributions to the emission
from both shock and precursor.  The `S' track in Figure~\ref{d243dgn2} shows
the range of values for shock velocities between 100 and 200 $\kms$,
with the `S' symbol denoting the locus of the 200 $\kms$ end.
Estimates for the expansion velocity of the optical remnant range from
110 $\kms$ (Dopita 1979) to $350\ \kms$ (Shull 1983); additional
estimates have been obtained and compiled by Chu \& Kennicutt (1988).
The presence of at least some radiative shocks with velocities above
200 $\kms$ is indicated by the presence of the photoionized precursor
mentioned above, and seen in Figure~\ref{figd243}$b$.  Dopita \&
Sutherland (1995) showed that, for radiative shocks, the velocity of
the ionization front produced by the EUV photons from the shock only
becomes greater than the shock velocity for velocities $\gtrsim 175\ \kms$.
%We note that the predicted intensity of \ion{He}{2} $\lambda4686$ is
%about 0.11 \Hb\ for a 200 $\kms$ shock.
While we did not detect \ion{He}{2} $\lambda4686$, Levenson
{\etal}(1995) report two detections of this line, with intensities of 
0.044 and 0.024 with respect to \Hb.  This is roughly consistent with the
prediction of 0.046 for a 200 $\kms$ shock, undiluted by a stellar
photoionized contribution.  We combined each of our shock models with
the zone of the photoionization model at $0.4 - 0.6\ R_{\rm S}$.
The cross in Figure~\ref{d243dgn2} shows line ratios
for the composite with the 200 $\kms$ model, and the
dotted line extends to the sum with the 100 $\kms$ model.

While the equal weighting, with respect to \Hb\ emission, in the sum is
arbitrary, it is 
apparent that a composite spectrum is in much better agreement with
the observations than either the `P' or `S' models alone.
Figure~\ref{d243dgn2} also demonstrates that it is the higher velocity
shock that more closely resembles many `P' line ratios, owing to the
higher ionization and emission in species like \oiii.
However, this enhanced emission does affect both the
\etap\ and $R23$ parameters (Figure~\ref{d243dgn2}$e$ and $f$).  Thus,
while it may be difficult to diagnose the presence of a 200 $\kms$
shock, it can cause misleading values in the diagnostic parameters.

This object has historically been identified as an SNR
within an \hii\ region, especially since the presence of the OB
association LH~83 increases the probability of Type~II SN events in
the region.  However, it might be surprising to find the
conventional optical signatures of elevated \sii\ and \oi\ in an
ionized environment where the dominant species of S and O are S$^{++}$
and O$^{++}$.  We suggest that the origin of the blastwave may be
located behind the nebula, and is impacting gas behind or near
the edge of the \hii\ region.  This is also supported by the presence
of only blueshifted velocity structure in the optical remnant (Shull 1983).
It has been noted by previous authors (e.g., Shull 1983) that the
western lobe of the bright optical SNR knot appears to be only
photoionized, as opposed to the adjacent lobes, which show the strong
shock signatures.  This is also consistent with the blastwave
impacting the region from behind, or with the SNR being a background object
altogether.  We note that our superimposed shock \+ photoionized
spectra are consistent with the nebular emission, without
incorporating combined radiative transfer of these processes.
When examining Figure~\ref{d243dgn2}, it should be
kept in mind that apertures D243.30S-13 and D243.30S-16 cross the
photoionized lobe.  

\subsection{DEM L199: Early WR}

DEM L199 is a large and luminous nebular complex that poses a greater
challenge for the idealized geometry of the photoionization models.
As such, it may be more 
representative of \hii\ regions observed in more distant galaxies.
The dominant stars are early-type WR binaries, Br~32, Br~33, and Br~34,
having spectral types WC4 + O6 V--III, WN3 + OB, and WN3: + B3 I
(see Table~\ref{sample}).  The WR atmospheres are much more difficult
to model than the main sequence spectral types, and the nebular results are
likewise more uncertain.  There is also an O3--4~V star present, but
it contributes only 10 -- 20\% of the observed \Ha\ luminosity.  In
addition, DEM~L199 has the most complicated morphology of any object
in our sample.  The \Ha\ image suggests a range of densities in
different regions (Figure~\ref{figd199}$a$), although density estimates
from the ratio \sii$\lambda6716/\lambda6731$ are again in the low-density
limit, suggesting $n \lesssim$ a few 10's $\cc$.  There is also a 
prominent network of shells and filamentary material.  The WR stars
are located on the interior of the overall shell structure.

\subsubsection{WR model atmospheres}

Comparing the nebular emission-line spectrum with models based on
synthetic atmospheres is a particularly important test for WR stars.
Crowther (1999) emphasizes the importance of using nebular analyses to
test the model atmospheres.  The emergent spectra result from
complicated radiation transfer in NLTE, supersonic expanding
atmospheres having density and opacity 
stratification and inhomogeneities.  A widely-used set of basic models
for WR atmospheres is the grid of Schmutz {\etal}(1992; hereafter
SLG92), which consists of pure He atmospheres incorporating NLTE
radiation transfer and a power-law wind density law.  More recently,
Hamann and colleagues have computed new grids that include the principal
metals and electron scattering in lines (Hamann \& Koesterke 1998;
Gr\"afener {\etal}1998).  We will examine photoionization models using
atmospheres from this group, and also SLG92.  Crowther (1999) and Schmutz
\& de Marco (1999) review the various WR models produced by different
groups. 

Only a few \hii\ regions containing WR stars have been examined in
conjunction with the model atmospheres
(Crowther {\etal}2000; Esteban {\etal}1993; Kingburgh \& Barlow 1995),
and most of these are WR ejecta nebulae.
DEM L199 is an interesting case for testing WR atmospheres since it is
not an ejecta nebula, but is a large \hii\ region whose principal
ionizing sources are WR stars.
%  Oey \& Kennicutt (1997) found a
%ratio of 1.22 for the observed to predicted \Ha\ emission based on the
%O star population.  
It should therefore be more analogous, and directly comparable, to
classical \hii\ regions than a strongly 
density-bounded ejecta shell with abundance anomalies.  

\begin{figure*}
\plotone{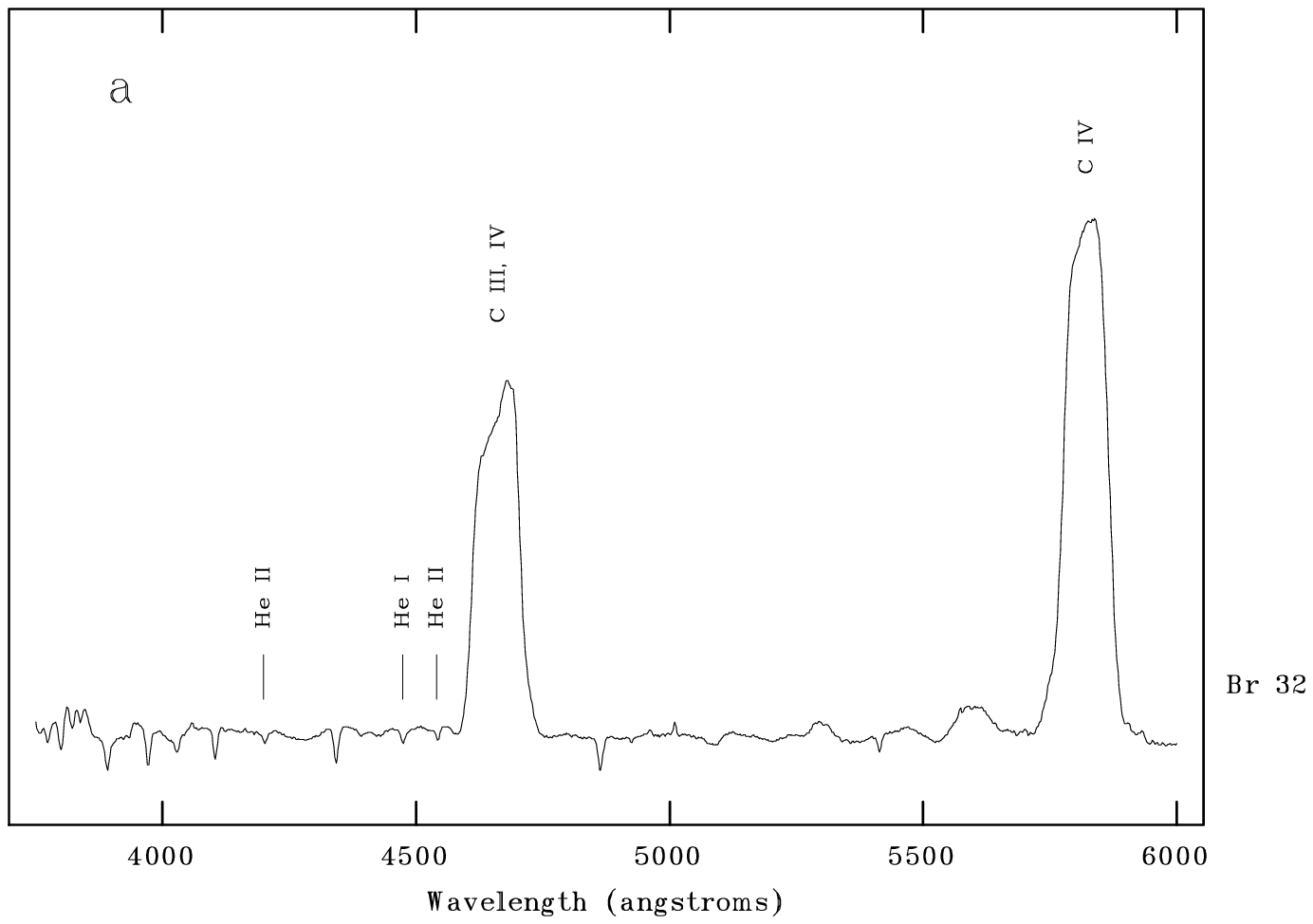}
\plotone{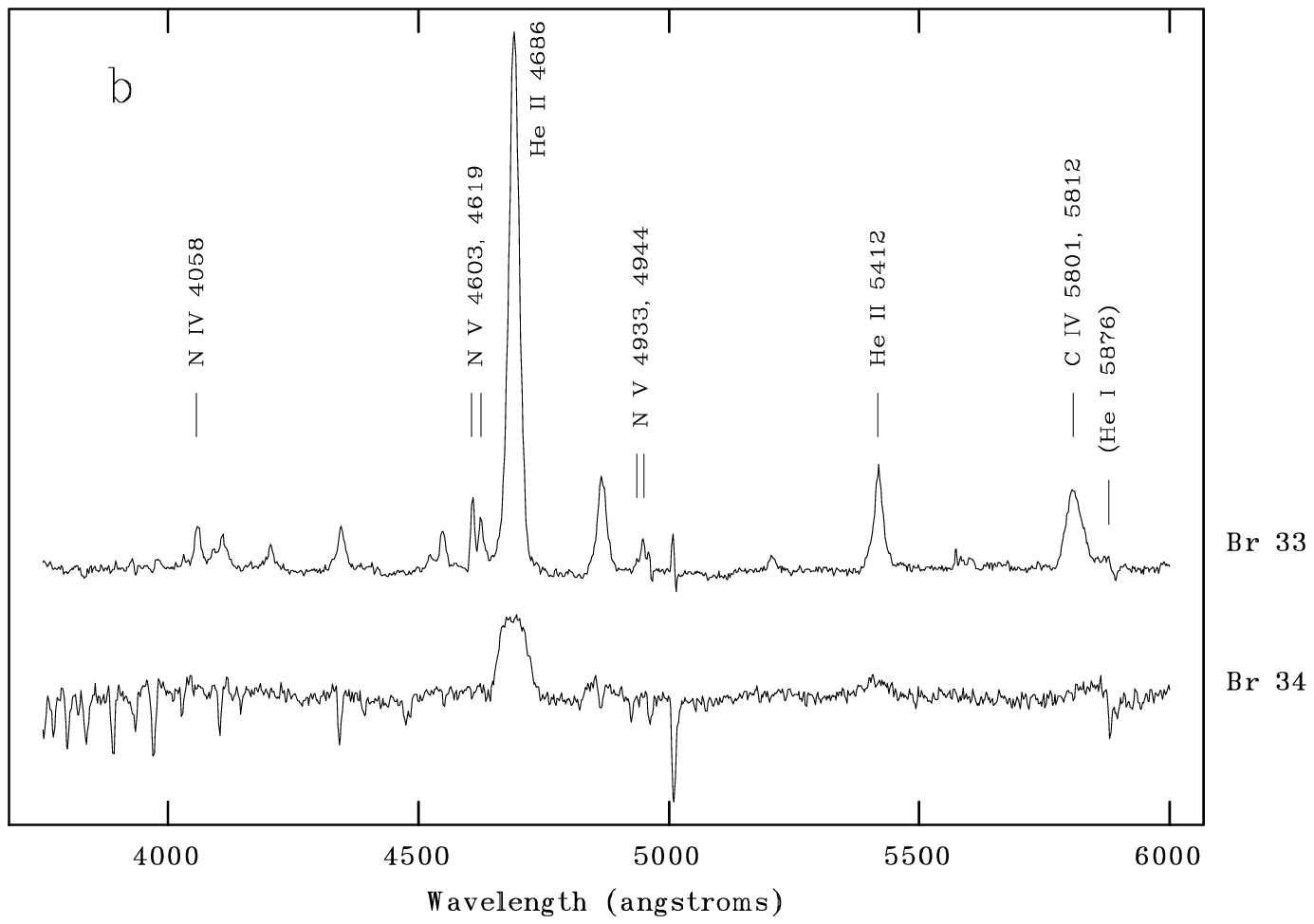}
\caption{Rectified blue spectra of the WR stars in DEM~L199.
Figure~\ref{WR}$a$ 
shows the WC4 + O6~V-III star, Br~32; Figure~\ref{WR}$b$ shows Br~33,
a WN3 star with possible OB companion, and Br~34 (WN3: + B3~I).  The
position of \protect\ion{He}{1} $\lambda5876$ is indicated; the absorption
feature is telluric water vapor.  Spectral types are as taken in
Table~\ref{sample}. 
\label{WR}}
\end{figure*}

Slit position D199.205 includes both Br~32 and Br~33, and our scanned
position D199.496W87 includes a transient exposure of Br~34
(Figure~\ref{figd199}$a$).  Our extracted spectra of these stars are
shown in Figure~\ref{WR}.  The WR models are computed for a grid of
$R_t$ and \tstar, where $R_t$ is the reference ``transformed'' stellar
radius, and \tstar\ is now the core effective temperature.
We use the spectral diagnostics of Br~33 to estimate its $R_t$ and \tstar\
from the contours of these parameters on the $R_t - $\tstar\ grid of
Hamann \& Koesterke (1998, their Figures 1 -- 3; hereafter HK98) for WN stars.
We measure equivalent widths of \heii\ \lam5412 $\sim 12$~\AA,
\ion{N}{4} \lam4058 $\sim 3.5$~\AA, and \ion{N}{5} \lam4933--44 $\sim2.7$~\AA.
There is no detectable \hei\ or \ion{N}{3} \lam 4640.  This is a
weak-lined WNE type, according to the the equivalent width of \heii\
\lam5412.  The line strengths suggest that
the HK98 WN models for $\log R_t/R_\odot = 0.1$ 
and 0.2, at $\log T_\star/{\rm K} = 5.1$ should correspond to the
observed WN3-w spectrum.  We caution that if the system is indeed a
binary as originally suggested by Breysacher (1981), then the
equivalent widths may be underestimated;  however, we
see no evidence of a companion in Figure~\ref{WR}.

\begin{figure*}
\plotone{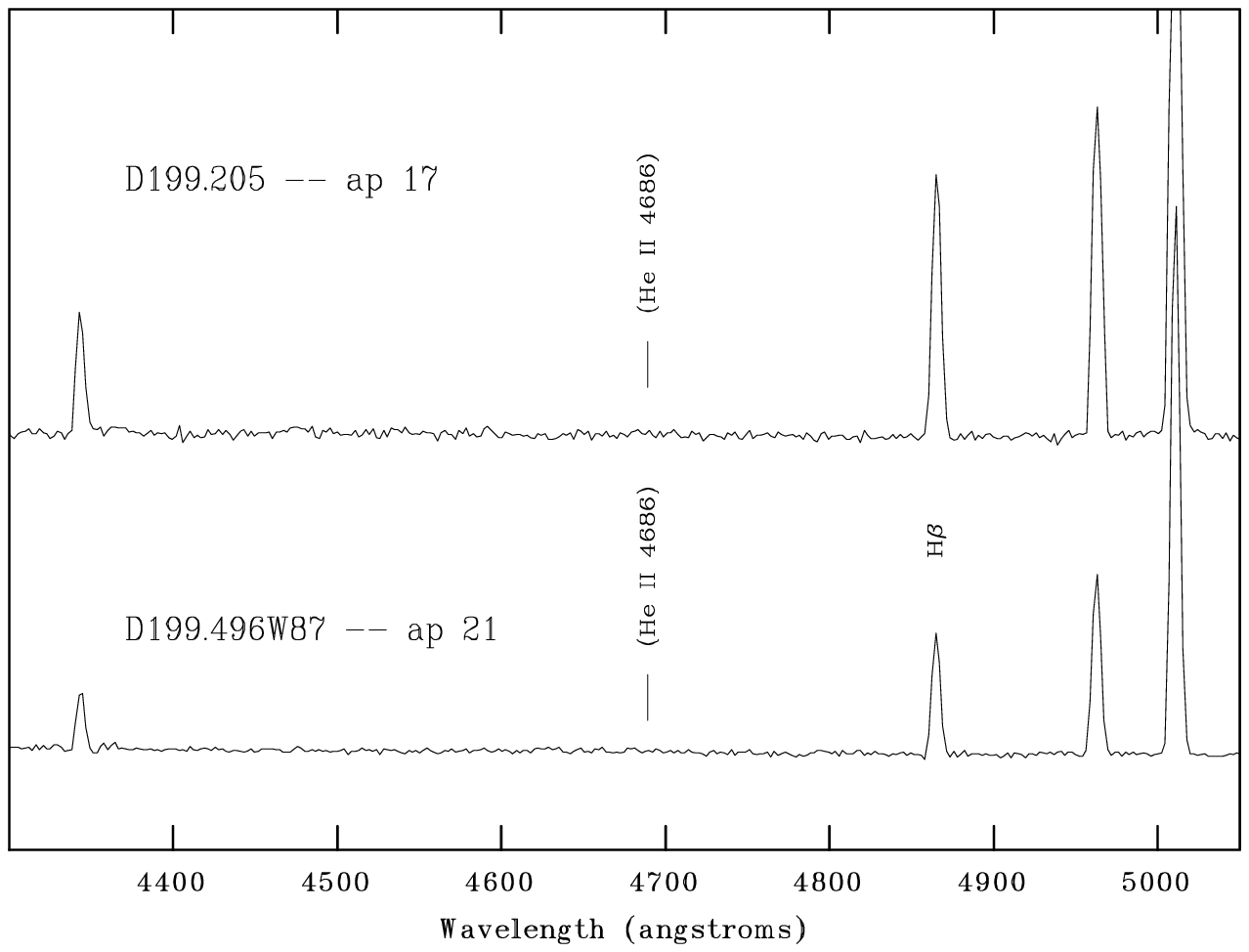}
\caption{Subsections of representative nebular spectra of DEM~L199 in
two apertures near the WR stars.  There is no detection of
\protect\ion{He}{2} $\lambda4686$ to $< 1$\% of \protect\Hb.
\label{figd199heii}}
\end{figure*}

There is also a simple and powerful constraint on the atmospheres from
the nebular data:  we find no detection of nebular \heii\ \lam4686 to $<
1$\% of the \Hb\ intensity.  Figure~\ref{figd199heii} shows apertures
D199.205-17 and 
D199.496W87-21 in the vicinity of this line.  D199.205-17 is adjacent
to Br~33, and D199.496W87-21 scans the region between both of the WN3
stars, Br~33 and Br~34 (Figure~\ref{figd199}$a$).  This clearly
constrains the models to those that produce no He$^+$-ionizing
emission.  However, it is also interesting that neither Br~33 nor
Br~34 show
detection of \hei\ \lam5876 (Figure~\ref{WR}$b$).  There is a telluric
absorption feature close to this position, but it is clear that any
emission in \hei\ must be minimal.  This lack of stellar \hei\
suggests that the stars are hot enough to fully ionize He$^0$ to
He$^+$.  We would therefore expect the
production of He$^+$-ionizing photons, yet this is apparently not the
case.  {\it Both} of these stars appear to be hot enough to fully
ionize He$^0$ in their atmospheres, yet not hot enough to produce
He$+$-ionizing emission.  The applicable stellar model should be one
that is essentially on the boundary between these two conditions, and
therefore corresponds to a very narrow range of stellar parameters.
The boundary contour can be determined on the $R_t - T_\star$ plane,
as for example, in Figure~1 of SLG92 and Figure~3 of HK98.  

Thus the HK98 models that we selected are adjacent to this
boundary, with \tstar\ determined by the observed stellar \ion{N}{4},
\ion{N}{5}, and \heii\ lines.  HK98 model \#15-22 has \tstar\ = 125.9
kK and $R_t = 0.794\ \Rsol$; model \#15-19 has \tstar\ = 125.9 kK,
$R_t = 1.58\ \Rsol$.  The corresponding model on the SLG92 grid has
$T_\star = 127.4$~kK and $R_t = 1.05\ \rm R_\odot$.  Note that the
atmospheres of the different groups are not strictly analogous since 
they assume different input physics; the difference in the resulting
emergent fluxes is apparent in Figure~\ref{atm}.  The SLG92 model has
a pure He composition, whereas the HK98 model includes N.  More
importantly, the SLG92 model
assumes a much steeper wind velocity law of $\beta = 2$, where 
$\beta$ is the power-law index of the wind velocity gradient.  In
contrast, the HK98 models assume $\beta = 1$.  A hydrodynamically
calculated wind velocity law for the WN5 star HD~50896 by Schmutz
(1997) suggests that the standard assumption of $\beta = 1$ may be a
substantial underestimate in these cases.  Also, the HK98 model
\#15-19 produces He$^+$-ionizing emission, while \#15-22 does not.

In comparing with the stellar spectra, model \#15-19 predicts no 
\ion{He}{1} $\lambda5876$ in the stellar spectrum, and matches the
optical spectrum of Br~33 well \break (W-R. Hamann, private communication).
It is a much better spectral fit than 
model \#15-22, which predicts a $\lambda5876$ equivalent width
$\gtrsim 10$ \AA, that is not seen in the stellar spectrum
(Figure~\ref{WR}$b$).  However, \#15-19 produces substantial
He$^+$-ionizing emission:  the nebular models predict \ion{He}{2}\
$\lambda4686$ in the range $0.2 - 0.3$ \Hb, but this is clearly not
seen in Figure~\ref{figd199heii}.

Br~34 has a similar spectral type to Br~33 (Table~\ref{sample}), so we
presume that it is described by the same model.  However, it would be
desireable to obtain a better spectrum of Br~34.
%  We note that
%the ionization of S and N suggests that Br~34 is actually not as
%hot as Br~33.  
Br~32 has spectral type WC4, and several stars of this
type have been modeled by the same group (Gr\"afener {\etal}1998).
Among these models, the WNE atmosphere identified above for Br~33
dominates the emission, so we do not consider the effect of the WC4
star in the energy distribution.  Likewise, star \#496 (Garmany
{\etal}1994) is an O3--4~V star, and we noted above that its ionizing
luminosity will also be strongly dominated by the WR stars.
It is interesting to note that in the \sii/\Ha\ ratio map of DEM
L199, there is a zone of low \sii/\Ha\ around each of 
the WR stars (Figure~\ref{figd199}$c$).  Two apertures,
D199.205-14 and D199.205N120-17, that traverse this zone for Br~32 show the
highest excitations in the sample for \oiii/\Hb, and among the
lowest for [\ion{N}{2}]/\Hb\ and \sii/\Hb\ (Table~2).  This suggests 
high ionization states for N and S, perhaps with non-negligible fractions of
N$^{+3}$ and S$^{+4}$ in this region, and in fact, the relevant IP's
are virtually the same at 47.61 and 47.30, respectively.  It may be
that the ionization is enhanced by stellar wind shocks, suppressing
recombination.
%The level of \sii/\Ha\ appears to be higher in the
%corresponding zone for Br~34, which is also about half the radius of that
%for Br~33.  
The strong \oiii/\Ha\ seen in Figure~\ref{figd199}$b$ in
the zone for Br~34, and measured for the zones around Br~33 and Br~32
are again consistent with these stars being too cool to doubly ionize He.    

\subsubsection{Photoionization models}

For the photoionization models, we take \break $\Llyc/\ergs = 38.9$, obtained
as usual from the observed \Ha\ luminosity (Oey \& Kennicutt 1997).
Based on measurements from Paper~II, we adopt abundances: $\log X$/H =
($-1.05, -4.1, -4.93, -3.64, -4.40,$ $-6.0, -5.3, -5.30, -5.90, -6.2$) for
(He, C, N, O, Ne, Mg, Si, S, Ar, and Fe).
Based on the relative positions of the stars with respect to the gas
in this complex, we run {\sc Mappings} with a central cavity of $0.5
R_{\rm S}$.  However, clearly the model geometry will be a
coarser approximation to reality than for the other objects in
the sample.  There are two stationary slit observations, and both of
these cut across both high and low-density regions, sampling areas
both interior and exterior to the shell.  We present these data in the
same manner as before, by superposing the slit observations; we
caution that these will not be expected to closely follow the geometry of the
models.  Instead, it may be more useful to compare the data with the
range in the predicted quantities.

Figure~\ref{d199dgn1} shows the same diagnostic parameters as
Figure~\ref{d323dgn1}.  The squares show data from the
stationary position D199.205 and the crosses show D199.205N120
(Figure~\ref{figd199}$a$).  The solid, dashed, and dotted lines show
models for $n = 100\ \cc$ using the WR atmospheres of SLG92, HK98
\#15-22, and HK98 \#15-19.  The dot-dashed line shows the results for
CoStar model E2 for a hot O star, thus it should
not be expected to yield diagnostics consistent with the data.
As mentioned above, HK98 \#15-19 has the best fit to
the stellar spectrum of Br~33, but is predicted to produce nebular
He~II $\lambda$4686.  The discrepancy is also apparent in 
Figure~\ref{d199dgn1}$f$, where the population of He$^{++}$ diminishes
the intensity of \ion{He}{1}\ $\lambda5876$ for the model using
atmosphere \#15-19.  The data are clearly consistent with no He$^{++}$
population.

\begin{figure*}
\plotone{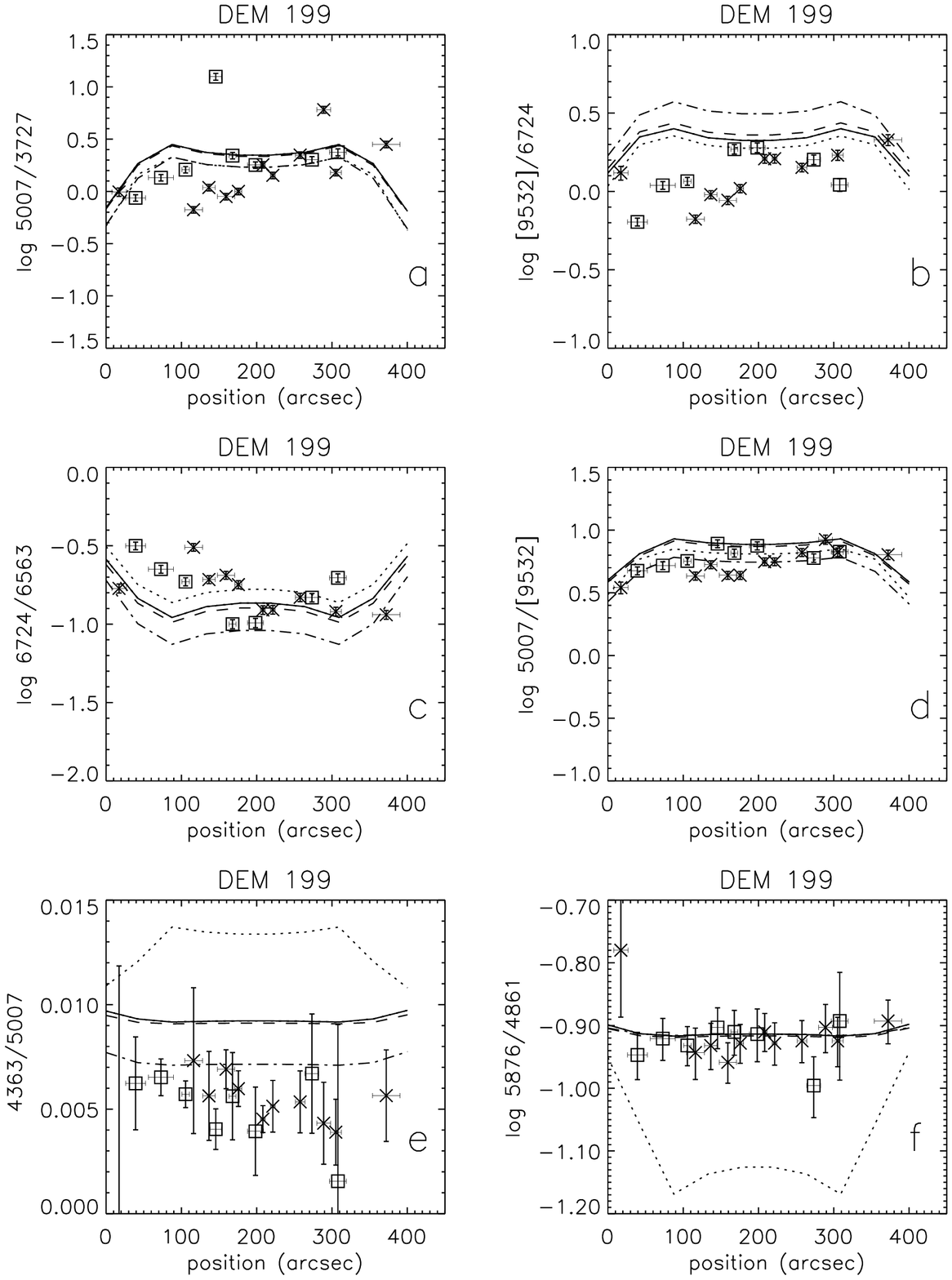}
\caption{Diagnostic parameters as a function of position across the
nebula for DEM~L199, following the same format as
Figure~\ref{d323dgn1}.  Solid, dashed, and dotted lines indicate
results using WR atmospheres from SLG92, HK98 \#15-22, and HK98
\#15-19, respectively.  The dot-dashed line shows results using 
CoStar atmosphere E2, for an O3 -- O4 star.  
\label{d199dgn1}}
\end{figure*}

\begin{figure*}
\figurenum{\ref{d199dgn1}}
\plotone{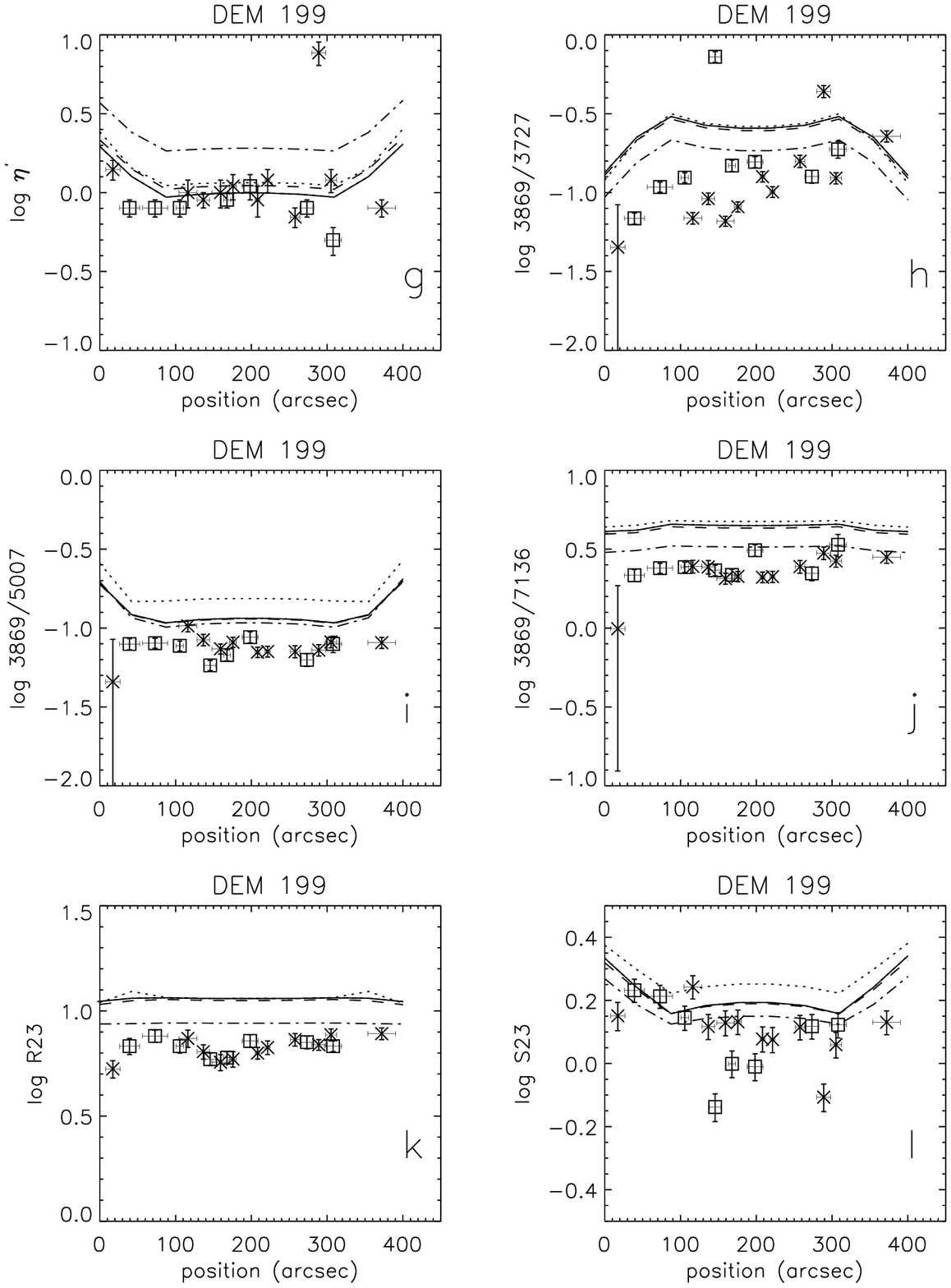}
\caption{{\it continued}}
\end{figure*}

The atmospheres of SLG92 and HK98 \#15-22 (solid and dashed lines)
generally show excellent agreement with the data, and produce
virtually indistinguishable nebular diagnostics in spite of the
noticeable differences in Figure~\ref{atm}.  Indeed, these 
line ratios are extremely robust to the WR atmospheres, except for the 
distinction between cases that do and do not produce He$^+$-ionizing
emission.  For example, the HK98 atmosphere for \tstar\ = 89.1 kK,
$R_t = 3.98\ \Rsol$ also produces diagnostics virtually identical to
HK98 \#15-22.
%  **WHY IS THIS?**

The $U$-tracer \oiii/\oii\ (Figure~\ref{d199dgn1}$a$) shows as good an
agreement with the data as was found for the O3-4 stars in DEM~L323
and DEM L243 (Figures~\ref{d323dgn1}$a$ and \ref{d243dgn1}$a$).
The data for \siii/\sii\ appear to be about 0.2 dex below the models,
however (Figure~\ref{d199dgn1}$b$).  This may be caused by the
fact that the dominant \sii-emitting zone corresponds to the
lower-density, filamentary material on the outer edges of the region.
This zone is particularly extensive in this object, and would have a
lower $U$ than assumed in these models.
%The agreement for \siii/\sii\ is
%again significant in light of previous difficulties matching
%\siii/\sii\ in the past, as mentioned above.
As in the previous
objects, the predicted \oiii\ electron temperature for the models is
hotter than indicated by the observations (Figure~\ref{d199dgn1}$e$),
this time by $\sim1500$ K.

The \etap\ parameter again matches the WR models beautifully,
as found for the O star nebulae.  (Figure~\ref{d199dgn1}$g$).  The
other \tstar\ diagnostics, \neiii/\oii, \neiii/\oiii, \neiii/\ariii,
and $R23$ are all still slightly overpredicted as before
(Figure~\ref{d199dgn1}$h-k$).  This suggests that the energy
distributions of the model WR atmospheres are slightly too hard.
\neiii/\oiii\ exhibits the same overprediction problem seen for all
the objects.  $S23$ is in good agreement (Figure~\ref{d199dgn1}$l$),
although the two apertures sampling the high-ionization zone around
Br~32, mentioned above, show anomalously low values.  From the discussion
of these apertures above, this demonstrates that
$S23$ does not adequately trace all ions of S in high-ionization regions,
which are especially relevant to spatially resolved 
observations or regions dominated by extremely hot stars.

It would appear to be a strange coincidence that {\it both} Br~33 and
Br~34, and perhaps even Br~32, exhibit the narrow condition of being
hot enough to fully ionize He$^0$, but not hot enough to produce any
He$^{++}$.  The He I $\lambda$5876 emission in modeled stellar
spectra is sensitive to even small amounts of neutral He in the
atmospheres (W-R. Hamann and W. Schmutz, private communications).  
One possible factor that may alleviate this situation is that all of
these stars are binaries.  Br~33 was suggested to be a binary by
Breysacher (1981), although there is currently no evidence of a
companion in Figure~\ref{WR}$b$.  Absorption of He~I $\lambda$5876 in
the companion could mask emission in the WR star, but for Br~33, at
least, the effect must be small in view of the lack of any other
evidence of a companion.  The fact that at least two of the stars in
DEM L199 seem to meet this very specific condition hints that we are
not seeing the full picture. 

\subsection{DEM L301:  O3 Shell}

DEM L301 (N70) is a well-known object with a remarkable, filamentary shell
structure (Figure~\ref{figd301}).  The dominant ionizing stars are SW3
and NW4, with spectral types O3~I and O5~III, respectively (Oey
1996a; Table~\ref{sample}).  Thus, the nebular emission-line
properties might be comparable to DEM~L323, except now varying the parameter
of morphology.  Dopita {\etal}(1981) found that the nebular emission is largely
consistent with photoionization as the dominant excitation mechanism,
rather than shocks.  However, Lasker (1977) and Skelton {\etal}(1999)
suggest a limited contribution by shock heating to resolve
discrepancies that are apparent in the lower-ionization species.
This is further supported by the existence of X-ray and kinematic
evidence of a recent SNR impact (Chu \& Mac Low 1990; Oey 1996b).  
Thus, the emission properties of this object may be more complex than
simple photoionization.  

Furthermore, Oey \& Kennicutt (1997) and Skelton {\etal}(1999) found that
the observed \Ha\ 
luminosity of the \hii\ region was much less than expected, based on
the stellar census present.  We therefore use the stellar parameters
instead of the \Ha\ luminosity to constrain the total ionizing flux.
From the bolometric magnitudes reported by Oey (1996a), we obtain
bolometric luminosities for SW3 and NW4 of $\log L_\star/\ergs =
39.07$ and 39.15, respectively.  Our
photoionization models therefore assume a total stellar bolometric
luminosity of $\log L_\star/\ergs = 39.4$.  We
also compute shock models for this object, again including
contributions to the emission from both shock and precursor.  For DEM
L301, we take an initial ionization balance with H 20\% neutral, since
this object has a low ionization parameter.  The results are
insensitive to details of the initial ionization balance, however.
Based on results in Paper~II, the adopted abundances are: $\log X$/H =
($-1.05, -4.1, -5.14, -3.90, -4.64, -6.0, -5.3, -5.45,$ \break 
$-6.23, -6.2$) for (He, C, N, O, Ne, Mg, Si, S, Ar, and Fe).

We have only stationary slit observations for this \hii\ region:
one position along the west edge, and one across a central
region (Figure~\ref{figd301}$a$).  As is apparent in
Figure~\ref{d301dgn1}, the apertures for 
both slit positions yield similar spectra, suggesting that the central
position, D301.SW1 (triangles), is sampling emission from essentially
the same radial zone as the edge position, D301.SW6 (diamonds).  This
suggests that the shell is indeed hollow and has very little emitting
gas in interior regions.  
Figure~\ref{d301dgn1} also shows the diagnostic ratios predicted from
four models.  The `P' symbol shows predictions for a photoionized, 
radiation-bounded model with inner and outer radii of 50 and 53 pc;
these geometric boundaries require a density $n=35\ \cc$.  The `D'
symbol shows a density-bounded model with the same inner radius of 50
pc, but density $n=10\ \cc$, and an outer radius truncating the
nebula at a radius of 53 pc.  This model includes roughly some 60\% of
the total volume of a radiation-bounded nebula at this gas density.
For the `P' and `D' models, the diagnostic ratios show little
variation between central and edge lines of sight.  We also
computed shock models for $n = 10\ \cc$ and
magnetic parameter $B/n^{0.5} = 3.16\ \rm\mu G\ cm^{3/2}$. 
These models are shown with the dashed line, with the
letter `S' indicating the locus of a 70 $\kms$ shock, and the opposite
end of the line corresponding to a 100 $\kms$ shock.  The choice of
these velocities is
motivated by Fabry-Perot observations of Rosado {\etal}(1981), whose
data for over 1000 positions across the nebula were most
consistent with an expansion velocity of 70 $\kms$.
Finally, the cross symbol shows a model that combines the
density-bounded model and the 70 $\kms$ shock model.

\begin{figure*}
\plotone{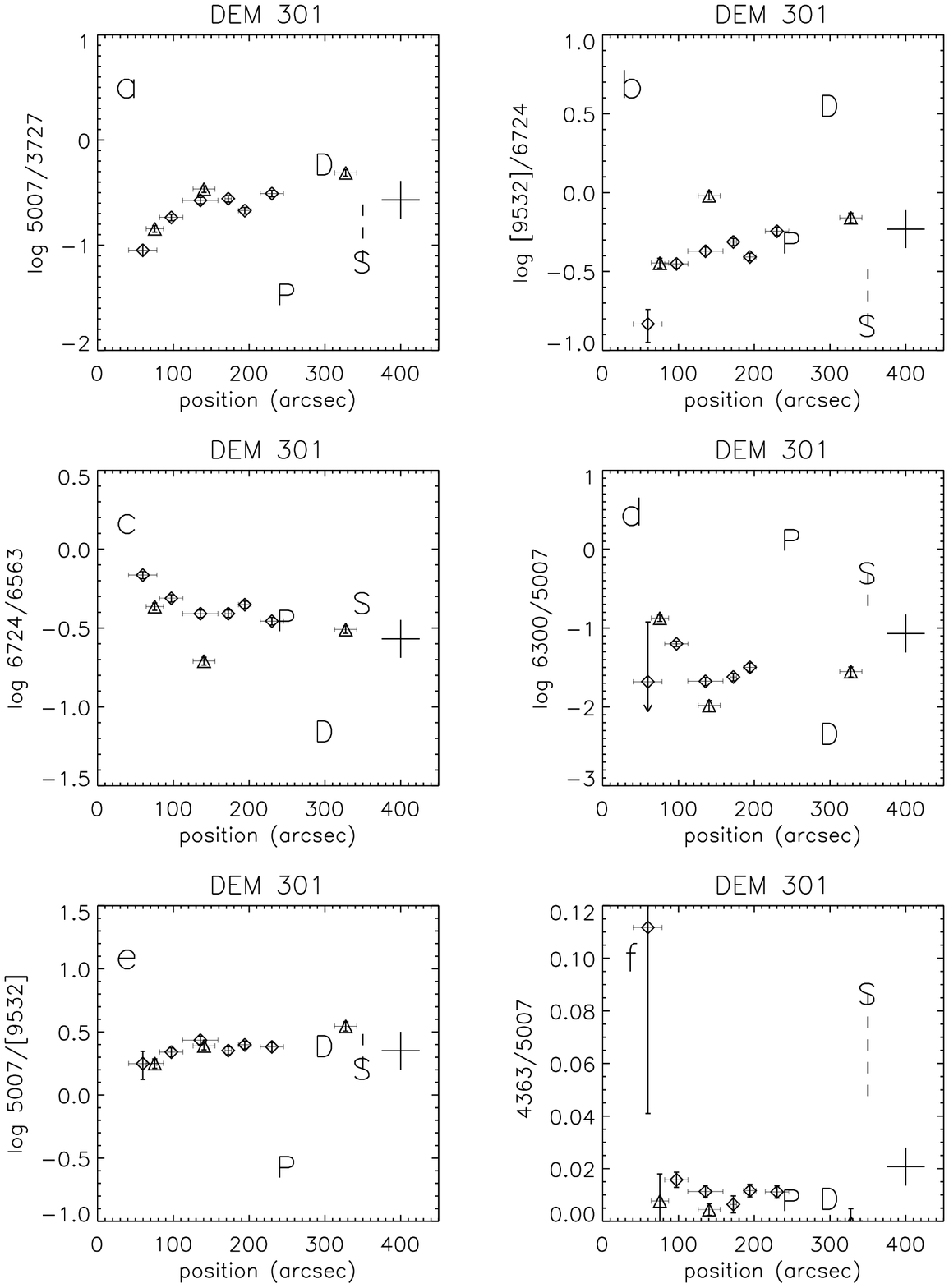}
\caption{Diagnostics for DEM L301, with triangles from slit position
D301.SW1 and diamonds from D301.SW6.  The `P' and `D' symbols show the
loci of the radiation- and density-bounded photoionized shell models;
the `S' symbol represents a 70 $\kms$ shock model, with the dashed
line extending to a 100 $\kms$ model.  The cross shows a composite of
the `D' and `S' models (see text for details on all the models).
The spatial positions of the models are arbitrary.
\label{d301dgn1}}
\end{figure*}

\begin{figure*}
\figurenum{\ref{d301dgn1}}
\plotone{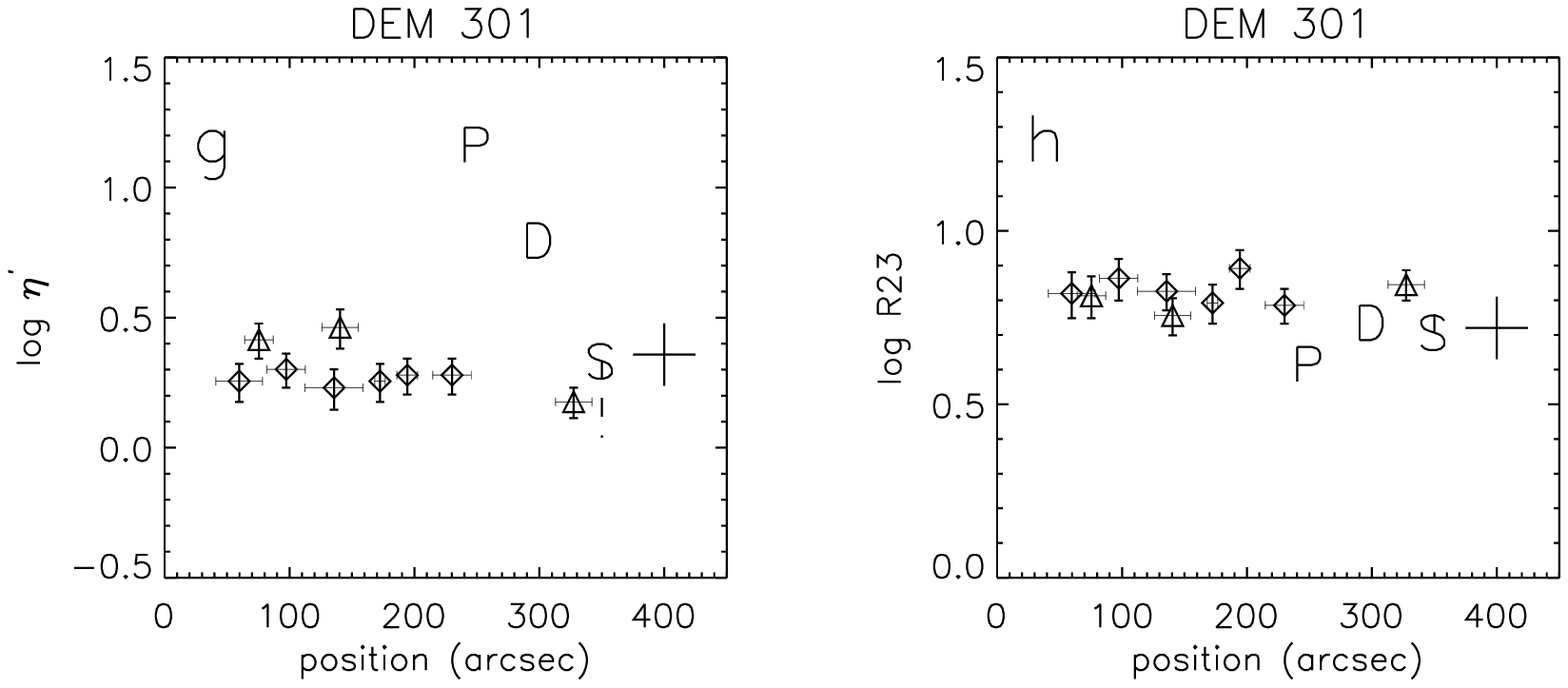}
\vspace*{-5in}
\caption{{\it continued}}
\end{figure*}

The diagnostics presented in Figure~\ref{d301dgn1} permit a detailed
evaluation of the candidate excitation processes.
Examination of the ratios \siii$\lambda9532$/\sii$\lambda6724$
(Figure~\ref{d301dgn1}$b$) and \oiii$\lambda4363/\lambda5007$
(Figure~\ref{d301dgn1}$f$) indeed shows the photoionization models in
much better agreement with the observations than the shock models.
The predicted stellar ionizing luminosity is a factor of $\sim 5$
greater than implied by the observed \Ha\ luminosity (Oey \& Kennicutt
1997), suggesting that a photoionized model of the nebula should be
density-bounded.  However, where the radiation-bounded (`P') and
density-bounded (`D') models differ, we see that neither can
satisfactorily account for all the diagnostics shown.  While the `P' model
is consistent with \sii$\lambda6724$/\Ha\ (Figure~\ref{d301dgn1}$c$),
line ratios such as \oi$\lambda6300$/\oiii$\lambda5007$
and \oiii$\lambda5007$/\siii$\lambda9532$ (panels $d$ and $e$) are in
better agreement with the `D' model.

We do find a satisfactory resolution to the puzzle by adding the
density-bounded, photoionized model with the shock model.  By simply
summing the `D' model with the 70 $\kms$ `S' model, using equal weighting
in \Hb\ line flux, we find remarkable agreement in all the
diagnostics shown in Figure~\ref{d301dgn1} (crosses).  As in the case
for DEM L243, the composite model implies
distinct, shock-dominated and photoionized components superposed.
%, not physically combined radiative transfer.  The predicted
\Hb\ flux for the 70 $\kms$ shock model is $5\times 10^{-5}\
\ergs\ \rm cm^{-2}$.  For the observed shell radius of 53 pc, this
would yield a total \Hb\ emission of $1.7\times 10^{37}\ \ergs$ if the
entire nebula were shock-excited.  This is exactly the same as
the observed value (Oey \& Kennicutt 1997), thereby demonstrating that
the shock model is easily capable of significantly contributing to the
emission as suggested by the composite model.  Our preferred model for
DEM L301 is a composite structure in which radiative shocks are
propagating outward, forming a system of dense filaments, while the
cool, post-shock gas in these filaments is photoionized on the inner
surface by radiation from the central stars.  

The ambiguity of this emission-line spectrum emphasizes the necessity
of a careful analysis in determining the excitation mechanisms.  It is
apparent that the \hii\ region is not strongly density-bounded, as
evidenced by the ratios of \oiii/\oii\ and \siii/\sii\
(Figure~\ref{d301dgn1}$a$ and $b$), and by the \sii/\Ha\ ratio map
(Figure~\ref{figd301}$c$).  The contribution of shock
heating to the lower-ionization species also partly counteracts the
loss of low-ionization emission in a photoionized, density-bounded
nebula, for example, in the ratios \sii/\Ha\ and \oi/\oiii\
(Figure~\ref{d301dgn1}$c$ and $d$).  These characteristics may lead to an
initial diagnosis suggesting predominantly radiation-bounded photoionization.
Furthermore, it is apparent that conventional shock indicators such as
\oi/\oiii\ and $\lambda\lambda 4363/5007$ (panels $d$ and $f$), 
for low-velocity shocks, are easily diluted in the presence of
photoionized gas, owing to low shock excitation
in higher-ionization species like \oiii.  

It is apparent that the ambiguous line ratios do not clearly delineate
DEM L301 as a density-bounded object.  As just described, the
\hii\ region is not strongly density-bounded, despite the fact that
only about 20\% of the emitted stellar ionizing radiation is absorbed
by the nebula (Oey \& Kennicutt 1997).  However, the highly filamentary
morphology apparent in Figure~\ref{figd301} suggests that the
majority of this ionizing radiation is escaping through ``windows''
between the filaments.   Thus apparently, where absorbing material
exists, it is thick enough to present some diagnostics resembling
radiation bounding, whereas most of the ionizing photons 
are in fact easily escaping.  This object therefore
demonstrates that strong spectral signatures of density bounding are not
required evidence for enormous photon leakage from \hii\ regions.

It is also worth noting the behavior of the parameters \etap\
and $R23$ (Figure~\ref{d301dgn1}$g - h$).  If the composite model is
correct, adding the
density-bounded photoionization and shocks, we see that \etap\ is
significantly underestimated, compared to a standard,
radiation-bounded situation.  A contribution by shocks will therefore
lead to an overestimate in the inferred stellar \tstar.  On the other
hand, $R23$ is largely insensitive to {\it any} of the modeled
conditions, and is a robust diagnostic in the presence of shocks with
velocities $\lesssim 100\ \kms$.  $R23$ does become sensitive to
higher shock velocities, however, as seen in
Figure~\ref{d243dgn2}$f$. 

\section{Spatially integrated spectra:  \tstar\ sequence}

Spectrophotometric observations of \hii\ region complexes in distant
galaxies inevitably result in the averaging of line ratios across an
appreciable fraction, or entirety, of an object.  Our
spatially integrated observations that were
obtained by scanning the slit across the \hii\ regions (Table~\ref{integ})
should approximate those obtained by conventional 1 --
1.5$\arcsec$ slit observations if the LMC were at a distance of
roughly 15 Mpc.  In Figures~\ref{figd323}$a$ -- \ref{figd199}$a$, the
two end positions of the scanned observations are shown in black.
(As indicated in these figures, we also extracted smaller apertures
from the scanned positions, but will not discuss these here.)

\begin{figure*}
\epsscale{1.7}
\plotone{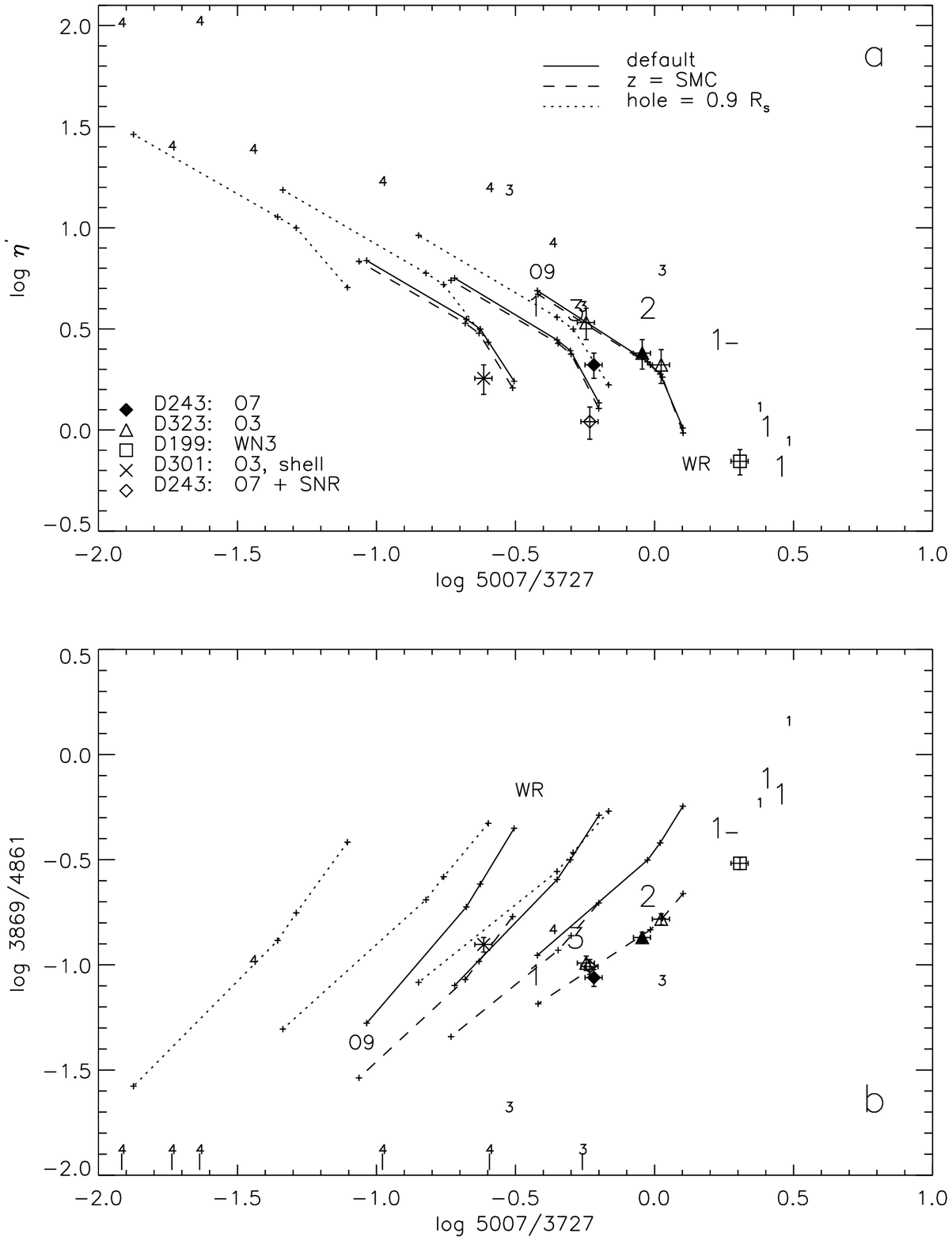}
\caption{Diagnostic parameters \protect\etap\ and \protect\neiii$\lambda3869$/\Hb\
vs. \protect\oiii$\lambda5007$/\protect\oii$\lambda3727$.  Each track of spatially
integrated models shows a range of \protect\tstar, with the 
high and low extremes indicated.  Default model parameters are
LMC metallicity measured from DEM~L323 and central hole size of 0.1
$R_{\rm S}$.  Different line types indicate variations from default,
as given by the key in Figure~\ref{integ1}$a$.  The three tracks for
each line type correspond to $n = 1,\ 10,$ and $100\ \cc$, with the
the last having the highest values in \protect\oiii/\protect\oii.  Spatially
integrated spectra for our
objects are shown with the symbols indicated by the key in
Figure~\ref{integ1}$a$.  The solid triangle corresponds to the scan
over the entire object for DEM~L323.  Numerical symbols show data from
BK that are {\it not} spatially integrated, with
LMC and Galactic objects shown with large and small font,
respectively.  The numbers (1, 2, 3, 4) correspond to types
(WR, O3--O4, O5--O8, O9 --B0), respectively.
\label{integ1}}
\end{figure*}

How well can we determine \tstar\ from the diagnostic line ratios of
the integrated spectra?  Figure~\ref{integ1} shows \etap\ and
\neiii/\Hb\ from the integrated spectra for our objects.  For
DEM~L243, we show values derived from the spectrum
with the SNR-contaminated region subtracted (solid diamond);
and we also show the total integrated region, including the SNR
(open diamond).  DEM~L323 has three scanned observations
(triangles):  one across the northern half of the object, one across
the southern half, and one across the entire nebula
(see Figure~\ref{figd323}$a$).  Diagnostics from the total integrated
spectrum across the entire nebula are indicated with the solid
triangle.  For spherical 
symmetry, these should all produce similar results, so the variation
between these gives an indication of the how well subsampling is
representative of the total spectrum.  The square and cross show
DEM~L199 and DEM~L301, respectively.  DEM~L199 has a larger angular
size, and our scanned position did not traverse the entire nebula
(Figure~\ref{figd199}$a$).
However, we included both low and high-surface brightness regions in
proportions that we hope are representative of the entire nebula.
Likewise, the scanned region includes areas both near and far from the
ionizing stars.  Again, the variation in the three observations of
DEM~L323 suggests the degree to which the single scan of DEM~L199
might be representative.  In lieu of a scanned observation of
DEM~L301, we use the spatially integrated data from aperture
D301.SW6.  Since the shell morphology of this \hii\ region is so thin,
the stationary observations are reasonably representative of the
integrated spectra, as indicated by the smaller range in values for the
diagnostic parameters (Figure~\ref{d301dgn1}).  

Each line in Figure~\ref{integ1} represents a track of
photoionization models that varies \tstar, using the same 
stellar atmosphere models as before:  O8--9 (B2), O6--O7 (C2), and
O3--O4 (E2) CoStar models, and the WR atmosphere of SLG92, with
$T_\star = 127.4$ kK and $R_t = 1.05\ \rm R_\odot$.  We take 
$\log L_{\rm LyC}/\ergs = 38.7$, which is the value
for DEM~L323, and a default central cavity of $0.1 R_{\rm S}$.
As before, the filling factor is 0.1, and the default LMC metallicity is
that for DEM~L323 (\S 3.1).  Tracks with the same line type
vary $U$ by changing the density over $n = 1, 10,$ and $100\ \cc$.  
The dotted lines show shell models with a central cavity of $0.9 R_{\rm
S}$; and the dashed lines show models with the default inner radius, at
SMC metallicity.  For the SMC abundances, we take He from 
Russell \& Dopita (1992), Mg and Fe depleted by $-2.0$ dex from the solar
values of Anders \& Grevesse (1989), and the remaining elements as
reported by Garnett (1999) from \hii\ region measurements, hence: $\log X$/H = 
($-1.09, -4.6, -5.5, -4.0, -4.8, -6.4, -5.7, -5.7, -6.1,$ and $-6.3$) for
(He, C, N, O, Ne, Mg, Si, S, Ar, and Fe).

At the time of this writing, Kennicutt et al. (ApJ, in
preparation; hereafter KBFM)
are also completing a study that investigates the behavior of \tstar\
with nebular diagnostics.  They have kindly made available their data
to us, and we overplot a subset of their nebular line ratios in
Figure~\ref{integ1} with numerical symbols.  Their data represent
stationary slit positions on Galactic (small symbols) and LMC (large
symbols) \hii\ regions.  For their LMC observations,
measurements for \lam5007 and blueward were obtained with the CTIO
2D-Frutti image tube, and the rest of their data were obtained with
CCD detectors.  Their line ratios appear to show an offset
from ours in Figure~\ref{integ1}, and this is probably caused
primarily by their selection of the brighter, central areas of 
the \hii\ regions, which have higher local ionization parameters, as
shown in their work and \S 3 above.  
Most of the well-known Galactic \hii\ regions subtend many 10's of
arcminutes or even degrees, and it is therefore more difficult for a
single slit position to characterize the emission-line properties.
These objects also present a greater challenge in cleanly identifying the
ionizing stellar population, owing to patchy extinction and distance
uncertainties.  To make the best comparison to our data, we therefore
take from their Galactic sample only those objects studied by Hunter
\& Massey (1990) and Esteban {\etal}(1993), plus the Orion nebula (M42).
Hunter \& Massey (1990) specifically examined a sample of
small Galactic \hii\ regions with the goal of spectroscopically
classifying the ionizing stars.  Any objects in the Hunter \& Massey
sample for which the dominant star was not spectroscopically
identified was also omitted.  The work by Esteban {\etal}(1993) was a
similar study, for WR nebulae.  We do include all of the KBFM LMC
objects that were not part of our study.
Table~\ref{tblKBFM} lists the objects we include from their sample,
along with the spectral types of the ionizing stars and references.
In Figure~\ref{integ1}, we show their objects ionized by (WR, O3--O4,
O5--O8, and O9 --B0) with the symbols (1, 2, 3, and 4), respectively.
The 30 Dor giant \hii\ region is indicated with a tick mark.

Figure~\ref{integ1}$a$ shows the conventional measures of the
radiation softness parameter $\eta$ vs. the ionization parameter $U$,
displaying the observables \etap\ vs. \oiii/\oii.  
%* NOTE IN FIG THAT THE LABELED DEFAULT MODELS FOR N=1, 10 ARE THE
%ONES THAT SHOULD BE COMPARED TO THE OTHERS.
We can see from the models that \etap\ is insensitive to metallicity, since the
numerator and denominator of equation~\ref{etap} correspond to the
ionization of single elements.  However, it is apparent that \etap\ is
signficantly affected by nebular morphology.  Stasi\'nska (1999) and
Garnett \& Kennicutt (1994)
demonstrated that changing the geometry between a filled and hollow
sphere produces different ionization properties than merely changing
the mean ionization parameter, although the local change in $U$
is the dominant effect.  The variation in \etap\ over the
parameter space for our models shows offsets by 0.3 -- 0.5 dex, which
is the same range in \etap\ displayed by the data itself.  

Empirically, however, Figure~\ref{integ1}$a$ shows that the data for
DEM~L323 and DEM~L301 do {\it not} exhibit the offset predicted in the
models.  These are the objects with similar O3 -- O4 ionizing stars,
but with Str\"omgren sphere vs. shell morphology.  DEM~L323
(triangles) shows values of \etap\ in the predicted range, and
DEM~L301 (cross) shows a similar 
value, rather than a higher one, as predicted for its shell
morphology.  As seen in Figure~\ref{d301dgn1}$g$, density-bounding and 
shock excitation can explain this behavior.  On the other hand,
DEM~L301 does exhibit the predicted strong decrease in \oiii/\oii. 

As found in \S 3 from the spatially resolved analysis, \etap\ is not
predicted to be sensitive to $T_\star\gtrsim 40$ kK for H-burning
stars, using the CoStar atmospheres.  Our spatially integrated data
also support this, since it is apparent that the data for DEM~L323
(triangles) and DEM~L243 (solid diamond) show values that are
difficult to distinguish.  Combined with the
KBFM data (2's and 3's), we do see a tendency for the later types to
fall to higher \etap, but it is impossible to empirically
differentiate the early and late types around \etap$\sim 0.5$.
The WR objects, however, do show significant differentiation from the
others.  DEM~L199 shows a typical observed value of
$\log\eta^\prime = -0.15$.  Although this \hii\ region includes
substantial shell structure (Figure~\ref{figd199}), the integrated
ionization parameter as indicated by \oiii/\oii\ is still the highest
of any of our objects.  This is also true in spite of the fact that the WR
stars are located at some distance from the high-surface brightness 
areas.  The KBFM data for 30 Dor do show a value closer to those of
early O star nebulae.  Massey \& Hunter (1998) suggest that most
of the WN stars in this object are actually core H-burning
stars that are so hot that they exhibit WR features.  The locus of the
KBFM nebular observations of 30 Dor are consistent with this
interpretation, although we again caution that they are taken from a
single slit position.  Turning to the KBFM objects dominated by O9 -- B0
stars (4's in Figure~\ref{integ1}$a$), we see that \etap\ indeed
recovers its sensitivity, and can reliably identify this stellar
population for \etap$\gtrsim 1$.

As expected, the integrated spectrum for DEM~L243 including the SNR
shows a hotter value of \etap\ than implied by the photoionized
region only.  The same appears to be be true for DEM L301.  It is
therefore important to test for the presence of superimposed SNRs or
shocks by some independent measure, such as the [\ion{O}{1}] 
line strengths.  Another strong discriminant would be the presence of
[\ion{Fe}{2}] and \ion{Ca}{2} lines in the visible or near IR. 
However, as we saw in the case of DEM L301, the presence of shocks may
be difficult to ascertain.  We also note that our nebular spectra
presented here do not resolve any velocity structure revealing the
shocks. 

In Figure~\ref{integ1}$b$ we show a similar diagram, now examining
whether \neiii/\Hb\ might be a useful indicator of \tstar.   
This diagnostic shows a larger range of values in
differentiating \tstar.  We initially investigated \neiii/\oiii\ and
\neiii/\ariii, but these line ratios did not improve the range of
sensitivity as much as simply using \neiii/\Hb.  In contrast to \etap,
\neiii/\Hb\ is fairly insensitive to the nebular morphology; the
variation with morphology follows the same track as that for
ionization parameter. 
However, \neiii/\Hb\ should of course be completely sensitive to
the metallicity, as can be seen from the effect of the 0.28 dex
variation in Ne abundance between the adopted LMC and SMC metallicity
in these models.  It is also apparent that the data are not falling on
the predicted locus of the models.  This is unsurprising, since the
\neiii\ intensity appears to be systematically somewhat overpredicted
in the models shown in \S 3.  Likewise, \oiii/\oii\ is often slightly
underpredicted by the models, resulting in the offset between the data
and predictions in Figure~\ref{integ1}$b$.
We therefore recommend using an empirical calibration if these
diagnostics are used to infer \tstar.

Nevertheless, the data form a fairly clean sequence in
Figure~\ref{integ1}$b$.  KBFM measurements with no \neiii\ detection are
shown along the bottom, at their measured \oiii/\oii\ values.
Surprisingly, the Galactic and LMC objects do not show the predicted
dependence on abundance; additional data are necessary to understand
the metallicity dependence.  The scatter in the points for DEM~L323
(triangles) and DEM~L243 (diamonds) shows  
that it is still difficult to reliably distinguish the O7 and O3
stellar temperatures, although the degeneracy is less
than for \etap.  Furthermore, it appears that this diagnostic is
less sensitive to the presence of shock excitation, since the loci for
DEM L243 with shock and DEM L301 do not show obviously anomalous
positions.  Presumably only faster shocks would be able to produce
enough Ne$^{++}$ to significantly affect the \neiii/\Hb\ ratio.
We do see an anomalous locus for one WR object in the KBFM LMC sample,
which corresponds to N51 D, a superbubble \hii\ region.  This object
does not follow the predicted track for shell objects, nor that for
WR-dominated objects.  The discrepancy may again be related to the
localized sampling of nebular conditions.  There are also a couple
of O9 -- B0 objects showing apparently significant \neiii/\Hb\
fluxes; the one with higher \oiii/\oii\ is S288, whose spectral type
is not as well-determined (Table~\ref{tblKBFM}).
%  The other object
%may be a spurious \neiii\ detection (***RCK or FB, private communication).

Based on Figure~\ref{integ1}$b$, we tentatively suggest an empirical
calibration for \neiii/\Hb\ and \etap, for LMC metallicity, given in
Table~\ref{calib}.  We do not assign \tstar\ in Table~\ref{calib},
since the evidence discussed above suggests that the spectral type --
\tstar\ conversion used with the CoStar models gives \tstar\ that are 
somewhat too hot. 
%\lam3869/\Hb\ $> -0.6$:  WR;  $-0.9<
%\lambda3869/\hb < -0.6$:  O3 -- O4; $ 0 < \lambda3869/\hb < -0.9$:  O5
%-- O8; and  $\lambda3869/\hb = 0$:  O9 and later.

We defer a full discussion of the abundance parameters, $R23$ and
$S23$ to Paper~II.  However, it is worth emphasizing their
\tstar-dependence here.  As can be seen in Table~\ref{integ}, the
observed values of $R23$ and $S23$ for the integrated spectra do
follow a sequence such that lower values correspond to cooler \tstar.
As seen in the spatially-resolved models of \S 3, the predicted values
are in good agreement with the data, although tending to be slight
overestimates. 

\section{Conclusion}

We have obtained both spatially resolved and integrated observations
of four \hii\ regions, with ionizing stellar spectral 
types of O7, O3--4, and WN3, and morphologies ranging from Str\"omgren
sphere to extreme shell.  Two of the objects show evidence of shock
excitation in addition to the dominant photoionization.
Empirically, we find that lower-ionization species such as \sii\
and \oii\ show a large degree of intrinsic scatter, spatially.  This
is likely to be caused by conditions such as stronger density
fluctuations in the outer regions where these species dominate.  Any
inferences extrapolated from spatially-resolved observations involving these
species should therefore be interpreted with care.  Likewise, our
scanned observations of the almost-spherical object DEM L323, show that
spatial subsampling of an \hii\ region can show significant
variation from the total integrated spectrum, even when the subsampled
region includes all ionization zones.

Overall, the spatially-resolved, optical emission-line diagnostics
are in good agreement with the current generation of hot star
atmospheres for our objects.  For O stars, we tested CoStar atmospheres
of Schaerer \& de Koter (1997), and for WNE stars, the models of
Schmutz {\etal}(1992) and Hamann \& Koesterke (1998).  In general, the
{\sc Mappings} photoionization models and the observations agree to
within $\lesssim 0.2$ dex.  The remaining discrepancies may 
be just as likely to be caused by inaccuracies in the photoionization
code as in the stellar atmosphere models.  There is also the
likelihood that systematic effects in density and geometry, affecting the local
ionization parameter, can systematically affect the observed diagnostics.

The primary trend of this remaining disagreement between observations and
models is that the nebular models appear to be too hot.  This discrepancy is
in the opposite sense of that expected from $T_e$ fluctuations
(Peimbert 1967).  In the two objects
with high-quality measurements of \oiii\lam4363/5007, the \oiii\ electron
temperature appears to be systematically overpredicted:  in DEM L323
by 850~K, and in DEM L199 by 1500~K.  The results are also consistent
with a similar discrepancy in DEM L243, but the standard deviation in
derived $T_e$ for this object is too high to constrain this.  There is
a similar problem with data quality for DEM L301, in addition to
complications with geometry and excitation mechanism.  The discrepancy
is clear for DEM L323 and DEM L199, however, and may be related to the
systematic slight overestimate in the high-ionization species \neiii. 
This problem is most pronounced in \neiii/\oiii.  Because of the
relatively small difference in IP of 5.85 eV required for these two
species, the deviation may be due in part to small-scale features in
the stellar energy distribution, or the depth of the Ne$^{++}$ edge in
the atmospheres.  It is interesting that the
more highly-resolved WNE energy distribution by Hamann \& Koesterke
(1998) yields an identical result for \neiii/\oiii\ as that of Schmutz
{\etal}(1992) (Figure~\ref{d199dgn1}$i$; the lines cannot be
distinguished).  Furthermore, the models produced with the Hummer \&
Mihalas (1970) atmospheres exhibit the same problem.
It is unclear whether the overestimated $T_e$ are
caused by the stellar atmosphere models or the photoionization codes.
A test run with {\sc Cloudy} (e.g., Ferland 1998) confirms
the result with that code.

As has been found by Sellmaier (1997) and Stasi\'nska \& Schaerer
(1997), we confirm that the earlier problem in ionizing Ne to
Ne$^{++}$ has been resolved in the new generation O star models.
Whereas with earlier generation atmospheres the \neiii/\oiii\ ratio was
underestimated by $\sim 0.6$ dex, our models now show an overestimate
of $\lesssim 0.2$ dex.  As mentioned above, the source of the
remaining discrepancy is unclear, and seems unlikely
to be caused by an overestimate in the Ne abundance.
There does seem to be a hint that
the stellar atmospheres for the O7 stars in DEM L243, at least, are
slightly too hard.  It appears that He is not uniformly ionized to
He$^+$, and S appears to be slightly over-ionized compared to the
data, as seen in the ratios of \siii/\sii\ and \sii/\Ha.  

We find no evidence of a problem with the ratio \siii/\sii,
as has occasionally been suggested in the past (e.g., Garnett 1989;
Dinerstein \& Shields 1986).  While our models and observations are
not always perfectly matched, the discrepancies are not large, and
show both over- and underprediction by the models.  Our data suggest
significant fractions of S$^{+3}$
and perhaps even S$^{+4}$, as seen in the regions closest to the WR
stars in DEM L199 and in the spatial behavior of the $S23$ parameter.
Thus for spatially-resolved nebular observations, the behavior of
the S lines might be another potentially useful probe of \tstar.

DEM L199 is an important test of WNE-w model atmospheres.  The
dominant stars are Br~33 and Br~34, of spectral type WN3-w, with a possible
significant contribution by Br~32 (WC4).  
%Br~34 has a similar spectral type,
%but the nebular imaging in \oiii\ and \sii\ suggests it might be
%cooler than Br~33.
The lack of \hei\ \lam5876 emission from either Br~33 or Br~34
suggests that they are hot enough to fully ionize He$^0$ in
their atmospheres.  However, peculiarly, there is no evidence of any
\ion{He}{2} \lam4686 emission from the nebula.  This corresponds to a
very confined, boundary range of conditions in the stellar atmosphere
models, and it is puzzling that both of these stars exhibit these
properties.  As mentioned above, both the pure He model atmosphere by Schmutz
{\etal}(1992) and the latest generation models by Hamann \& Koesterke
(1998) yield photoionization models of the \hii\ complex that are in
excellent agreement with the observations, and show the same patterns
with respect to the data as the CoStar atmospheres.  For WR
atmospheres that do not produce He$^+$-ionizing photons, the optical
nebular diagnostics are actually quite insensitive to the energy
distributions; models from the two different groups produce virtually
identical results. 

We find that DEM L301 shows excellent agreement with the composite
spectrum of a density-bounded, shell \hii\ region and a $70\ \kms$ shock.
The object is not strongly-density bounded, in spite of predictions
that the \Ha\ emission of the nebula accounts for only $\sim 20$\% of
the ionizing emission from the stars.  The filamentary structure of
this extreme shell object suggests that these shock-produced filaments
trap the stellar ionizing radiation, producing spectral properties roughly
resembling radiation-bounded photoionization.  Nevertheless, the majority
of ionizing photons are likely to be escaping through the holes between the
filaments.  This implies that enormous leakage of ionizing radiation
can take place without producing signatures of strong density-bounding.

The two objects in our sample with evidence of shock excitation do
indeed show spectral diagnostics that agree well with composite models
including both photoionization and shock components.  DEM L243 (N63
A) has an embedded or superimposed SNR, and its emission-line spectrum is
consistent with a shock velocity of $\sim 200\ \kms$.  DEM L301 (N70)
shows evidence of a recent SNR impact, and its spectrum is consistent
with a lower shock velocity of $\sim 70\ \kms$.  For both objects, the
shock velocities agree with kinematic observations.  We note that
the existence of the shocks in both objects is difficult to discern
in the spatially integrated spectra.  The higher-velocity
shock in DEM L243 shows optical line ratios that are closer to those
from photoionization by hot stars, and are therefore easily
diluted.  However, the lower-velocity shock in DEM L301 has
such weak emission in \oiii\lam5007 that line ratios such as
\lam4363/\lam5007 are also extremely diluted in combination with
photoionized emission.  The historical confusion over the excitation
mechanism in DEM L301 demonstrates the subtleties of these effects.
Contamination by shock emission clearly decreases the observed value
of \etap\ in the spatially integrated spectra.  This results in an
inferred \tstar\ that is spuriously high.  Given the difficulty in
diagnosing the presence of a contaminating shock, this would be an
additional uncertainty in the application of \etap.  $R23$ can also be
affected, especially by higher-velocity shocks.

Previous authors (e.g., Skillman 1989; Bresolin {\etal}1999) have
noted difficulties in using the \etap\ parameter for estimating
\tstar.  In addition to its sensitivity to $U$, the
value of \etap\ also tends to saturate for O stars above a certain \tstar.  
Empirically, \etap\ has a similar a value, around 2, for our O star \hii\
regions, which have dominant stellar spectral types ranging from O3 to O7.
While for earlier generations of stellar atmosphere models,
the regime of sensitivity to \tstar\ extended into O star spectral
types, we show that for the CoStar atmospheres, \etap\ is an ineffective
discriminant for O star $T_\star \gtrsim 40$ kK.  Objects dominated by
WR stars, however, and O9 -- B0 stars, are still clearly
differentiated by \etap.  We also 
show that, in contrast to spatially resolved observations of, e.g., 30
Dor (V\'\i lchez \& Pagel 1988), there is often a strong spatial variation in
\etap, for example in our data for DEM L323 and DEM L243.  Finally, we
demonstrate that \etap\ is sensitive to nebular morphology, with shell
structures having significantly higher values.

Given the limitations of \etap\ as an indicator of \tstar, we suggest
that examination of the simple line strength of \neiii\lam3869 can be an
additional useful constraint in estimating \tstar.  While \neiii/\Hb\
is dependent on metallicity, it nevertheless
offers some advantages over \etap.  In the first instance, it is
more sensitive than \etap\ in differentiating O star spectral types.
For our photoionization models, \neiii/\Hb\ varies by about 0.75 dex
between types O9 and O3, as opposed to 0.35 dex for \etap.
Unfortunately, the modeled difference between O3 and O7 spectral types is
still $\lesssim 0.1$ dex.  We note, however, that the spatially
integrated data for our O3 nebula, DEM 323, and our O7 nebula, DEM
243, appear better differentiated in \neiii/\Hb\ than \etap.  This
stronger differentiation is supported by data for additional objects
from the forthcoming sample of Kennicutt et al.  Another
advantage of \neiii/\Hb\ is its lower sensitivity to the presence of
contaminating shocks, owing to the high IP required for Ne$^{++}$.
This is also demonstrated by our spatially integrated data.  Finally,
\neiii/\Hb\ is almost independent of nebular morphology, and somewhat
less sensitive to $U$.  The primary disadvantage, aside from presumable
metallicity sensitivity, is that the modeled values are not in
agreement with the data, and hence an empirical calibration is
necessary.  The observations presented here provide some initial
data points for this purpose, suggesting a rough empirical calibration
at LMC metallicity given by Table~\ref{calib}.

%\placetable{tbl-3}  |
%\placefigure{fig1}

\acknowledgments

We have enjoyed discussions of this work with many people, including
Paul Crowther, Evan Skillman, Wolf-Rainer Hamann, Henny Lamers, John
Mathis, Werner Schmutz, and Bill Blair's SNR lunch group.  We are
especially grateful to Don Garnett for discussions and comments on the
manuscript.  Special thanks to Rob Kennicutt and Fabio Bresolin for
providing their data in advance of publication, and also to
Wolf-Rainer Hamann for providing model WR atmospheres from his
group.  We are grateful to Mark Allen, Daniel Schaerer, and Grazyna
Stasi\'nska for help in sorting out photoionization codes, and also
thank Ken Freeman for assistance at the telescope.  Finally, thanks to the
referee, Jim Kingdon, for helpful comments.  Some of
this work was carried out by MSO while at the Institute of
Astronomy, Cambridge.

% Two options are available to the author for producing tables:  the
% "deluxetable" environment provided by the AASTeX package or the LaTeX
% "table" environment.  The AASTeX "deluxetable" environment is preferred
% by the Production Offices.  Only short tables should be included in the
% body of the text.  If tables extend over a page they should be generated
% using either the apjpt4 or aj_pt4 style file; these styles also use the 
% "deluxetable" environment - but these tables will be produced as
% "camera-ready".
%
% The caption contains only the caption text.  The "Table N." identification
% is generated by the \tablecaption command on its own.  It is necessary to 
% \label tables and figures *after* the caption has been specified because 
% the table/figure number is generated by the caption, not by \begin{whatever}.
% The column headings are specified within a \colhead command and all the
% column headings are included within a single \tablehead command.  The
% \enddata command comes at the end of the data, and the table is closed with 
% an \end{deluxetable} command.  It the table is too wide for the page, \small
% (11pt), \footnotesize (10pt), or \scriptsize (8pt) may be used inside
% the deluxetable environment - the table will still be double-spaced.  For
% even wider tables see the AASTeX guide.

\onecolumn
\clearpage

\begin{deluxetable}{clllll}
%\scriptsize
\footnotesize
\tablecaption{\hii\ region sample and ionizing stars \label{sample}}
\tablewidth{0 pt}
\tablehead{
\colhead{DEM} & \colhead{Henize}   & \colhead{LH}   & \colhead{Star ID} & 
\colhead{Sp. Type}  &  \colhead{Reference} 
} 
\startdata
DEM L199 & N144 & LH 58    & Br 32 & WC4 + O6 V-III & Moffat {\etal}(1990) \nl
 & & &                       Br 33 & WN3 + OB     & Breysacher (1981) \nl
 & & &                       Br 34 & WN3: + B3 I  & Breysacher (1981) \nl
 & & &                       496   & O3-4 V   & Garmany {\etal}(1994) \nl
\hline
DEM L243 & N63 A & LH 83   & 2  & O7 V((f)) &  Oey (1996a) \nl
 & & &                       5  & O7 If     &  Oey (1996a) \nl
\hline
DEM L301 & N70 & LH 114    & SW3      & O3 If*    &  Oey (1996a) \nl
 & & &                       NW4      & O5: III:e &  Oey (1996a) \nl
\hline
DEM L323 & N180 B & LH 117 & LH 117--118 & O4: III(f) & Massey {\etal}(1989) \nl
 & & &                       LH 117--140 & O3-4 (f*)  & Massey {\etal}(1989) \nl
 & & &                       LH 117--214 & O3-4       & Massey {\etal}(1989) \nl
\enddata
\end{deluxetable}

\begin{deluxetable}{crrrrrrrrrrrrrr}
\scriptsize
\tablecaption{Dereddened line intensities for stationary 
	positions\tablenotemark{a} \label{lines}}
%\tablenum{2---{\it continued}}
\tablewidth{0 pt}
\tablehead{\multicolumn{15}{c}{D199.205     } \\
\colhead{line (\AA)}      & \colhead{ap10} & \colhead{err}
      & \colhead{ap11} & \colhead{err}
      & \colhead{ap13} & \colhead{err}
      & \colhead{ap14} & \colhead{err}
      & \colhead{ap15} & \colhead{err}
      & \colhead{ap16} & \colhead{err}
      & \colhead{ap17} & \colhead{err}
}
\startdata
3727   & 314.5& 15.8 & 272.0& 13.6 & 216.9& 10.9 &  33.6&  1.9 & 153.7&  7.8 & 212.7& 10.8 & 193.0&  9.9 \\
3869   &  21.6&  1.4 &  29.5&  1.6 &  27.0&  1.4 &  24.4&  1.3 &  22.8&  1.5 &  33.3&  2.1 &  24.4&  1.9 \\
4069   &   3.4&  0.9 &   2.2&  0.5 &   2.3&  0.4 &   0.0&  0.6 &   0.6&  0.9 &   2.4&  1.3 &   2.6&  1.5 \\
4076   &   0.5&  0.9 &   0.9&  0.5 &   0.5&  0.4 &   0.1&  0.6 &   0.0&  0.9 &   0.5&  1.3 &   0.5&  1.5 \\
4101   &  23.5&  1.5 &  23.8&  1.3 &  25.6&  1.3 &  24.8&  1.4 &  23.4&  1.5 &  20.1&  1.6 &  21.6&  1.8 \\
4340   &  45.4&  2.3 &  45.4&  2.3 &  46.1&  2.3 &  46.1&  2.3 &  44.5&  2.3 &  43.5&  2.3 &  42.1&  2.4 \\
4363   &   1.7&  0.6 &   2.4&  0.3 &   2.0&  0.2 &   1.7&  0.4 &   1.9&  0.7 &   1.5&  0.8 &   2.6&  1.1 \\
4471   &   3.6&  0.6 &   3.5&  0.5 &   4.0&  0.5 &   3.8&  0.6 &   2.7&  0.5 &   2.4&  0.7 &   2.9&  0.8 \\
4861   & 100.0&  5.0 & 100.0&  5.0 & 100.0&  5.0 & 100.0&  5.0 & 100.0&  5.0 & 100.0&  5.1 & 100.0&  5.1 \\
4959   &  91.4&  4.6 & 122.9&  6.2 & 117.8&  5.9 & 140.1&  7.0 & 111.8&  5.6 & 128.0&  6.4 & 129.4&  6.5 \\
5007   & 272.7& 13.6 & 367.9& 18.4 & 350.0& 17.5 & 420.8& 21.0 & 338.1& 16.9 & 380.5& 19.0 & 388.0& 19.4 \\
5200   &   1.2&  0.6 &   0.7&  0.5 &   0.2&  0.4 &   0.0&  0.6 &   0.0&  0.5 &   0.0&  0.7 &   0.0&  0.8 \\
5755   &   0.6&  0.6 &   0.6&  0.4 &   0.5&  0.2 &   0.4&  0.3 &   0.7&  0.5 &   1.1&  0.8 &   0.0&  0.9 \\
5876   &  11.3&  0.8 &  12.0&  0.7 &  11.7&  0.6 &  12.5&  0.7 &  12.3&  0.8 &  12.2&  1.0 &  10.1&  1.0 \\
6300   &   5.2&  0.6 &   4.6&  0.4 &   2.6&  0.2 &   0.0&  0.4 &   0.0&  0.5 &   0.0&  0.8 &   0.0&  0.9 \\
6312   &   1.6&  0.6 &   1.5&  0.4 &   1.3&  0.2 &   1.5&  0.3 &   1.2&  0.5 &   2.6&  0.8 &   1.2&  0.9 \\
6548   &  11.6&  0.7 &   9.5&  0.5 &   8.1&  0.4 &   1.4&  0.2 &   5.7&  0.4 &   5.1&  0.4 &   6.9&  0.6 \\
6563   & 285.3& 14.3 & 287.7& 14.4 & 285.2& 14.3 & 286.1& 14.3 & 276.7& 13.8 & 261.9& 13.1 & 274.7& 13.7 \\
6584   &  36.4&  1.9 &  29.2&  1.5 &  24.4&  1.2 &   3.0&  0.2 &  18.6&  1.0 &  16.3&  0.9 &  20.3&  1.1 \\
6678   &   3.7&  0.4 &   3.5&  0.3 &   3.5&  0.2 &   3.9&  0.3 &   3.9&  0.4 &   3.8&  0.3 &   3.6&  0.5 \\
6716   &  52.7&  2.7 &  37.8&  1.9 &  31.2&  1.6 &   0.0&  0.2 &  16.5&  0.9 &  15.6&  0.8 &  24.0&  1.3 \\
6731   &  37.4&  1.9 &  26.7&  1.4 &  22.0&  1.1 &   0.0&  0.2 &  11.2&  0.6 &  11.1&  0.6 &  16.6&  1.0 \\
7065   &   1.9&  0.4 &   2.1&  0.2 &   2.1&  0.2 &   2.1&  0.2 &   2.1&  0.3 &   2.2&  0.3 &   2.1&  0.5 \\
7136   &  10.0&  0.6 &  12.3&  0.6 &  11.1&  0.6 &  10.5&  0.6 &  10.5&  0.6 &  10.7&  0.6 &  11.0&  0.7 \\
9069   &  23.0&  1.2 &  28.2&  1.4 &  24.7&  1.2 &  21.6&  1.1 &  20.6&  1.1 &  20.3&  1.1 &  25.9&  1.5 \\
 \etap  &   0.8&  0.1 &   0.8&  0.1 &   0.8&  0.1 &   0.0& -0.3 &   0.9&  0.1 &   1.1&  0.2 &   0.8&  0.1 \\
$R23$   &   6.8&  0.6 &   7.6&  0.5 &   6.8&  0.5 &   5.9&  0.4 &   6.0&  0.4 &   7.2&  0.5 &   7.1&  0.5 \\
$S23$   &  1.71& 0.11 &  1.63& 0.11 &  1.40& 0.09 &  0.73& 0.06 &  1.00& 0.07 &  0.98& 0.07 &  1.31& 0.09 \\
$c$\tablenotemark{b} & 0.00 & 0.05 & 0.00 & 0.05 & 0.00 & 0.05 & 0.00 & 0.05 & 0.11 & 0.05 & 0.11 & 0.05 & 0.11 & 0.05 \\
\enddata
\tablenotetext{a}{Relative to \Hb.}
\tablenotetext{b}{Adopted reddening correction.}
\tablenotetext{c}{Affected by SNR.}
\end{deluxetable}

\topsep=0.1in
\textheight=22cm
\topmargin=-0.5cm

\begin{deluxetable}{crr|rrrrrrrrrrrr}
\scriptsize
\tablecaption{}
\tablenum{2---{\it continued}}
\tablewidth{0 pt}
\tablehead{& \multicolumn{2}{c}{D199.205 }& \multicolumn{12}{c}{D199.205N120 } \\
\colhead{line (\AA)}      & \colhead{ap18} & \colhead{err}
      & \colhead{ap 9} & \colhead{err}
      & \colhead{ap10} & \colhead{err}
      & \colhead{ap11} & \colhead{err}
      & \colhead{ap12} & \colhead{err}
      & \colhead{ap13} & \colhead{err}
      & \colhead{ap14} & \colhead{err}
}
\startdata
3727   & 163.1&  9.2 & 230.8& 17.1 & 389.0& 19.5 & 260.6& 13.1 & 259.4& 13.0 & 250.8& 12.6 & 184.6&  9.3 \\
3869   &  30.7&  3.3 &  10.4&  8.9 &  26.8&  1.8 &  23.9&  1.6 &  17.1&  0.9 &  20.4&  1.1 &  23.3&  1.3 \\
4069   &   0.0&  3.0 &   0.0&  8.9 &   2.9&  1.2 &   1.6&  1.1 &   2.3&  0.4 &   1.9&  0.5 &   1.5&  0.5 \\
4076   &   0.6&  3.0 &   2.7&  8.9 &   0.0&  1.2 &   0.0&  1.1 &   0.0&  0.4 &   0.5&  0.5 &   0.4&  0.5 \\
4101   &  17.6&  3.1 &  17.8&  9.0 &  26.2&  1.7 &  23.5&  1.6 &  26.0&  1.4 &  26.0&  1.4 &  25.1&  1.3 \\
4340   &  38.2&  3.5 &  49.6&  4.1 &  47.5&  2.5 &  47.0&  2.4 &  47.9&  2.4 &  48.4&  2.4 &  46.6&  2.3 \\
4363   &   0.6&  2.9 &   0.0&  3.3 &   1.9&  0.9 &   1.6&  0.6 &   1.6&  0.2 &   1.5&  0.2 &   1.5&  0.2 \\
4471   &   2.8&  2.0 &   7.1&  5.5 &   5.1&  1.2 &   4.0&  0.7 &   3.6&  0.3 &   3.9&  0.4 &   4.3&  0.4 \\
4861   & 100.0&  5.4 & 100.0&  7.4 & 100.0&  5.1 & 100.0&  5.0 & 100.0&  5.0 & 100.0&  5.0 & 100.0&  5.0 \\
4959   & 131.0&  6.8 &  74.9&  6.6 &  89.2&  4.6 &  96.4&  4.9 &  77.6&  3.9 &  84.4&  4.2 & 110.3&  5.5 \\
5007   & 387.1& 19.5 & 227.8& 12.6 & 259.6& 13.0 & 284.0& 14.2 & 231.6& 11.6 & 250.8& 12.5 & 331.7& 16.6 \\
5200   &   0.0&  2.0 &   1.5&  5.5 &   1.3&  1.1 &   1.1&  0.6 &   0.9&  0.3 &   0.7&  0.3 &   0.6&  0.3 \\
5755   &   2.5&  2.4 &   0.0&  3.3 &   0.8&  0.7 &   1.0&  0.6 &   0.4&  0.2 &   0.4&  0.2 &   0.1&  0.2 \\
5876   &  12.8&  2.4 &  16.6&  3.4 &  11.4&  0.9 &  11.7&  0.8 &  11.0&  0.6 &  11.8&  0.6 &  12.3&  0.6 \\
6300   &   3.7&  2.4 &  13.5&  3.3 &   5.6&  0.8 &   2.0&  0.6 &   2.8&  0.2 &   2.4&  0.3 &   1.7&  0.2 \\
6312   &   0.0&  2.4 &   1.7&  3.3 &   2.0&  0.7 &   1.1&  0.6 &   1.0&  0.2 &   1.1&  0.3 &   1.1&  0.2 \\
6548   &   5.2&  1.1 &   8.8&  1.9 &  13.0&  0.8 &   9.2&  0.6 &   9.8&  0.5 &   9.5&  0.5 &   6.5&  0.4 \\
6563   & 264.5& 13.3 & 295.1& 14.9 & 292.7& 14.6 & 291.4& 14.6 & 294.4& 14.7 & 308.3& 15.4 & 298.0& 14.9 \\
6584   &  19.6&  1.4 &  23.0&  2.2 &  40.6&  2.1 &  26.7&  1.4 &  29.3&  1.5 &  28.2&  1.4 &  19.3&  1.0 \\
6678   &   4.2&  1.1 &   3.4&  1.9 &   3.0&  0.4 &   3.0&  0.5 &   3.2&  0.2 &   3.5&  0.3 &   3.8&  0.2 \\
6716   &  30.9&  1.9 &  29.8&  2.4 &  53.6&  2.7 &  33.0&  1.7 &  35.5&  1.8 &  32.4&  1.6 &  21.6&  1.1 \\
6731   &  21.3&  1.5 &  20.1&  2.1 &  36.9&  1.9 &  23.1&  1.2 &  24.9&  1.3 &  22.7&  1.2 &  15.1&  0.8 \\
7065   &   2.7&  1.0 &   1.5&  1.8 &   1.8&  0.4 &   2.0&  0.5 &   1.8&  0.2 &   2.0&  0.2 &   2.3&  0.2 \\
7136   &   9.1&  1.1 &  10.5&  1.9 &  10.9&  0.7 &   9.8&  0.7 &   8.3&  0.5 &   9.6&  0.5 &  11.1&  0.6 \\
9069   &  23.0&  2.0 &  26.2&  2.4 &  24.1&  1.3 &  21.4&  1.1 &  21.2&  1.1 &  23.0&  1.2 &  23.7&  1.2 \\
 \etap  &   0.5&  0.1 &   1.4&  0.2 &   1.0&  0.2 &   0.9&  0.1 &   1.0&  0.2 &   1.1&  0.2 &   0.9&  0.2 \\
$R23$   &   6.8&  0.4 &   5.3&  0.5 &   7.4&  0.7 &   6.4&  0.5 &   5.7&  0.5 &   5.9&  0.5 &   6.3&  0.4 \\
$S23$   &  1.33& 0.09 &  1.42& 0.13 &  1.75& 0.11 &  1.31& 0.08 &  1.35& 0.09 &  1.36& 0.09 &  1.20& 0.08 \\
$c$ & 0.20 & 0.05 & 0.20 & 0.05 & 0.11 & 0.05 & 0.11 & 0.05 & 0.20 & 0.05 & 0.29 & 0.05 & 0.11 & 0.05 \\
\enddata
\end{deluxetable}

\begin{deluxetable}{crrrrrrrrrr|rrrr}
\scriptsize
\tablecaption{}
\tablenum{2---{\it continued}}
\tablewidth{0 pt}
\tablehead{& \multicolumn{10}{c}{D199.205N120 }& \multicolumn{ 4}{c}{D243.2S      } \\
\colhead{line (\AA)}      & \colhead{ap15} & \colhead{err}
      & \colhead{ap16} & \colhead{err}
      & \colhead{ap17} & \colhead{err}
      & \colhead{ap18} & \colhead{err}
      & \colhead{ap19} & \colhead{err}
      & \colhead{ap 7} & \colhead{err}
      & \colhead{ap 6} & \colhead{err}
}
\startdata
3727      & 232.5& 11.7 & 184.5&  9.5 &  76.4&  4.3 & 254.2& 12.8 & 163.4&  8.4 & 497.1& 25.7 & 250.9& 12.7 \\
3869      &  23.5&  1.4 &  29.2&  2.1 &  33.5&  2.2 &  31.4&  2.0 &  37.2&  2.3 &   0.0&  4.5 &   6.7&  1.5 \\
4069      &   1.5&  0.7 &   1.4&  1.5 &   0.1&  1.4 &   0.9&  1.2 &   0.4&  1.3 &   4.9&  4.5 &   2.0&  1.4 \\
4076      &   0.0&  0.7 &   0.0&  1.5 &   0.0&  1.4 &   0.0&  1.2 &   0.3&  1.3 &   0.0&  4.5 &   0.0&  1.4 \\
4101      &  26.4&  1.5 &  24.0&  1.9 &  23.1&  1.8 &  24.9&  1.8 &  24.4&  1.8 &  28.6&  4.7 &  26.6&  2.0 \\
4340      &  47.0&  2.4 &  47.0&  2.4 &  48.5&  2.6 &  47.8&  2.5 &  46.5&  2.5 &  53.8&  3.2 &  49.7&  2.6 \\
4363      &   1.7&  0.4 &   2.2&  0.6 &   2.0&  0.9 &   1.5&  0.6 &   2.6&  1.0 &   4.0&  1.8 &   2.6&  0.8 \\
4471      &   4.3&  0.5 &   4.8&  0.5 &   5.0&  0.8 &   4.0&  0.6 &   5.2&  0.6 &   0.0&  1.6 &   3.7&  0.6 \\
4861      & 100.0&  5.0 & 100.0&  5.0 & 100.0&  5.1 & 100.0&  5.0 & 100.0&  5.0 & 100.0&  5.2 & 100.0&  5.0 \\
4959      & 110.1&  5.5 & 138.2&  6.9 & 155.0&  7.8 & 129.1&  6.5 & 154.3&  7.7 &  13.2&  1.7 &  58.1&  3.0 \\
5007      & 330.8& 16.5 & 411.4& 20.6 & 462.3& 23.1 & 384.9& 19.3 & 460.5& 23.0 &  37.7&  2.5 & 175.5&  8.8 \\
5200      &   0.5&  0.4 &   0.0&  0.5 &   0.0&  0.7 &   0.0&  0.5 &   0.0&  0.6 &   0.6&  1.6 &   0.0&  0.6 \\
5755      &   0.4&  0.3 &   0.5&  0.4 &   0.8&  0.7 &   0.8&  0.6 &   0.4&  0.5 &   0.1&  0.9 &   0.0&  0.4 \\
5876      &  11.8&  0.7 &  11.9&  0.7 &  12.5&  0.9 &  11.9&  0.9 &  12.8&  0.8 &   4.9&  0.9 &  10.5&  0.7 \\
6300      &   1.4&  0.3 &   0.6&  0.4 &   0.0&  0.7 &   0.0&  0.6 &   2.3&  0.5 &   3.5&  0.9 &   0.0&  0.4 \\
6312      &   1.2&  0.3 &   1.6&  0.4 &   1.8&  0.7 &   1.5&  0.6 &   1.9&  0.5 &   0.0&  0.9 &   0.5&  0.4 \\
6548      &   7.3&  0.4 &   6.7&  0.4 &   1.0&  0.4 &   5.9&  0.5 &   3.0&  0.5 &  16.5&  1.4 &   7.2&  0.5 \\
6563      & 295.3& 14.8 & 294.1& 14.7 & 286.2& 14.3 & 281.4& 14.1 & 295.4& 14.8 & 320.4& 16.1 & 299.8& 15.0 \\
6584      &  21.9&  1.1 &  19.8&  1.0 &   3.8&  0.5 &  19.6&  1.0 &  16.4&  0.9 &  46.4&  2.6 &  19.9&  1.1 \\
6678      &   3.7&  0.3 &   3.6&  0.3 &   3.1&  0.4 &   3.6&  0.4 &   3.1&  0.5 &   1.9&  1.1 &   3.6&  0.4 \\
6716      &  21.3&  1.1 &  25.6&  1.3 &   0.8&  0.4 &  20.0&  1.1 &  20.2&  1.1 &  47.6&  2.6 &  12.8&  0.7 \\
6731      &  15.2&  0.8 &  18.0&  0.9 &   0.4&  0.4 &  13.9&  0.8 &  13.7&  0.8 &  33.2&  2.0 &   8.8&  0.6 \\
7065      &   2.2&  0.3 &   2.1&  0.3 &   2.1&  0.4 &   2.1&  0.4 &   2.2&  0.5 &   1.0&  1.1 &   1.9&  0.4 \\
7136      &  11.1&  0.6 &  11.9&  0.7 &  11.2&  0.7 &  11.8&  0.7 &  13.2&  0.8 &   2.5&  1.1 &   8.0&  0.5 \\
9069      &  23.6&  1.2 &  24.8&  1.3 &  22.0&  1.5 &  23.1&  1.4 &  28.9&  2.0 &  15.8&  1.4 &  19.9&  1.1 \\
 \etap  &   1.2&  0.2 &   0.7&  0.1 &   7.7&  1.3 &   1.2&  0.2 &   0.8&  0.1 &   6.7&  1.1 &   3.5&  0.6 \\
$R23$   &   6.7&  0.5 &   7.3&  0.5 &   6.9&  0.4 &   7.7&  0.5 &   7.8&  0.5 &   5.5&  0.8 &   4.8&  0.4 \\
$S23$   &  1.19& 0.08 &  1.30& 0.09 &  0.78& 0.07 &  1.15& 0.08 &  1.35& 0.10 &  1.36& 0.09 &  0.91& 0.06 \\
$c$ & 0.11 & 0.05 & 0.11 & 0.05 & 0.00 & 0.05 & 0.11 & 0.05 & 0.11 & 0.05 & 0.08 & 0.05 & 0.08 & 0.05 \\
\enddata
\end{deluxetable}
 
\begin{deluxetable}{crrrrrrrrrr|rrrr}
\scriptsize
\tablecaption{}
\tablenum{2---{\it continued}}
\tablewidth{0 pt}
\tablehead{& \multicolumn{10}{c}{D243.2S      }& \multicolumn{ 4}{c}{D243.5S      } \\
\colhead{line (\AA)}      & \colhead{ap 9} & \colhead{err}
      & \colhead{ap12\tablenotemark{c}} & \colhead{err}
      & \colhead{ap13\tablenotemark{c}} & \colhead{err}
      & \colhead{ap10\tablenotemark{c}} & \colhead{err}
      & \colhead{ap14} & \colhead{err}
      & \colhead{ap 3} & \colhead{err}
      & \colhead{ap 4} & \colhead{err}
}
\startdata
3727      & 222.0& 11.2 & 179.3&  9.3 & 187.0&  9.7 & 271.1& 13.7 & 341.7& 17.2 & 341.0& 17.2 & 277.4& 15.4 \\
3869      &   9.1&  1.1 &  22.2&  2.0 &  20.2&  2.2 &  17.3&  1.4 &  13.8&  1.3 &   0.7&  1.4 &   4.0&  4.8 \\
4069      &   2.6&  1.1 &   1.6&  1.7 &   1.5&  1.9 &   2.5&  1.2 &   2.5&  1.1 &   2.1&  1.4 &   0.9&  4.8 \\
4076      &   0.7&  1.0 &   0.2&  1.7 &   1.0&  1.9 &   0.1&  1.2 &   0.6&  1.1 &   0.7&  1.4 &   0.2&  4.8 \\
4101      &  26.6&  1.7 &  26.5&  2.1 &  25.8&  2.3 &  26.1&  1.7 &  26.4&  1.7 &  28.9&  2.0 &  24.8&  4.9 \\
4340      &  49.4&  2.5 &  50.3&  2.6 &  48.9&  2.6 &  49.4&  2.5 &  48.9&  2.6 &  50.1&  2.6 &  50.9&  2.9 \\
4363      &   2.0&  0.6 &   4.5&  0.7 &   4.3&  1.0 &   3.3&  0.6 &   3.0&  0.9 &   0.3&  0.7 &   0.0&  1.4 \\
4471      &   2.4&  0.4 &   2.6&  0.6 &   2.3&  0.9 &   2.2&  0.4 &   1.6&  0.5 &   1.5&  0.5 &   3.4&  1.3 \\
4861      & 100.0&  5.0 & 100.0&  5.0 & 100.0&  5.1 & 100.0&  5.0 & 100.0&  5.0 & 100.0&  5.0 & 100.0&  5.2 \\
4959      &  57.2&  2.9 & 113.1&  5.7 & 108.1&  5.5 &  87.1&  4.4 &  69.1&  3.5 &   5.4&  0.5 &  42.2&  2.5 \\
5007      & 172.3&  8.6 & 339.4& 17.0 & 324.3& 16.2 & 260.0& 13.0 & 205.5& 10.3 &  16.7&  1.0 & 131.1&  6.7 \\
5200      &   0.3&  0.4 &   0.0&  0.5 &   0.0&  0.8 &   0.3&  0.4 &   0.6&  0.5 &   0.0&  0.5 &   0.0&  1.3 \\
5755      &   0.3&  0.3 &   0.2&  0.5 &   0.3&  0.6 &   0.3&  0.3 &   0.4&  0.3 &   0.2&  0.5 &   1.2&  1.3 \\
5876      &   8.0&  0.5 &   8.7&  0.6 &   8.5&  0.7 &   6.5&  0.4 &   5.9&  0.4 &   3.4&  0.5 &  10.3&  1.4 \\
6300      &   0.0&  0.3 &   0.0&  0.5 &   0.0&  0.6 &   0.3&  0.3 &   0.2&  0.3 &   2.2&  0.5 &   1.9&  1.3 \\
6312      &   0.4&  0.3 &   0.6&  0.5 &   1.2&  0.6 &   0.3&  0.3 &   0.0&  0.3 &   0.0&  0.5 &   0.0&  1.3 \\
6548      &   8.0&  0.5 &   3.7&  0.5 &   6.4&  0.5 &   7.9&  0.5 &  10.0&  0.6 &  12.7&  0.8 &  10.1&  1.3 \\
6563      & 299.0& 15.0 & 292.8& 14.6 & 303.7& 15.2 & 293.9& 14.7 & 291.9& 14.6 & 309.6& 15.5 & 332.4& 16.7 \\
6584      &  22.2&  1.1 &  16.0&  0.9 &  18.4&  1.0 &  26.1&  1.3 &  32.2&  1.7 &  37.3&  1.9 &  29.1&  1.9 \\
6678      &   3.0&  0.3 &   3.0&  0.4 &   2.8&  0.5 &   2.4&  0.3 &   2.5&  0.4 &   1.0&  0.4 &   3.8&  1.2 \\
6716      &  23.1&  1.2 &  20.2&  1.1 &  23.5&  1.3 &  39.1&  2.0 &  49.8&  2.5 &  44.6&  2.3 &  27.6&  1.9 \\
6731      &  16.5&  0.9 &  15.2&  0.9 &  17.8&  1.0 &  28.2&  1.4 &  35.6&  1.8 &  31.0&  1.6 &  20.8&  1.6 \\
7065      &   1.7&  0.3 &   1.7&  0.4 &   1.6&  0.4 &   1.3&  0.3 &   1.0&  0.4 &   0.7&  0.4 &   2.3&  1.2 \\
7136      &   6.2&  0.4 &   7.4&  0.6 &   7.7&  0.6 &   5.5&  0.4 &   4.9&  0.5 &   2.0&  0.4 &   8.5&  1.3 \\
9069      &  15.1&  0.9 &  18.0&  1.0 &  20.4&  1.2 &  15.0&  0.9 &  13.7&  0.8 &  10.0&  0.8 &  21.0&  2.1 \\
 \etap  &   1.3&  0.2 &   0.7&  0.1 &   0.7&  0.1 &   0.6&  0.1 &   0.7&  0.1 &   7.2&  1.2 &   2.4&  0.4 \\
$R23$   &   4.5&  0.4 &   6.3&  0.4 &   6.2&  0.4 &   6.2&  0.5 &   6.2&  0.6 &   3.6&  0.6 &   4.5&  0.5 \\
$S23$   &  0.92& 0.06 &  0.98& 0.07 &  1.13& 0.08 &  1.20& 0.07 &  1.33& 0.08 &  1.11& 0.07 &  1.22& 0.09 \\
$c$ & 0.08 & 0.05 & 0.08 & 0.05 & 0.00 & 0.05 & 0.08 & 0.05 & 0.08 & 0.05 & 0.16 & 0.05 & 0.33 & 0.05 \\
\enddata
\end{deluxetable}
 
\begin{deluxetable}{crrrrrrrr|rrrrrr}
\scriptsize
\tablecaption{}
\tablenum{2---{\it continued}}
\tablewidth{0 pt}
\tablehead{& \multicolumn{ 8}{c}{D243.5S      }& \multicolumn{ 6}{c}{D243.30S     } \\
\colhead{line (\AA)}      & \colhead{ap 6} & \colhead{err}
      & \colhead{ap 5} & \colhead{err}
      & \colhead{ap 7} & \colhead{err}
      & \colhead{ap 8} & \colhead{err}
      & \colhead{ap 8} & \colhead{err}
      & \colhead{ap 9} & \colhead{err}
      & \colhead{ap10} & \colhead{err}
}
\startdata
3727      & 273.5& 13.7 & 173.3&  8.7 & 246.6& 12.5 & 272.8& 13.9 & 367.1& 20.1 & 309.5& 16.9 & 266.6& 13.8 \\
3869      &   7.5&  0.5 &  13.2&  1.1 &   7.5&  1.4 &   7.3&  1.9 &   0.0&  5.8 &   3.2&  4.7 &   8.8&  2.5 \\
4069      &   2.3&  0.4 &   1.1&  0.8 &   1.9&  1.4 &   2.8&  1.8 &   1.7&  5.8 &   2.5&  4.7 &   0.5&  2.5 \\
4076      &   0.3&  0.4 &   1.3&  0.8 &   1.6&  1.4 &   3.1&  1.8 &   0.0&  5.8 &   0.0&  4.7 &   0.0&  2.5 \\
4101      &  27.4&  1.4 &  27.8&  1.6 &  26.6&  1.9 &  27.4&  2.3 &  30.4&  6.0 &  28.5&  4.9 &  30.2&  2.9 \\
4340      &  48.5&  2.5 &  48.4&  2.4 &  47.8&  2.5 &  48.7&  2.6 &  45.5&  3.2 &  49.9&  4.1 &  49.4&  2.6 \\
4363      &   0.9&  0.4 &   1.0&  0.2 &   0.4&  0.6 &   0.9&  1.0 &   0.6&  2.2 &   2.5&  3.3 &   1.8&  0.9 \\
4471      &   2.8&  0.3 &   3.9&  0.4 &   4.1&  0.7 &   3.6&  1.0 &   1.0&  2.8 &   2.7&  2.4 &   4.8&  1.0 \\
4861      & 100.0&  5.0 & 100.0&  5.0 & 100.0&  5.1 & 100.0&  5.1 & 100.0&  5.8 & 100.0&  5.5 & 100.0&  5.1 \\
4959      &  55.0&  2.8 &  81.1&  4.1 &  55.6&  2.9 &  61.8&  3.2 &   0.0&  2.9 &  38.1&  3.1 &  62.7&  3.3 \\
5007      & 165.9&  8.3 & 246.3& 12.3 & 166.0&  8.3 & 182.3&  9.2 &   0.0&  3.0 & 107.4&  5.9 & 188.0&  9.5 \\
5200      &   0.4&  0.3 &   0.1&  0.4 &   0.0&  0.7 &   0.0&  0.9 &   7.8&  2.9 &   3.7&  2.4 &   1.4&  1.0 \\
5755      &   0.4&  0.2 &   0.3&  0.2 &   0.5&  0.6 &   0.3&  0.7 &   1.2&  2.7 &   2.6&  2.5 &   0.8&  1.0 \\
5876      &   8.1&  0.4 &  11.1&  0.6 &   8.7&  0.7 &   9.2&  0.8 &   0.0&  2.7 &   6.8&  2.5 &   9.0&  1.1 \\
6300      &   1.4&  0.2 &   1.0&  0.2 &   0.9&  0.6 &   0.8&  0.7 &   0.0&  2.6 &   1.9&  2.5 &   1.6&  1.0 \\
6312      &   0.5&  0.2 &   0.4&  0.2 &   0.0&  0.6 &   0.0&  0.7 &   0.0&  2.7 &   1.5&  2.5 &   0.8&  1.0 \\
6548      &   8.3&  0.4 &   4.8&  0.3 &   7.9&  0.6 &   7.6&  0.9 &  14.2&  1.2 &  12.1&  1.1 &   7.6&  0.6 \\
6563      & 297.0& 14.8 & 310.2& 15.5 & 297.0& 14.9 & 285.5& 14.3 & 307.5& 15.4 & 313.2& 15.7 & 307.9& 15.4 \\
6584      &  24.1&  1.2 &  13.8&  0.7 &  23.3&  1.3 &  21.4&  1.3 &  48.4&  2.6 &  34.3&  2.0 &  24.6&  1.3 \\
6678      &   2.3&  0.2 &   3.7&  0.3 &   3.3&  0.5 &   3.0&  0.8 &   0.0&  1.0 &   0.7&  0.9 &   1.8&  0.4 \\
6716      &  31.4&  1.6 &  13.9&  0.7 &  25.8&  1.4 &  24.9&  1.5 &  69.5&  3.6 &  36.5&  2.0 &  29.7&  1.5 \\
6731      &  22.0&  1.1 &  10.0&  0.5 &  18.1&  1.0 &  17.8&  1.2 &  50.3&  2.7 &  25.4&  1.6 &  21.2&  1.1 \\
7065      &   1.3&  0.1 &   1.9&  0.2 &   1.6&  0.5 &   1.4&  0.8 &   0.0&  1.0 &   1.0&  0.9 &   1.5&  0.4 \\
7136      &   5.8&  0.3 &   8.2&  0.4 &   6.8&  0.6 &   7.2&  0.9 &   0.8&  1.0 &   6.2&  1.0 &   7.7&  0.6 \\
9069      &  17.6&  0.9 &  18.8&  1.0 &  16.4&  1.3 &  18.1&  1.5 &   0.0&  1.1 &  15.7&  1.6 &  19.5&  1.3 \\
 \etap  &   1.4&  0.2 &   1.5&  0.2 &   1.5&  0.2 &   1.7&  0.3 &   0.6&  0.1 &   1.9&  0.3 &   1.4&  0.2 \\
$R23$   &   4.9&  0.5 &   5.0&  0.4 &   4.7&  0.4 &   5.2&  0.5 &   3.5&  0.6 &   4.6&  0.5 &   5.2&  0.5 \\
$S23$   &  1.15& 0.07 &  0.90& 0.06 &  1.01& 0.07 &  1.06& 0.07 &  1.16& 0.08 &  1.17& 0.08 &  1.19& 0.08 \\
$c$ & 0.16 & 0.05 & 0.16 & 0.05 & 0.08 & 0.05 & 0.08 & 0.05 & 0.16 & 0.05 & 0.08 & 0.05 & 0.08 & 0.05 \\
\enddata
\end{deluxetable}
 
\begin{deluxetable}{crrrrrrrrrrrrrr}
\scriptsize
\tablecaption{}
\tablenum{2---{\it continued}}
\tablewidth{0 pt}
\tablehead{\multicolumn{15}{c}{D243.30S     } \\
\colhead{line (\AA)}      & \colhead{ap11} & \colhead{err}
      & \colhead{ap12\tablenotemark{c}} & \colhead{err}
      & \colhead{ap13\tablenotemark{c}} & \colhead{err}
      & \colhead{ap16\tablenotemark{c}} & \colhead{err}
      & \colhead{ap17\tablenotemark{c}} & \colhead{err}
      & \colhead{ap14} & \colhead{err}
      & \colhead{ap15} & \colhead{err}
}
\startdata
3727   & 228.6& 12.0 & 181.2&  9.6 & 269.0& 13.7 & 470.8& 23.6 & 108.7& 10.1 & 201.8& 11.1 & 213.6& 12.6 \\
3869   &   8.7&  2.7 &  17.0&  2.5 &   9.8&  1.8 &   3.9&  0.7 &  13.1&  6.1 &   7.5&  3.3 &  12.7&  4.7 \\
4069   &   2.5&  2.7 &   3.5&  2.3 &   5.6&  1.7 &   3.1&  0.7 &   0.0&  6.0 &   0.7&  3.3 &   0.0&  4.7 \\
4076   &   0.4&  2.7 &   0.0&  2.3 &   0.3&  1.7 &   1.7&  0.6 &   1.2&  6.0 &   1.1&  3.3 &   4.3&  4.7 \\
4101   &  26.7&  3.0 &  22.4&  2.6 &  26.6&  2.2 &  26.7&  1.5 &  22.1&  6.1 &  28.0&  3.6 &  27.2&  4.9 \\
4340   &  49.6&  2.9 &  48.7&  2.8 &  47.9&  2.5 &  49.4&  2.5 &  46.5&  3.6 &  48.4&  3.1 &  50.5&  3.4 \\
4363   &   0.4&  1.5 &   3.9&  1.4 &   1.5&  0.6 &   1.0&  0.3 &   1.8&  2.7 &   2.9&  1.9 &   0.2&  2.2 \\
4471   &   3.5&  1.0 &   3.2&  1.5 &   3.8&  0.5 &   1.8&  0.4 &   2.8&  2.2 &   2.6&  8.6 &   5.6& 22.5 \\
4861   & 100.0&  5.1 & 100.0&  5.2 & 100.0&  5.0 & 100.0&  5.0 & 100.0&  5.5 & 100.0&  9.9 & 100.0& 23.1 \\
4959   &  73.9&  3.8 &  94.5&  4.9 &  62.1&  3.1 &  26.5&  1.4 & 104.5&  5.7 &  55.0&  9.0 &  63.4& 22.7 \\
5007   & 220.5& 11.1 & 285.7& 14.4 & 186.2&  9.3 &  80.0&  4.0 & 306.4& 15.5 & 161.4& 11.8 & 190.9& 24.4 \\
5200   &   1.6&  0.9 &   0.0&  1.5 &   1.5&  0.5 &   1.4&  0.4 &   1.8&  2.2 &   0.5&  8.6 &   0.0& 22.5 \\
5755   &   1.6&  0.8 &   1.6&  1.0 &   1.1&  0.4 &   0.7&  0.1 &   0.4&  1.7 &   1.0&  1.0 &   0.0&  1.3 \\
5876   &   9.5&  0.9 &   8.9&  1.1 &  10.3&  0.6 &   6.8&  0.4 &  11.2&  1.8 &   7.0&  1.0 &  10.0&  1.4 \\
6300   &   0.8&  0.8 &   2.0&  1.1 &  16.8&  0.9 &   7.7&  0.4 &   0.0&  1.7 &   1.6&  1.0 &   0.0&  1.3 \\
6312   &   0.0&  0.8 &   0.9&  1.0 &   1.1&  0.4 &   0.9&  0.1 &   0.0&  1.7 &   0.7&  1.0 &   0.0&  1.3 \\
6548   &   7.1&  0.5 &   4.7&  0.7 &  10.1&  0.6 &  15.9&  0.8 &   3.2&  1.1 &   6.8&  0.7 &   5.6&  1.2 \\
6563   & 295.2& 14.8 & 289.7& 14.5 & 302.8& 15.1 & 320.8& 16.0 & 290.4& 14.6 & 296.0& 14.8 & 300.6& 15.1 \\
6584   &  20.2&  1.1 &  16.5&  1.1 &  31.3&  1.6 &  46.5&  2.3 &   9.2&  1.2 &  22.6&  1.3 &  23.3&  1.7 \\
6678   &   2.6&  0.4 &   2.3&  0.7 &   2.9&  0.3 &   2.3&  0.2 &   3.3&  1.1 &   2.1&  0.6 &   4.2&  1.2 \\
6716   &  21.0&  1.1 &  17.9&  1.1 &  37.2&  1.9 &  43.2&  2.2 &  11.7&  1.2 &  31.5&  1.7 &  28.8&  1.9 \\
6731   &  15.0&  0.8 &  13.0&  0.9 &  37.7&  1.9 &  36.7&  1.8 &  10.2&  1.2 &  22.7&  1.3 &  22.9&  1.6 \\
7065   &   1.5&  0.4 &   1.9&  0.7 &   2.2&  0.3 &   1.4&  0.2 &   1.2&  1.1 &   0.7&  0.6 &   2.1&  1.2 \\
7136   &   7.2&  0.5 &   8.0&  0.8 &   7.1&  0.4 &   5.2&  0.3 &   9.5&  1.2 &   5.3&  0.6 &   7.3&  1.2 \\
9069   &  16.5&  1.1 &  16.6&  1.4 &  14.0&  0.7 &  15.0&  0.8 &  13.4&  1.5 &  13.3&  1.7 &  17.4&  2.2 \\
 \etap  &   1.3&  0.2 &   0.9&  0.1 &   0.7&  0.1 &   2.9&  0.5 &   0.6&  0.1 &   0.8&  0.1 &   1.0&  0.2 \\
$R23$   &   5.2&  0.4 &   5.6&  0.4 &   5.2&  0.5 &   5.8&  0.8 &   5.2&  0.3 &   4.2&  0.5 &   4.7&  0.9 \\
$S23$   &  0.94& 0.06 &  0.89& 0.06 &  1.24& 0.07 &  1.32& 0.08 &  0.69& 0.05 &  1.01& 0.11 &  1.13& 0.27 \\
$c$ & 0.16 & 0.05 & 0.08 & 0.05 & 0.33 & 0.05 & 0.16 & 0.05 & 0.16 & 0.05 & 0.08 & 0.05 & 0.08 & 0.05 \\
\enddata
\end{deluxetable}
 
\begin{deluxetable}{crrrrrr|rrrrrrrr}
\scriptsize
\tablecaption{}
\tablenum{2---{\it continued}}
\tablewidth{0 pt}
\tablehead{& \multicolumn{ 6}{c}{D301.SW1     }& \multicolumn{ 8}{c}{D301.SW6     } \\
\colhead{line (\AA)}      & \colhead{ap 6} & \colhead{err}
      & \colhead{ap 5} & \colhead{err}
      & \colhead{ap 7} & \colhead{err}
      & \colhead{ap 6} & \colhead{err}
      & \colhead{ap 7} & \colhead{err}
      & \colhead{ap 8} & \colhead{err}
      & \colhead{ap 9} & \colhead{err}
}
\startdata
3727      & 543.2& 27.3 & 393.0& 19.7 & 425.1& 21.5 & 589.1& 31.3 & 587.1& 29.4 & 496.6& 24.8 & 451.1& 22.6 \\
3869      &  11.8&  1.8 &   6.6&  0.6 &  14.0&  2.6 &   0.0&  7.6 &  10.9&  1.0 &  10.3&  0.9 &  10.7&  0.8 \\
4069      &   6.2&  1.7 &   2.7&  0.6 &   4.9&  2.5 &   0.9&  7.6 &   6.0&  0.9 &   4.5&  0.7 &   4.0&  0.6 \\
4076      &   0.2&  1.7 &   1.0&  0.6 &   0.5&  2.5 &   7.3&  7.6 &   1.8&  0.8 &   2.1&  0.7 &   1.5&  0.5 \\
4101      &  27.5&  2.2 &  25.6&  1.4 &  25.8&  2.8 &  30.4&  7.7 &  26.5&  1.6 &  26.3&  1.5 &  24.4&  1.3 \\
4340      &  45.8&  2.4 &  46.1&  2.3 &  46.4&  2.6 &  52.6&  4.5 &  47.2&  2.4 &  46.6&  2.3 &  46.0&  2.3 \\
4363      &   0.6&  0.8 &   0.6&  0.3 &   0.0&  1.1 &   5.9&  3.7 &   1.7&  0.3 &   1.5&  0.3 &   0.8&  0.4 \\
4471      &   4.3&  1.2 &   4.1&  0.6 &   3.7&  1.0 &   1.5&  3.7 &   3.6&  0.6 &   3.8&  0.6 &   3.5&  0.4 \\
4861      & 100.0&  5.1 & 100.0&  5.0 & 100.0&  5.1 & 100.0&  6.2 & 100.0&  5.0 & 100.0&  5.0 & 100.0&  5.0 \\
4959      &  25.9&  1.8 &  44.4&  2.3 &  67.8&  3.5 &  13.9&  3.8 &  35.7&  1.9 &  43.7&  2.3 &  41.5&  2.1 \\
5007      &  77.9&  4.1 & 134.3&  6.7 & 207.1& 10.4 &  52.8&  4.6 & 107.9&  5.4 & 132.6&  6.7 & 125.0&  6.3 \\
5200      &   0.7&  1.2 &   0.0&  0.5 &   0.6&  1.0 &   0.9&  3.7 &   0.4&  0.5 &   0.5&  0.5 &   0.3&  0.4 \\
5755      &   1.0&  0.6 &   0.3&  0.2 &   0.2&  0.7 &   1.2&  5.2 &   0.4&  0.3 &   0.8&  0.3 &   0.9&  0.2 \\
5876      &   9.5&  0.7 &  10.9&  0.6 &  11.4&  0.9 &  11.9&  5.2 &  11.3&  0.6 &  12.7&  0.7 &  11.6&  0.6 \\
6300      &  10.4&  0.8 &   1.4&  0.2 &   5.8&  0.8 &   1.1&  5.2 &   6.8&  0.4 &   2.8&  0.3 &   3.0&  0.3 \\
6312      &   0.1&  0.6 &   1.2&  0.2 &   1.9&  0.7 &   0.0&  5.2 &   1.8&  0.3 &   1.6&  0.3 &   1.4&  0.2 \\
6548      &  17.0&  1.0 &  10.8&  0.6 &  11.1&  0.7 &  22.5&  2.0 &  17.0&  0.9 &  15.2&  0.8 &  14.7&  0.8 \\
6563      & 282.6& 14.1 & 293.5& 14.7 & 274.9& 13.8 & 295.4& 14.9 & 285.3& 14.3 & 293.5& 14.7 & 291.9& 14.6 \\
6584      &  49.5&  2.5 &  31.3&  1.6 &  34.5&  1.8 &  50.4&  3.0 &  49.6&  2.5 &  44.7&  2.2 &  43.3&  2.2 \\
6678      &   2.3&  0.4 &   3.0&  0.2 &   3.1&  0.4 &   2.8&  1.7 &   3.1&  0.3 &   3.0&  0.3 &   3.2&  0.3 \\
6716      &  72.0&  3.6 &  33.8&  1.7 &  50.1&  2.5 & 119.4&  6.2 &  81.8&  4.1 &  67.1&  3.4 &  66.7&  3.3 \\
6731      &  50.3&  2.6 &  23.5&  1.2 &  35.0&  1.8 &  83.0&  4.5 &  57.5&  2.9 &  47.4&  2.4 &  47.2&  2.4 \\
7065      &   1.9&  0.4 &   1.8&  0.2 &   1.9&  0.4 &   3.2&  1.7 &   2.0&  0.3 &   2.1&  0.3 &   2.1&  0.3 \\
7136      &   6.7&  0.5 &   8.5&  0.5 &   9.5&  0.6 &   6.3&  1.7 &   9.3&  0.6 &   9.7&  0.5 &   9.4&  0.5 \\
9069      &  17.5&  1.2 &  21.9&  1.1 &  23.6&  1.6 &  11.9&  2.8 &  19.7&  1.2 &  19.5&  1.1 &  22.2&  1.1 \\
 \etap  &   2.6&  0.4 &   2.9&  0.5 &   1.5&  0.2 &   1.8&  0.3 &   2.0&  0.3 &   1.7&  0.3 &   1.8&  0.3 \\
$R23$   &   6.5&  0.9 &   5.7&  0.7 &   7.0&  0.7 &   6.6&  1.0 &   7.3&  1.0 &   6.7&  0.8 &   6.2&  0.8 \\
$S23$   &  1.84& 0.11 &  1.34& 0.09 &  1.68& 0.11 &  2.44& 0.18 &  2.08& 0.12 &  1.83& 0.11 &  1.92& 0.12 \\
$c$ & 0.00 & 0.05 & 0.00 & 0.05 & 0.00 & 0.05 & 0.00 & 0.05 & 0.00 & 0.05 & 0.00 & 0.05 & 0.00 & 0.05 \\
\enddata
\end{deluxetable}
 
\begin{deluxetable}{crrrr|rrrrrrrrrr}
\scriptsize
\tablecaption{}
\tablenum{2---{\it continued}}
\tablewidth{0 pt}
\tablehead{& \multicolumn{ 4}{c}{D301.SW6     }& \multicolumn{10}{c}{D323.C1    } \\
\colhead{line (\AA)}      & \colhead{ap10} & \colhead{err}
      & \colhead{ap11} & \colhead{err}
      & \colhead{ap13} & \colhead{err}
      & \colhead{ap 6} & \colhead{err}
      & \colhead{ap 7} & \colhead{err}
      & \colhead{ap 8} & \colhead{err}
      & \colhead{ap 9} & \colhead{err}
}
\startdata
3727   & 603.8& 30.2 & 434.0& 21.7 & 348.0& 17.5 & 369.5& 18.5 & 146.5&  7.3 & 119.7&  6.0 & 241.4& 12.1 \\
3869   &  14.0&  0.9 &   9.7&  1.1 &  13.5&  1.6 &   8.7&  0.5 &  14.6&  0.8 &  17.6&  0.9 &  17.1&  0.9 \\
4069   &   5.3&  0.7 &   3.7&  1.0 &   2.5&  1.4 &   1.6&  0.3 &   0.7&  0.2 &   0.7&  0.3 &   1.2&  0.3 \\
4076   &   1.2&  0.6 &   2.1&  1.0 &   0.1&  1.4 &   0.2&  0.3 &   0.1&  0.2 &   0.2&  0.3 &   0.3&  0.2 \\
4101   &  27.6&  1.5 &  25.8&  1.6 &  24.2&  1.9 &  25.0&  1.3 &  25.4&  1.3 &  25.1&  1.3 &  24.9&  1.3 \\
4340   &  48.2&  2.4 &  47.2&  2.4 &  44.3&  2.6 &  44.3&  2.2 &  45.1&  2.3 &  44.5&  2.2 &  44.7&  2.2 \\
4363   &   1.5&  0.3 &   1.5&  0.3 &   1.1&  1.4 &   0.6&  0.2 &   1.4&  0.2 &   1.5&  0.2 &   1.5&  0.1 \\
4471   &   4.1&  0.6 &   3.4&  0.6 &   4.5&  1.5 &   2.9&  0.4 &   4.1&  0.5 &   3.7&  0.4 &   3.7&  0.5 \\
4861   & 100.0&  5.0 & 100.0&  5.0 & 100.0&  5.2 & 100.0&  5.0 & 100.0&  5.0 & 100.0&  5.0 & 100.0&  5.0 \\
4959   &  42.5&  2.2 &  44.5&  2.3 &  84.1&  4.5 &  59.2&  3.0 &  96.4&  4.8 & 108.0&  5.4 &  96.6&  4.9 \\
5007   & 128.8&  6.5 & 134.5&  6.7 & 250.0& 12.6 & 176.4&  8.8 & 288.2& 14.4 & 321.9& 16.1 & 288.5& 14.4 \\
5200   &   0.3&  0.5 &   0.0&  0.5 &   0.0&  1.5 &   0.2&  0.4 &   0.0&  0.4 &   0.0&  0.4 &   0.0&  0.5 \\
5755   &   0.7&  0.5 &   0.5&  0.4 &   0.0&  1.3 &   0.2&  0.2 &   0.0&  0.1 &   0.0&  0.1 &   0.1&  0.1 \\
5876   &  12.1&  0.8 &  12.6&  0.7 &  11.0&  1.4 &   9.4&  0.5 &  12.0&  0.6 &  12.2&  0.6 &  11.4&  0.6 \\
6300   &   4.1&  0.5 &   0.0&  0.4 &   0.0&  1.3 &   0.4&  0.2 &   0.0&  0.2 &   0.0&  0.1 &   0.2&  0.1 \\
6312   &   1.9&  0.5 &   2.2&  0.4 &   0.3&  1.3 &   0.9&  0.2 &   1.0&  0.2 &   1.2&  0.1 &   1.3&  0.2 \\
6548   &  18.6&  1.0 &  13.1&  0.7 &   8.5&  0.9 &   9.5&  0.5 &   3.2&  0.2 &   2.9&  0.2 &   5.0&  0.3 \\
6563   & 295.9& 14.8 & 280.2& 14.0 & 276.9& 13.9 & 289.3& 14.5 & 289.1& 14.5 & 298.0& 14.9 & 283.6& 14.2 \\
6584   &  50.8&  2.6 &  38.5&  1.9 &  23.3&  1.4 &  27.0&  1.4 &   8.2&  0.4 &   7.5&  0.4 &  14.1&  0.7 \\
6678   &   5.1&  0.4 &   3.1&  0.3 &   3.5&  0.8 &   2.5&  0.2 &   3.3&  0.2 &   3.4&  0.2 &   3.1&  0.2 \\
6716   &  77.4&  3.9 &  57.6&  2.9 &  20.5&  1.3 &  22.7&  1.1 &   5.3&  0.3 &   5.6&  0.3 &  13.9&  0.7 \\
6731   &  54.3&  2.7 &  40.4&  2.0 &  13.7&  1.0 &  16.1&  0.8 &   3.7&  0.2 &   4.0&  0.2 &   9.9&  0.5 \\
7065   &   3.7&  0.4 &   2.0&  0.3 &   1.3&  0.8 &   1.5&  0.2 &   1.9&  0.1 &   1.9&  0.2 &   1.8&  0.2 \\
7136   &  11.4&  0.7 &   9.1&  0.5 &  10.2&  0.9 &   8.2&  0.4 &   9.6&  0.5 &  10.0&  0.5 &  10.1&  0.5 \\
9069   &  20.6&  1.1 &  22.3&  1.3 &  26.9&  1.6 &  26.2&  1.3 &  24.1&  1.2 &  24.2&  1.2 &  23.6&  1.2 \\
 \etap  &   1.9&  0.3 &   1.9&  0.3 &   2.9&  0.5 &   3.7&  0.6 &   3.6&  0.6 &   2.5&  0.4 &   2.2&  0.4 \\
$R23$   &   7.8&  1.0 &   6.1&  0.7 &   6.8&  0.6 &   6.1&  0.6 &   5.3&  0.3 &   5.5&  0.3 &   6.3&  0.5 \\
$S23$   &  2.04& 0.12 &  1.76& 0.11 &  1.28& 0.09 &  1.30& 0.09 &  0.93& 0.07 &  0.94& 0.07 &  1.06& 0.08 \\
$c$ & 0.00 & 0.05 & 0.00 & 0.05 & 0.08 & 0.05 & 0.08 & 0.05 & 0.16 & 0.05 & 0.08 & 0.05 & 0.16 & 0.05 \\
\enddata
\end{deluxetable}
 
\begin{deluxetable}{crrrrrr|rrrrrrrr}
\scriptsize
\tablecaption{}
\tablenum{2---{\it continued}}
\tablewidth{0 pt}
\tablehead{& \multicolumn{ 6}{c}{D323.C1      }& \multicolumn{ 8}{c}{D323.C2      } \\
\colhead{line (\AA)}      & \colhead{ap10} & \colhead{err}
      & \colhead{ap11} & \colhead{err}
      & \colhead{ap12} & \colhead{err}
      & \colhead{ap 5} & \colhead{err}
      & \colhead{ap 6} & \colhead{err}
      & \colhead{ap 7} & \colhead{err}
      & \colhead{ap 8} & \colhead{err}
}
\startdata
3727      & 327.2& 16.4 & 393.8& 19.7 & 458.9& 23.1 & 584.6& 29.4 & 495.4& 24.9 & 271.9& 13.6 & 270.1& 13.5 \\
3869      &   9.9&  0.7 &   5.2&  0.5 &   7.7&  1.7 &   1.5&  2.0 &   5.7&  1.7 &  10.4&  0.7 &  12.1&  0.8 \\
4069      &   1.7&  0.5 &   1.8&  0.4 &   3.0&  1.7 &   3.9&  2.0 &   2.8&  1.6 &   1.6&  0.5 &   1.1&  0.5 \\
4076      &   0.2&  0.4 &   0.7&  0.4 &   0.3&  1.7 &   0.9&  2.0 &   0.0&  1.6 &   0.1&  0.5 &   0.3&  0.4 \\
4101      &  23.8&  1.3 &  24.4&  1.3 &  23.7&  2.0 &  26.4&  2.4 &  26.2&  2.1 &  25.8&  1.4 &  25.4&  1.3 \\
4340      &  43.9&  2.2 &  44.2&  2.2 &  45.8&  2.4 &  43.8&  2.5 &  44.8&  2.3 &  45.0&  2.3 &  44.2&  2.2 \\
4363      &   1.0&  0.2 &   0.9&  0.1 &   0.8&  0.6 &   1.1&  1.1 &   1.3&  0.6 &   1.0&  0.3 &   1.4&  0.3 \\
4471      &   3.5&  0.5 &   3.7&  0.6 &   3.9&  0.9 &   2.8&  0.9 &   5.0&  0.9 &   4.0&  0.5 &   3.9&  0.4 \\
4861      & 100.0&  5.0 & 100.0&  5.0 & 100.0&  5.1 & 100.0&  5.1 & 100.0&  5.1 & 100.0&  5.0 & 100.0&  5.0 \\
4959      &  63.2&  3.2 &  48.4&  2.5 &  37.8&  2.1 &  16.0&  1.2 &  34.4&  1.9 &  70.2&  3.5 &  76.1&  3.8 \\
5007      & 188.7&  9.4 & 145.2&  7.3 & 111.9&  5.7 &  48.3&  2.6 & 104.2&  5.3 & 209.5& 10.5 & 228.5& 11.4 \\
5200      &   0.3&  0.4 &   0.0&  0.5 &   0.0&  0.9 &   1.3&  0.9 &   0.0&  0.8 &   0.0&  0.5 &   0.2&  0.4 \\
5755      &   0.2&  0.2 &   0.4&  0.1 &   0.4&  0.4 &   0.6&  0.7 &   0.7&  0.7 &   0.4&  0.2 &   0.3&  0.2 \\
5876      &  11.2&  0.6 &  11.5&  0.6 &  10.5&  0.7 &   5.1&  0.7 &   9.4&  0.8 &  11.3&  0.6 &  10.8&  0.6 \\
6300      &   0.2&  0.2 &   0.7&  0.1 &   1.8&  0.5 &   5.4&  0.7 &   0.0&  0.7 &   0.3&  0.2 &   0.6&  0.2 \\
6312      &   1.1&  0.2 &   1.3&  0.1 &   0.5&  0.4 &   1.0&  0.7 &   2.0&  0.7 &   1.2&  0.2 &   1.4&  0.2 \\
6548      &   7.7&  0.4 &   9.4&  0.5 &  10.8&  0.6 &  17.6&  1.0 &  13.5&  0.8 &   6.3&  0.3 &   6.6&  0.4 \\
6563      & 295.6& 14.8 & 304.3& 15.2 & 275.2& 13.8 & 299.2& 15.0 & 283.3& 14.2 & 291.3& 14.6 & 289.6& 14.5 \\
6584      &  22.4&  1.1 &  27.5&  1.4 &  31.6&  1.6 &  50.6&  2.6 &  37.1&  1.9 &  17.7&  0.9 &  18.4&  0.9 \\
6678      &   3.0&  0.2 &   3.2&  0.2 &   2.7&  0.4 &   1.5&  0.5 &   2.5&  0.4 &   3.2&  0.2 &   3.0&  0.2 \\
6716      &  21.6&  1.1 &  26.8&  1.3 &  33.2&  1.7 &  57.7&  2.9 &  30.5&  1.6 &  16.7&  0.8 &  18.4&  0.9 \\
6731      &  15.3&  0.8 &  19.5&  1.0 &  23.7&  1.2 &  40.1&  2.1 &  21.0&  1.1 &  11.8&  0.6 &  13.1&  0.7 \\
7065      &   1.8&  0.2 &   2.0&  0.2 &   1.3&  0.4 &   1.0&  0.5 &   1.7&  0.4 &   1.8&  0.1 &   1.8&  0.2 \\
7136      &   9.2&  0.5 &  10.4&  0.5 &   7.9&  0.5 &   3.8&  0.5 &   7.5&  0.5 &   9.1&  0.5 &   8.9&  0.5 \\
9069      &  23.5&  1.2 &  28.9&  1.5 &  24.1&  1.3 &  19.4&  1.1 &  23.1&  1.2 &  24.8&  1.2 &  22.1&  1.1 \\
 \etap  &   2.9&  0.5 &   4.4&  0.7 &   4.5&  0.7 &   6.3&  1.0 &   5.6&  0.9 &   2.9&  0.5 &   2.2&  0.4 \\
$R23$   &   5.8&  0.6 &   5.9&  0.7 &   6.1&  0.8 &   6.5&  1.0 &   6.3&  0.8 &   5.5&  0.5 &   5.7&  0.5 \\
$S23$   &  1.19& 0.08 &  1.47& 0.10 &  1.41& 0.09 &  1.66& 0.10 &  1.32& 0.09 &  1.15& 0.08 &  1.09& 0.07 \\
$c$ & 0.16 & 0.05 & 0.16 & 0.05 & 0.16 & 0.05 & 0.08 & 0.05 & 0.16 & 0.05 & 0.16 & 0.05 & 0.16 & 0.05 \\
\enddata
\end{deluxetable}
 
\begin{deluxetable}{crrrrrrrrrrrrrr}
\scriptsize
\tablecaption{}
\tablenum{2---{\it continued}}
\tablewidth{0 pt}
\tablehead{\multicolumn{15}{c}{D323.C2      } \\
\colhead{line (\AA)}      & \colhead{ap 9} & \colhead{err}
      & \colhead{ap10} & \colhead{err}
      & \colhead{ap11} & \colhead{err}
      & \colhead{ap12} & \colhead{err}
      & \colhead{ap13} & \colhead{err}
      & \colhead{ap 0} & \colhead{err}
      & \colhead{ap 0} & \colhead{err}
}
\startdata
3727   & 261.7& 13.1 & 239.4& 12.0 & 131.6&  6.6 & 161.9&  8.1 & 225.3& 11.4 &   0.0&  0.0 &   0.0&  0.0 \\
3869   &  16.4&  0.9 &  19.7&  1.0 &  21.8&  1.2 &  16.8&  1.0 &  15.3&  1.5 &   0.0&  0.0 &   0.0&  0.0 \\
4069   &   1.3&  0.2 &   1.1&  0.3 &   0.6&  0.5 &   0.8&  0.5 &   0.0&  1.3 &   0.0&  0.0 &   0.0&  0.0 \\
4076   &   0.4&  0.2 &   0.3&  0.3 &   0.5&  0.5 &   0.2&  0.5 &   0.0&  1.3 &   0.0&  0.0 &   0.0&  0.0 \\
4101   &  25.0&  1.3 &  25.0&  1.3 &  25.5&  1.4 &  25.0&  1.3 &  27.2&  1.9 &   0.0&  0.0 &   0.0&  0.0 \\
4340   &  43.9&  2.2 &  44.0&  2.2 &  44.4&  2.2 &  43.7&  2.2 &  44.3&  2.3 &   0.0&  0.0 &   0.0&  0.0 \\
4363   &   1.4&  0.2 &   1.9&  0.2 &   1.9&  0.3 &   1.5&  0.2 &   2.4&  0.6 &   0.0&  0.0 &   0.0&  0.0 \\
4471   &   3.9&  0.4 &   3.8&  0.4 &   4.1&  0.5 &   3.8&  0.5 &   4.2&  0.6 &   0.0&  0.0 &   0.0&  0.0 \\
4861   & 100.0&  5.0 & 100.0&  5.0 & 100.0&  5.0 & 100.0&  5.0 & 100.0&  5.0 &   0.0&  0.0 &   0.0&  0.0 \\
4959   &  90.6&  4.5 & 109.6&  5.5 & 122.0&  6.1 & 102.0&  5.1 &  93.7&  4.7 &   0.0&  0.0 &   0.0&  0.0 \\
5007   & 272.2& 13.6 & 328.5& 16.4 & 364.3& 18.2 & 307.0& 15.4 & 279.6& 14.0 &   0.0&  0.0 &   0.0&  0.0 \\
5200   &   0.1&  0.4 &   0.1&  0.4 &   0.0&  0.5 &   0.0&  0.5 &   0.2&  0.6 &   0.0&  0.0 &   0.0&  0.0 \\
5755   &   0.3&  0.1 &   0.3&  0.1 &   0.1&  0.2 &   0.3&  0.1 &   0.5&  0.6 &   0.0&  0.0 &   0.0&  0.0 \\
5876   &  11.4&  0.6 &  11.6&  0.6 &  12.1&  0.6 &  11.8&  0.6 &  11.8&  0.9 &   0.0&  0.0 &   0.0&  0.0 \\
6300   &   0.7&  0.1 &   0.1&  0.1 &   0.0&  0.2 &   0.0&  0.1 &   0.0&  0.6 &   0.0&  0.0 &   0.0&  0.0 \\
6312   &   1.2&  0.1 &   1.4&  0.1 &   1.5&  0.2 &   1.0&  0.1 &   0.7&  0.6 &   0.0&  0.0 &   0.0&  0.0 \\
6548   &   6.4&  0.3 &   5.6&  0.3 &   3.2&  0.2 &   4.1&  0.3 &   5.2&  0.5 &   0.0&  0.0 &   0.0&  0.0 \\
6563   & 284.4& 14.2 & 298.8& 14.9 & 293.6& 14.7 & 295.1& 14.8 & 295.2& 14.8 &   0.0&  0.0 &   0.0&  0.0 \\
6584   &  18.0&  0.9 &  15.7&  0.8 &   8.5&  0.4 &  11.3&  0.6 &  15.4&  0.9 &   0.0&  0.0 &   0.0&  0.0 \\
6678   &   3.1&  0.2 &   3.3&  0.2 &   3.2&  0.2 &   3.3&  0.2 &   2.6&  0.4 &   0.0&  0.0 &   0.0&  0.0 \\
6716   &  18.3&  0.9 &  15.7&  0.8 &   8.7&  0.5 &  10.3&  0.5 &  17.0&  0.9 &   0.0&  0.0 &   0.0&  0.0 \\
6731   &  13.0&  0.7 &  11.1&  0.6 &   6.2&  0.3 &   7.3&  0.4 &  12.6&  0.7 &   0.0&  0.0 &   0.0&  0.0 \\
7065   &   1.9&  0.2 &   2.0&  0.1 &   2.0&  0.2 &   1.9&  0.2 &   1.8&  0.4 &   0.0&  0.0 &   0.0&  0.0 \\
7136   &   9.7&  0.5 &  11.0&  0.6 &  10.5&  0.5 &   9.9&  0.5 &  10.0&  0.6 &   0.0&  0.0 &   0.0&  0.0 \\
9069   &  23.4&  1.2 &  26.9&  1.4 &  24.1&  1.2 &  26.2&  1.3 &  26.9&  1.5 &   0.0&  0.0 &   0.0&  0.0 \\
 \etap  &   1.9&  0.3 &   1.9&  0.3 &   1.5&  0.2 &   2.1&  0.3 &   1.9&  0.3 &   0.0&  0.0 &   0.0&  0.0 \\
$R23$   &   6.2&  0.5 &   6.8&  0.5 &   6.2&  0.4 &   5.7&  0.4 &   6.0&  0.4 &   0.0&  0.0 &   0.0&  0.0 \\
$S23$   &  1.13& 0.08 &  1.21& 0.09 &  0.99& 0.07 &  1.09& 0.08 &  1.24& 0.09 &0.00   &0.00   &0.00   &0.00   \\
$c$ & 0.08 & 0.05 & 0.08 & 0.05 & 0.08 & 0.05 & 0.08 & 0.05 & 0.00 & 0.05 & 0.00 & 0.00 & 0.00 & 0.00 \\
\enddata
\end{deluxetable}

\begin{deluxetable}{crrrrrrrrrrrrrr}
\scriptsize
\tablecaption{Dereddened line intensities for spatially integrated
	positions\tablenotemark{a} \label{integ}}
\tablewidth{0 pt}
\tablehead{& \multicolumn{2}{c}{D199.496W240} & 
	\multicolumn{2}{c}{D243.2(tot)} & \multicolumn{2}{c}{D243.2(no SNR)} & 
	\multicolumn{2}{c}{D301.SW6\tablenotemark{b}} &
	\multicolumn{2}{c}{D323.140} & 
	\multicolumn{2}{c}{D323.140N30N} & \multicolumn{2}{c}{D323.140N30S} \\
%  MAKE SURE TO EDIT D243 APS TO 11; AND ALSO CORRECT THE Hd AND Hg VALUES
\colhead{line (\AA)}      & \colhead{ap24} & \colhead{err}
      & \colhead{ap 1} & \colhead{err}
      & \colhead{ap 2} & \colhead{err}
      & \colhead{ap12} & \colhead{err}
      & \colhead{ap16} & \colhead{err}
      & \colhead{ap16} & \colhead{err}
      & \colhead{ap13} & \colhead{err}
}
\startdata
3727   & 185.6&  9.4 & 265.6& 13.3 & 261.2& 13.1 & 539.1& 27.0 & 277.0& 13.9 & 332.0& 16.6 & 256.9& 12.9 \\
3869   &  30.4&  2.0 &   9.8&  0.8 &   8.7&  0.8 &  12.5&  1.0 &  13.5&  0.8 &  10.2&  0.8 &  16.5&  1.0 \\
4069   &   0.2&  1.2 &   4.4&  0.7 &   2.1&  0.7 &   5.0&  0.8 &   1.1&  0.5 &   0.9&  0.7 &   1.2&  0.5 \\
4076   &   1.1&  1.2 &   1.3&  0.6 &   0.7&  0.7 &   1.5&  0.7 &   0.5&  0.5 &   1.0&  0.7 &   0.4&  0.5 \\
4101   &  27.7&  1.9 &  23.9&  1.3 &  24.2&  1.4 &  25.2&  1.5 &  24.8&  1.3 &  24.2&  1.4 &  24.7&  1.3 \\
4340   &  48.3&  2.5 &  45.7&  2.3 &  46.0&  2.3 &  46.4&  2.3 &  44.2&  2.2 &  43.8&  2.3 &  44.5&  2.2 \\
4363   &   1.1&  0.7 &   1.7&  0.4 &   1.5&  0.5 &   1.8&  0.3 &   1.4&  0.3 &   1.1&  0.5 &   1.1&  0.2 \\
4471   &   0.7&  0.6 &   2.0&  0.5 &   1.8&  0.6 &   3.6&  0.5 &   3.3&  0.4 &   3.4&  0.6 &   3.5&  0.4 \\
4861   & 100.0&  5.0 & 100.0&  5.0 & 100.0&  5.0 & 100.0&  5.0 & 100.0&  5.0 & 100.0&  5.0 & 100.0&  5.0 \\
4959   & 125.0&  6.3 &  51.8&  2.6 &  52.7&  2.7 &  43.5&  2.2 &  83.5&  4.2 &  63.1&  3.2 &  90.8&  4.6 \\
5007   & 376.5& 18.8 & 155.4&  7.8 & 158.1&  7.9 & 131.0&  6.6 & 250.2& 12.5 & 188.5&  9.4 & 271.5& 13.6 \\
5200   &   0.0&  0.6 &   0.9&  0.5 &   0.0&  0.5 &   0.1&  0.4 &   0.0&  0.4 &   0.0&  0.6 &   0.2&  0.4 \\
5755   &   0.5&  0.4 &   0.6&  0.3 &   0.5&  0.3 &   0.6&  0.3 &   0.1&  0.2 &   0.7&  0.4 &   0.1&  0.2 \\
5876   &   9.6&  0.7 &   8.8&  0.5 &   9.1&  0.6 &  12.1&  0.7 &  11.2&  0.6 &  11.4&  0.7 &  11.4&  0.6 \\
6300   &   0.0&  0.5 &  13.0&  0.7 &   1.9&  0.3 &   3.1&  0.3 &   0.0&  0.2 &   0.2&  0.4 &   0.1&  0.2 \\
6312   &   0.7&  0.4 &   0.5&  0.3 &   0.4&  0.3 &   1.4&  0.3 &   1.0&  0.2 &   0.9&  0.4 &   1.1&  0.2 \\
6548   &   5.8&  0.5 &   9.2&  0.6 &   7.7&  0.7 &  14.7&  0.8 &   7.0&  0.4 &   7.8&  0.4 &   6.2&  0.3 \\
6563   & 272.0& 13.6 & 299.4& 15.0 & 295.6& 14.8 & 285.5& 14.3 & 300.9& 15.0 & 303.1& 15.2 & 297.7& 14.9 \\
6584   &  18.7&  1.0 &  27.6&  1.4 &  23.2&  1.3 &  43.6&  2.2 &  19.2&  1.0 &  23.8&  1.2 &  17.6&  0.9 \\
6678   &   3.0&  0.4 &   2.8&  0.5 &   2.9&  0.6 &   3.2&  0.3 &   3.0&  0.2 &   3.1&  0.3 &   3.2&  0.2 \\
6716   &  22.8&  1.2 &  36.3&  1.9 &  22.0&  1.2 &  67.7&  3.4 &  18.2&  0.9 &  20.5&  1.0 &  16.8&  0.8 \\
6731   &  15.4&  0.9 &  35.3&  1.8 &  16.8&  1.0 &  47.8&  2.4 &  12.8&  0.7 &  14.6&  0.8 &  11.8&  0.6 \\
7065   &   2.0&  0.4 &   1.6&  0.4 &   1.5&  0.5 &   2.3&  0.3 &   1.8&  0.2 &   1.4&  0.2 &   2.0&  0.1 \\
7136   &   9.9&  0.6 &   6.6&  0.5 &   7.4&  0.6 &   9.5&  0.5 &   9.7&  0.5 &   8.8&  0.5 &  10.1&  0.5 \\
9069   &  20.8&  1.9 &  17.0&  1.1 &  18.5&  1.2 &  19.2&  1.1 &  25.1&  1.4 &  26.2&  1.4 &  24.8&  1.3 \\
 \etap  &   0.7&  0.1 &   1.1&  0.2 &   2.1&  0.3 &   1.8&  0.3 &   2.4&  0.4 &   3.4&  0.6 &   2.1&  0.4 \\
$R23$   &   6.9&  0.4 &   4.7&  0.5 &   4.7&  0.5 &   7.1&  0.9 &   6.1&  0.5 &   5.8&  0.6 &   6.2&  0.5 \\
$S23$   &  1.11& 0.08 &  1.31& 0.08 &  1.04& 0.07 &  1.83& 0.11 &  1.19& 0.08 &  1.27& 0.09 &  1.15& 0.08 \\
$c$ & 0.11 & 0.05 & 0.08 & 0.05 & 0.08 & 0.05 & 0.07 & 0.05 & 0.16 & 0.05 & 0.16 & 0.05 & 0.16 & 0.05 \\
\enddata
\tablenotetext{a}{Relative to \Hb.}
\tablenotetext{b}{Spatial integration over stationary position.}
\end{deluxetable}

\begin{deluxetable}{lllcl}
\footnotesize
\tablecaption{KBFM \protect\hii\ regions \label{tblKBFM}}
\tablewidth{0 pt}
\tablehead{
\colhead{Object} & \colhead{Alt. ID} & \colhead{Sp. Type} &
\colhead{symbol} & \colhead{Reference\tablenotemark{a}}
}
\startdata 
\cutinhead{Galaxy}
S212    & &  O7 f & 3 & HM90 \\
S237    & &  B0 V & 4 & HM90 \\
M42	& NGC 1976 &  O7  & 3 & Oudmaijer {\etal}(1997) \\
S255    & IC 2162 &  B0 V & 4 & HM90 \\
S271    & &  B0 V & 4 & HM90 \\
S288    & &  O9: V & 4 & HM90 \\
RCW 5   & S298 &  WN4	& 1 & Esteban {\etal}(1993) \\
RCW 8   & S305 &  B0 V & 4 & HM90 \\
S307    & &  B0 V & 4 & HM90 \\
RCW 48  & NGC 3199 &  WN5	& 1 & Esteban {\etal}(1993) \\
S148    &  & B0 V & 4 &  HM90 \\
S152    &  & O9 V	& 4 &  HM90 \\
S156    & IC 1470 & O6.5 V: & 3 & HM90 \\
\cutinhead{LMC}
DEM L34 & N11 B     & O3 & 2 & Parker {\etal}(1992) \\
DEM L152 & N44      & O5 & 3 & Oey \& Massey (1995) \\
DEM L174 & N138 D,B  & WN4p & 1 & Breysacher (1981) \\
DEM L192 & N51 D    & WC5, O4 III	& 1 & Oey \& Smedley (1998) \\
DEM L231 & N57 C     & WN4	& 1 & Breysacher (1981) \\
DEM L263 & 30 Dor   & WN4.5 & 1 & Massey \& Hunter (1998) \\
\enddata
\tablenotetext{a}{HM90 = Hunter \& Massey (1990)}
\end{deluxetable}

\begin{deluxetable}{lcc}
\footnotesize
\tablecaption{Empirical calibration of  \protect\tstar\ 
	diagnostics\tablenotemark{a} \label{calib}}
\tablewidth{0 pt}
\tablehead{
%\colhead{Sp. type} & \tstar\tablenotemark{b}(K) & 
\colhead{Sp. type} & 
\colhead{log \neiii\lam3869/\Hb} & \colhead{log \etap} 
}
\startdata 
WR &  $> -0.6$ & $< 0.2$ \\
O3 -- O4 & --0.9 to --0.6 & \nodata \\
O5 -- O8 &  $ > 0$, to --0.9 & \nodata \\
O9 and later & 0 & $> 1.0$ \\

%WR & $\gtrsim 55,000$ & $> -0.6$ & $< 0.2$ \\
%O3 -- O4 & 47,000 -- 54,000   & --0.9 to --0.6 & \nodata \\
%O5 -- O8 & 36,000 -- 47,000 &  $ > 0$, to --0.9 & \nodata \\
%O9 and later & $< 36,000$ & 0 & $> 1.0$ \\
\enddata
\tablenotetext{a}{For LMC metallicity, based on Figure~\protect\ref{integ1}.}
\end{deluxetable}

% Now comes the reference list.  In this document, we used \cite to call
% out citations, so we must use \bibitem in the reference list, which
% means we use the LaTeX thebibliography environment.  Please note that
% \begin{thebibliography} is followed by a null argument.  If you forget
% this, mayhem ensues, and LaTeX will say "Perhaps a missing item?" when
% you run it.  Do not call us, do not send mail when this happens.  Put
% the silly {} after the \begin{thebibliography}.
%
% Each reference has a \bibitem command to define the citation format
% to be placed in the text (in []) and the symbolic tag used for 
% cross referencing (in {}).
%
% See sample1.tex, or the AASTeX guide, for an alternative to the \cite-
% \bibitem command.

\clearpage

\textheight=8.4in
\topmargin=0in
\headheight=.15in

% \twocolumn
% |

% And finally, we must deal with the figures.  There are three figures
% associated with this manuscript; two figures are Encapsulated
% PostScript (EPS) files.  The third figure is a grey scale figure that does
% not exist in EPS form.
%
% Authors have three options for including figure information within a 
% manuscript.  Not all the options may be acceptable by the target Journal - be
% sure to look at the appropriate submission instructions, electronic or 
% otherwise.
%
% Option 1.  Using this option, only the figure captions are included in the
% main body of the manuscript.  The figure captions must start on a new page.
% The captions are generated with the \figcaption[]{} command: the first 
% argument is optional, if you put something in there, put the name of the 
% EPS file that goes with the caption; the second argument is the figure 
% caption itself, and may include a \label command.  The \figcaption command
% generates the figure numbers.  This option is acceptable for all manuscript
% submissions.

\end{document}